\documentclass[prd,aps,floats,twocolumn,nofootinbib]{revtex4}

\usepackage{commath}
\usepackage{appendix}
\pdfoutput=1
\usepackage[usenames, dvipsnames]{color}
\usepackage{graphicx, array}
\usepackage{multirow}
\usepackage[mathscr]{euscript}
\usepackage{graphics}
\usepackage{graphics,epsfig}
\usepackage{amssymb}
\usepackage{amsmath}
\usepackage{mathtools}
\usepackage{ulem}
\usepackage{multirow}
\usepackage{subfigure}
\usepackage{bm}
\usepackage{txfonts}
\usepackage{cancel}
\usepackage{url}
\usepackage{hyperref}
\usepackage{ulem}

\newcolumntype{C}[1]{>{\centering\arraybackslash}m{#1}}

\def\be{\begin{equation}}
\def\ee{\end{equation}}
\def\bi{\begin{itemize}}
\def\ei{\end{itemize}}
\def\ben{\begin{enumerate}}
\def\een{\end{enumerate}}
\def\bt{\begin{tabular}}
\def\et{\end{tabular}}
\def\bc{\begin{center}}
\def\ec{\end{center}}

\def\bea{\begin{eqnarray}}
\def\eea{\end{eqnarray}}

\def\sO{\mathcal{O}}
\def\M{M_{\nu}}
\def\s8{\sigma_{8}}
\def\Om{\Omega_{m}}

\let\oldhat\hat
\renewcommand{\vec}[1]{\boldsymbol{\mathbf{#1}}}
\renewcommand{\hat}[1]{\oldhat{\boldsymbol{\mathbf{#1}}}}

\begin{document}

\input{epsf}

\title{Fisher's Mirage: Noise Tightening of Cosmological Constraints in Simulation-Based  Inference}
\author{Christopher Wilson and Rachel Bean}

\affiliation{Department of Physics, Cornell University, Ithaca, New York 14853, USA.}
\affiliation{Department of Astronomy, Cornell University, Ithaca, New York 14853, USA.}

\begin{abstract}
We systematically analyze the implications of statistical noise within numerical derivatives on simulation-based Fisher forecasts for large scale structure surveys. Noisy numerical derivatives resulting from a finite number of simulations, $N_{sims}$, act to bias the associated Fisher forecast such that the resulting marginalized constraints can be significantly tighter than the noise-free limit. We show the source of this effect can be traced to the influence of the noise on the marginalization process. Parameters such as the neutrino mass, $\M$, for which higher-order forward differentiation schemes are commonly used, are more prone to noise; the predicted constraints can be akin to those purely from a random instance of statistical noise even using $(1\mathrm{Gpc}/h)^{3}$ simulations with $N_{sims}=500$ realizations. We demonstrate how derivative noise can artificially reduce parameter degeneracies and seemingly null the effects of adding nuisance parameters to the forecast, such as HOD fitting parameters. We mathematically characterize these effects through a full statistical analysis, and demonstrate how confidence intervals for the true noise-free, $N_{sims} \rightarrow \infty$, Fisher constraints can be recovered even when noise comprises a consequential component of the measured signal. The findings and approaches developed here are important for ensuring simulation-based analyses can be accurately used to assess upcoming survey capabilities.

\end{abstract}

\maketitle

\section{Introduction}

In the last few decades, cosmology has experienced a transformative period, with remarkable advancements in our understanding of the universe's fundamental components and dynamics. This era has been marked by increasingly precise constraints on the Cosmological Standard Model \cite{plank2020}, revealing deeper insights into the nature of primordial inflation and the intricate properties of matter, including the elusive presence of dark energy. These developments have been driven by a plethora of cosmological observations, providing distinct constraints not only on dark matter and baryonic matter but also on the intriguing properties of both relativistic and massive neutrinos. 

Neutrinos uniquely present unresolved questions for both the Standard Model of cosmology and the Standard Model of particle physics \cite{Gerbino:2022nvz}. Of particular interest in both the cosmology and particle physics communities is the sum of the neutrino masses, $\M = \sum m_{\nu}$, and the neutrino mass hierarchy. Within particle physics, any non-zero neutrino masses are an indication of physics beyond the Standard Model, with the exact value of $\M$ providing clues on how the Standard Model should be extended \cite{Arg_elles_2023,Qian_2015}. Within cosmology, the neutrino mass sum (henceforth just neutrino mass) is a key cosmological parameter as massive neutrinos impact the evolution of structure growth, and leave signatures in both the cosmic microwave background (CMB) and large scale structure (LSS), in galaxies and clusters of galaxies e.g. \cite{dvorkin2019neutrino,Lesgourgues_2012,Shoji:2010hm}. Ground based neutrino oscillation experiments have determined a lower bound of $\M \simeq 0.06\ \mathrm{eV}$, but are insensitive to the absolute neutrino mass scale \cite{Fukuda_1998,Ahmad_2002}. From the perspective of the particle physics community, tighter constraints on $M_{\nu}$ could differentiate between the normal and inverted neutrino mass hierarchies. Given that the Standard Model currently lacks a definitive mechanism for how neutrinos acquire mass, experimentally distinguishing between these hierarchies is crucial \cite{Arg_elles_2023,Qian_2015}. Determining the correct mass hierarchy is of paramount importance to the theoretical physics community as it could reveal new insights into the fundamental properties of neutrinos and facilitate extensions to the Standard Model.

Cosmological constraints on neutrinos provide a
complement to ground-based oscillation experiments by allowing researchers to probe the absolute neutrino mass scale \cite{TopicalConvenersKNAbazajianJECarlstromATLee:2013bxd}. Specifically for large scale structure, various observables such as power spectra \cite{Saito_2008,PhysRevLett.126.011301,Tanseri_2022}, Lyman-alpha forest measurements \cite{Palanque_Delabrouille_2015,Yeche:2017upn,Y_che_2017,ivanov2024fundamental} 21cm line observations \cite{Oyama_2013}, weak lensing \cite{DES:2022ccp,Abbott_2023}, and cosmic void related statistics such as the void size function \cite{Bayer_2021} may be employed to constrain $\M$ alongside an array of other cosmological parameters, such as $\Om$ or $\s8$. CMB temperature and polarization (B-mode) correlations provide complementary constraints  \cite{Ivanov:2019hqk,SPT:2023jql,ACT:2023kun}. Planck CMB in combination with SDSS/BOSS BAO measurements currently constrains $\M < 0.12\ \mathrm{eV}$ with $95\%$ confidence, while simultaneously constraining $\Om = 0.315 \pm 0.007$ and $\s8 = 0.810 \pm 0.006$ with $68\%$ confidence \cite{plank2020}.

We are entering an unprecedented era in which new surveys are commencing with greater volume, depth and precision;  multiple, complementary tracers will be observed causing datasets to grow dramatically in the coming decade. For LSS surveys this includes the Dark Energy Spectroscopic Instrument (DESI)\footnote{\url{https://www.desi.lbl.gov/}}\citep{DESICollaboration2016}, the Vera C. Rubin Observatory Legacy Survey of Space and Time (LSST)\footnote{\url{https://www.lsst.org/}}\citep{Ivezic2019}, ESA/NASA Euclid Space Telescope \footnote{\url{https://sci.esa.int/web/euclid}} \citep{Laureijs2011},  SPHEREx \footnote{\url{http://spherex.caltech.edu/}}\cite{Dore2014}, the Prime Focus Spectrograph (PFS)\footnote{\url{https://pfs.ipmu.jp/}}\citep{Takada2014}, and  the Nancy Grace Roman Space Telescope \footnote{\url{https://roman.gsfc.nasa.gov/}}\citep{Spergel2015}. Complementary to this are next-generation CMB experiments: the Simons Observatory (SO)\footnote{\url{https:/simonsobservatory.org/}}\citep{SimonsObservatory2019} and CMB-S4\footnote{\url{https://cmb-s4.org/}} \citep{Abazajian2016} which will provide higher precision temperature and B-mode polarization data with high fidelity CMB lensing. Together, these will truly allow cosmology and astronomy more broadly to enter the era of big data with the promise of rapid, potential order of magnitude, improvements in cosmological constraints including on the neutrino mass sum e.g. \cite{Font-Ribera:2013rwa,Pan:2015bgi,Mishra-Sharma:2018ykh,Chudaykin:2019ock, Euclid:2024imf}. 

This era presents both opportunities and challenges for cosmologists. Large and precise datasets, with complementary observables and covering multiple cosmic epochs, hold the promise of revealing new insights about the universe. Concurrently, the approaches to both predict survey capabilities and extract valuable information from the observed data need to have comparable requisite precision. The sheer volume and complexity of the data also necessitates efficient and robust methods for extracting valuable information, and then further extracting meaningful understanding. In many cases, these latter steps prove quite challenging. For many observables of cosmological interest, analytical models are insufficient to incorporate all relevant physical effects and survey properties to make accurate predictions pertaining to real world data, and thus large scale computer simulations must be used instead. 

The landscape of large scale cosmological simulations has evolved rapidly  (see e.g. \cite{Angulo:2021kes} for a review). Examples of N-body simulation suites and their derived products include:  mock galaxy catalogues designed to specifically mirror next generation surveys such as the Euclid Flagship simulations \cite{Euclid:2024few} and Rubin LSST Data Challenge 2 (DC2) simulations \cite{LSSTDarkEnergyScience:2020oya}; simulations that incorporate the effects of neutrinos e.g. \cite{Castorina:2015bma,Wright:2017dkw,Liu:2017now,Villaescusa-Navarro:2017mfx,Euclid:2022qde,Banerjee_2016}; and those which include multiple realizations over extended cosmological parameter ranges to discern differences across cosmological models e.g. HADES \cite{Villaescusa-Navarro:2017mfx}, to enable emulator development, e.g. Mira-Titan simulations \cite{Heitmann:2015xma} and to facilitate Fisher matrix analyses and machine learning inference e.g. the Quijote/Molino simulations \cite{Villaescusa_Navarro_2020,Makinen_2022}.

Whether computing simple cosmological observables analytically from Boltzmann code such as CAMB \cite{Lewis_2000} or using N-body simulations to address the complexities of more realistic cosmological data, Fisher forecasting has become a standard approach within the field of cosmology and astronomy more broadly \cite{PhysRevD.105.103534,Joachimi_2010,Bailoni_2017,fisherpaper2023,bhandari2021fisher,refId0,coulton2023estimate,Coulton_2023,PhysRevD.77.042001,Park_2023}. This method leverages the Fisher information matrix to estimate the precision with which model parameters can be measured, offering a balance between computational efficiency and analytical insight. Although more accurate, Monte Carlo Markov Chains (MCMC) are significantly more labor-intensive, requiring extensive computational resources to thoroughly explore and characterize the multi-parameter space. Fisher forecasting, on the other hand, allows for quick and reliable assessments, making it a valuable tool for preliminary analyses and guiding more detailed studies with MCMC when higher precision is necessary.

One risk with Fisher analyses is their apparent simplicity. Given a set of observables, a covariance matrix, and a set of numerical derivatives of each observable with respect to each parameter, the Fisher information matrix may be simply calculated and subsequently inverted to derive constraints on parameters of interest. While extensive theoretical work has established criteria for the convergence of the observable covariance matrix, as well as workarounds for cases of ill-convergence e.g. \cite{PhysRevD.77.042001,Hartlap_2006,Taylor_2013,PhysRevD.88.063537,Sellentin_2015,sellentin2016quantifying},  few studies have focused on the sensitivity of results to the convergence of numerical derivatives when using simulations \cite{coulton2023estimate}. 

In this paper we show that significant care and attention needs to be given when using simulations in Fisher analyses.  Specifically, we outline the challenges associated with statistical noise within numerical derivatives, and establish well-motivated convergence criteria for determining when Fisher forecast parameter constraints may be considered as cosmological in origin, rather than the results of statistical noise.

The paper is structured as follows: In section \ref{sec:Summary}, we provide a summary of our main results. In section \ref{sec:bkg}, we provide a summary of the simulations and cosmological observables considered, a brief technical review of the Fisher information matrix and Fisher forecasting, and an overview on the various numerical differentiation schemes considered and how we characterize their associated statistical noise. In section \ref{sec:noise_Fisher}, we apply our anaylsis first to observables based the underlying particles and dark matter halos, and then to HOD derived mock galaxies, focusing first on $\M$ and later extending our analysis to non-$\M$ parameters. In section \ref{sec:conclusion}, we summarize our findings, and discuss the implications of our results for analyses related to upcoming surveys. 

\section{Summary Of Results}
\label{sec:Summary}

We provide a powerful analytical framework to understand how statistical noise, arising from averaging and differencing realizations over a finite number of simulations ($N_{sims}$), can significantly and systematically tighten marginalized parameter constraints compared to their true, noise-free cosmological values, in the limit of $N_{sims}\rightarrow\infty$. 

Our framework characterizes both the severity of the effects of noise, including a focus on commonly used observables such as the real space dark matter halo power spectrum, and proposes mitigation techniques to reduce its impact.

We propagate noise, well-modelled as a multivariate Gaussian random variable, through the Fisher forecast formalism. The effect on the marginalized Fisher constraints for a parameter $\theta$, can be expressed as a random variable, $\mathscr{F}^{Noise}_{\theta \theta}$, which explicitly provides the difference between the true ($N_{sims} \rightarrow \infty$) and measured ($N_{sims}$ finite) inverse square constraints, $\sigma^{-2}_{\theta}$ and $\bar{\sigma}^{-2}_{\theta}$ respectively,   
\begin{equation}
\bar{\sigma}^{-2}_{\theta} = \sigma^{-2}_{\theta} + \mathscr{F}^{Noise}_{\theta \theta}.
\label{eq:introEq}
\end{equation}
While noise on the simulation-based numerical derivatives is 0-mean, $\mathscr{F}^{Noise}_{\theta \theta}$ is almost exclusively positive and can provide a significant or dominant contribution to $\bar{\sigma}^{-2}_{\theta}$. As a result, the measured constraints are 
artificially noise-tightened compared to their true cosmological values, potentially with $\bar{\sigma}_{\theta} \ll \sigma_{\theta}$.

While the assessment of noise in the derivatives is an important consideration for all parameters, we find it is most acute for parameters,  such as $\M$, for which higher-order differentiation schemes are commonly used to calculate one-sided derivatives, i.e., those that difference multiple values to only one side of the fiducial value, rather than differencing values symmetrically to either side. The higher-order schemes are used to provide more systematically accurate derivative estimates, thereby compensating for the lack of symmetry in the one-sided calculation.
We find, however, that this must be balanced against a far greater susceptibility to statistical noise than simple first-order finite difference derivatives. We find this to be true for both two and three-point observables, such as the halo or HOD mock galaxy derived power spectrum and bispectrum.
 
We provide a complete analysis of the effects of $\mathscr{F}^{Noise}_{\mu \nu}$, the noise correction to the marginalized Fisher information matrix, on both $1D$ constraints and $2D$ joint confidence ellipses between pairs of cosmological parameters $\theta^{\mu}$ or $\theta^{\nu}$. Analytic expressions to compute both the expectation and variance of $\mathscr{F}^{Noise}_{\mu \nu}$
are provided in limits of physical interest: one in which the noise is exclusively concentrated in the parameters to be constrained, and a more general perturbative expansion to leading order in $1/N_{sims}$. We show that in addition to tightening 1D confidence intervals, statistical noise also reduces the observed parameter degeneracies relative to the intrinsic cosmological values, which is relevant for understanding the total influence of noise on both 1D confidence intervals for other non-noisy parameters and joint 2D confidence ellipses. For noise dominated constraints, with $\bar{\sigma}^{-2}_{\theta} \simeq \mathscr{F}^{Noise}_{\theta \theta}$, the degeneracy breaking can provide the false illusion that the forecasted constraint $\bar{\sigma}_{\theta}$ is completely insensitive to the effects of additional nuisance parameter marginalization. 

Our analysis provides an explicit paradigm to distinguish between when the value of a constraint may be confidently ascribed to the cosmological signal, or remains in a noise-dominated or noise-influenced regime. We provide a statistically well-grounded approach to determine the $68\%$ confidence interval within which the true cosmological constraint may be assumed to lie, even in regimes where the noise contribution to the measured constraint remains on roughly equal footing with the cosmological signal. Since we quantify all of these effects of noise as a function of $N_{sims}$, our work also allows for the estimation of the number of simulations needed before any arbitrary level of desired precision within the parameter constraints may be attained for a given observable. 

Finally, our methods to access the effects of noise on parameter constraints are completely general, and independent of the specific parameters, differentiation schemes, and observables being considered. While our work is often highly mathematical, we intend to provide an intuitive geometric understanding of the varied effects of statistical noise through the framework of analytic marginalization. The findings could have implications for simulation-based inference more broadly.

\section{Background}
\label{sec:bkg}

The cosmological simulations and the observational probes used in this work are described in sections \ref{sec:sims} and \ref{sec:cosmoprobes} respectively. An overview of the Fisher forecasting approach is given in section  \ref{sec:Fisherintro}. A summary of the numerical differentiation schemes employed and how uncertainties in the derivatives are characterized are given sections \ref{sec:num_derivs} and \ref{sec:deriv_uncert}.

\subsection{The Quijote Simulations}
\label{sec:sims}

We employ the Quijote simulations \cite{Villaescusa_Navarro_2020}, and the related Molino suite of HOD mock galaxy catalogs \cite{bispectrumII}, to study various aspects of simulation-based Fisher forecasts.

Each realization of the Quijote simulations is comprised of $512^{3}$ dark matter particles (and massive neutrinos when relevant) in a box of side length $L=$1Gpc$/h$ evolved from initial conditions at a redshift of $z$=127 down to $z$=0. These simulations feature 15000 realizations of the fiducial cosmology $\theta^A$ = $\{\M (\mathrm{eV}),\Om,\s8,\Omega_{b},h,n_{s}\} = \{0,0.3175,0.834,0.049,0.6711,0.9624\}$ which may be used to calculate the observable covariance matrices. Quijote also features smaller sets of simulations with modified cosmological parameters so that numerical derivatives for Fisher analyses may be calculated. Excluding the neutrino mass, Quijote features sets of 500 simulations each where a single parameter $\theta^{A}$ has been either increased or decreased by an amount $\Delta \theta^{A}$, so that two-sided finite difference numerical derivatives of observables with respect to cosmological parameters may be calculated. For the non-$\M$ parameters, we have $\Delta\theta^A = \{\Delta\Om,\Delta\s8,\Delta\Omega_{b},\Delta h,\Delta n_{s}\}=\{0.01,0.015,0.002,0.02,0.02\}$. Since the fiducial cosmology features $\M=0$, and a negative neutrino mass is unphysical, Quijote also features 500 simulations for each of three scenarios with $\M>0$, specifically $\M$(eV) = $\{0.1,0.2,0.4\}$, so that $\M$ numerical derivatives may be calculated using a variety of one-sided, or forward-differentiation, schemes.

All simulations involving the fiducial cosmology or adjustments to non-$\M$ cosmological parameters have been created using 2LPT initial conditions \cite{Villaescusa_Navarro_2020}. Simulations involving changes to $\M$, along with a further 500 simulations at the $\M=0$ fiducial cosmology, have initial conditions generated using the Zel'dovich approximation so that $\M$ numerical derivatives may be calculated in a consistent way. Thus, derivatives featuring $\M$ are obtained by differencing results from simulations using the Zel'dovich approximation while the fiducial covariance and observable derivatives using non-$\M$ parameters are calculated from the 2LPT suite of sims.

From the dark matter particles, halo catalogs are created using a Friends-of-Friends algorithm \cite{1985ApJ...292..371D} with the linking length parameter $b$=0.2. Only CDM particles, and not neutrinos, are counted when creating the Quijote halo catalogs  \cite{Villaescusa_Navarro_2020}. Due to their limited resolution, the fiducial resolution Quijote halo catalogs are complete only down to $M_{min} \simeq 3 \times 10^{13} h^{-1} M_{\odot}$. Since some of our cosmological parameters implicitly change the mass of the dark matter particles within the simulations, we impose a minimum dark matter particle number rather than a mass cutoff on the halo catalogs. Thus, we consider only halos of $N_{parts}\geq 46$ in our halo based analysis, which corresponds to $M_{cut}=3.0\times 10^{13}  h^{-1} M_{\odot}$ for the fiducial cosmology. We have verified that none of our core results depend on the exact $N_{parts}$ cutoff used.

From the Quijote halo catalogs, the Molino suite of mock galaxy catalogs is constructed by applying the 5-parameter HOD fitting function introduced by Zheng et al. \cite{Zheng_2007},
\bea
\left< N_{\mathrm{cen}}(M) \right> &=& \frac{1}{2}\left[ 1+\mathrm{erf}\left(\frac{\mathrm{log} M - \mathrm{log} M_{min}}{\sigma_{\mathrm{log}M}} \right)     \right]\\
\left<N_{\mathrm{sat}} (M) \right> &=& \left< N_{\mathrm{cen}} \right> \left(     \frac{M-M_{0}}{M_{1}}    \right)^{\alpha}.
\label{eq:HOD}
\eea
Here $\left< N_{\mathrm{cen}}(M) \right>$ and $\left<N_{\mathrm{sat}} (M) \right>$ denote, respectively, the average numbers of central and satellite galaxies that a halo of mass $M$ usually accommodates. The model parameters are typically obtained by fitting the resulting galaxy catalog's 2-point function against observational survey data. For the Molino galaxy catalogs, the fiducial HOD parameters are chosen to align with the high luminosity samples from the Sloan Digital Sky Survey \cite{bispectrumII}. Explicitly, the fiducial HOD parameters are given as $\{\alpha, \mathrm{log}M_{0},\mathrm{log}M_{1},\mathrm{log}M_{Min},\sigma_{\mathrm{log}M}\} = \{1.1,14,14, 13.65,0.2\}$.

Within the Molino suite, the fiducial HOD model is applied a single time to each of the 15000 fiducial cosmology Quijote simulations so that any observable covariance matrix will reflect the probabilistic nature of (\ref{eq:HOD}). For the simulations used to calculate numerical derivatives, the HOD model is applied 5 times to each simulation so that $N_{sims}=\mathrm{500}$ results in 2500 mock galaxy catalogs.
The suite also features 2500 mock galaxy catalogs at the fiducial cosmology with increased or decreased values of each HOD parameter so that they may be marginalized over as part of any Fisher forecast. Explicitly, the changes to these HOD parameters are $\Delta \theta^{A} = \{\Delta \alpha, \Delta \mathrm{log} M_{0},\Delta \mathrm{log} M_{1},\Delta\mathrm{log}M_{Min},\Delta\sigma_{\mathrm{log}M}\}$ = \{0.2,0.2,0.2,0.05,0.2\}. For more information, we refer readers to \cite{Villaescusa_Navarro_2020} and \cite{bispectrumII} for the Quijote and Molino suites, respectively. 

\subsection{Cosmological Probes}
\label{sec:cosmoprobes}

In this paper, we focus on commonly considered cosmological probes resulting from 2-point or 3-point correlation functions with a variety of tracers in either real or redshift space. We first consider the real space matter power spectrum as calculated from the underlying particle distribution and halos, $P_{parts}(k)$ and $P_{halos}(k)$ respectively. We use the usual definitions
\begin{align}
P(\vec{k})&=\delta(\vec{k}) \delta(-\vec{k})\\
P(k)&=\left<\delta(\vec{k}) \delta(-\vec{k}) \right> 
\label{eq:RSPS}
\end{align}
where $\delta(\vec{k})$ is the Fourier transform of the real space density contrast $\delta(\vec{x})=\left(n(\vec{x})-1\right)/\bar{n}$ for tracer number density $n(\vec{x})$ and average tracer number density $\bar{n}$. For both the halos or particles as tracers, $n(\vec{x})$ is calculated on a grid with $N_{grid}=360$ points per side with the Piecewise Cubic Spline (PCS) mass assignment scheme, which has been shown to reduce the aliasing contribution to these quantities \cite{Sefusatti_2016}. The $\left<...\right>$ in (\ref{eq:RSPS}) refers to an average in $k$-space of all modes falling within the $k=|\vec{k}|$ bin of width $\Delta k$ equal to the fundamental k-mode $k_{f} = 2 \pi /L$ where $L=1\ \mathrm{Gpc}/h$ is our simulation box size. We consider a maximum $k_{max}=0.5\ h/\mathrm{Mpc}$, which is a common choice in the literature \cite{Bayer_2021,bispectrumI,bispectrumII}, and amounts to $79$ evenly spaced $k$ bins for our real space power spectrum. Additionally in order to facilitate a more direct comparison between $P_{parts}(k)$ and $P_{halos}(k)$, all $P_{parts}(k)$ from simulations for both derivative and covariance matrix calculations are scaled by a fixed bias factor of $b^{2}=2$.

Using HOD-derived mock galaxies as tracers from the Molino suite of catalogs, we extend our analysis to observables aligned with spectroscopic surveys: the monopole and quadrupole of the redshift space power spectrum $P_{0}^{g}(k)\ \&\ P_{2}^{g}(k)$, and the monopole of the redshift space galaxy bispectrum $B_{0}^{g}$. For the monopole ($\ell=0)$ and quadrupole $(\ell=2)$, we use \cite{Sefusatti:2015aex}
\begin{equation}
P^{g}_{\ell}(k) = \frac{2\ell +1}{2} \int^{1}_{-1} P^{g}(k,\mu) L_{\ell}(\mu) \mathrm{d}\mu 
\end{equation}
Where here $\mu$ is the cosine of the angle $\vec{k}$ makes with the axis along which the redshift space distortions have been implemented, and $L_{\ell}(\mu)$ is the $\ell^{th}$ Legendre polynomial. The superscript $g$ is used to indicate that HOD-derived mock galaxies have been used as tracers. We again use $\Delta k=k_{f}$ sized bins out to $k_{max}=0.5\ h/\mathrm{Mpc}$ for both the monopole and the quadrupole which again gives $79$ evenly spaced bins for each. When calculating either the real space power spectrum or the multipole moments of the redshift space power spectrum, we use the publicly available code package PYLIANS \footnote{https://github.com/franciscovillaescusa/Pylians}, with the aforementioned Piecewise Cubic Spline (PCS) mass assignment scheme.  

For the redshift space galaxy bispectrum $B_{0}^{g}$, we use
\cite{Sefusatti:2015aex}
\begin{align}
B_0^{g}(k_{1},k_{2},k_{3}) = &\frac{1}{V_{B}} \int_{k_{1}} \mathrm{d}^{3} q_{1} \int_{k_{2}} \mathrm{d}^{3} q_{2} \int_{k_{3}} \mathrm{d}^{3} q_{3} \bigg(\delta_{D}(\vec{q}_{123})\nonumber \\ 
 &\times \delta(\vec{q}_{1})\delta(\vec{q}_{2})\delta(\vec{q}_{3})\bigg) - B^{SN}_{0}
\end{align}
Where $\delta_{D}$ is the 3-dimensional Dirac-delta function, $\delta(\vec{q})$ is the Fourier transform of $\delta(\vec{x})$ described above still using $N_{grid}=360$ , $\vec{q}_{123} = q_{1} + q_{2} + q_{3}$, and each integral is over a spherical shell in $k$-space centered at $k_{i}$ with width $\delta k = 3k_{f}$. $V_{B}$ is a normalization factor, and $B^{SN}_{0}$ is a correction term for Poission shot noise, both given respectively by 
%
\begin{align}
V_{B} &= \int_{k_{1}} \mathrm{d}^{3} q_{1} \int_{k_{2}} \mathrm{d}^{3} q_{2} \int_{k_{3}} \mathrm{d}^{3} q_{3} \delta_{D}(\vec{q}_{123})\\
B^{SN}_{0}&= \frac{1}{\bar{n}}(P^{g}_{0}(k_{1}) + P^{g}_{0}(k_{2}) + P^{g}_{0}(k_{3})) + \frac{1}{\bar{n}^{2}}
\end{align}
We use the publicly available code pySpectrum\footnote{https://github.com/changhoonhahn/pySpectrum}
and consider for all $(k_{1},k_{2},k_{3})$ triangle configurations with $k_{max} < 0.5 h/\mathrm{Mpc}$ for a total of 1,898 configurations, as in \cite{bispectrumI,bispectrumII}.

We average over $N_{sims}=500$ simulated realizations for each cosmological model to calculate to calculate the various numerical derivatives $P_{parts}(k)$ and $P_{halos}(k)$. For  $P_{0}^{g}(k)$, $P_{2}^{g}(k)$ and $B^{g}_{0}$ for each of the 500 underlying simulations we also average over the 5 HOD instances and 3 independent lines of sight in the Molino simulations.

For all of the observables, $\sO^{\ i}$, listed above, the observable covariance matrix matrix, $C^{ij}$, from the simulations is calculated as 
\begin{equation}
C^{ij} =\frac{1}{N_{cov}} \sum_{n=1}^{N_{cov}}\left(\sO^{\ i}_{n}- \overline{\sO}{}^{\ i}\right) \left( \sO^{\ j}_{n} - \overline{\sO}{}^{\ j}\right).
\label{eq:covDef}
\end{equation}
where $\sO^{\ i}_{n}$ is the value of the $i^{th}$ observable as calculated from the $n^{th}$ simulation used for the covariance matrix, and $\overline{\sO}{}^{\ i}$ is the mean value of the $i^{th}$ observable as calculated from all $N_{cov}= 15000$ simulations available from either the Quijote or Molino suite. Note, we implicitly take into account, but do not explicitly write factors of $N-1$ instead of $N$ related to covariance calculations; $N$ is always large and therefore this difference is negligible. For redshift space observables, only one line of sight and one HOD application is used in the computation of $C^{ij}$ so as not to include artificial correlations between the observables. When $P_{parts}(k)$, $P_{halos}(k)$, $P_{0}^{g}(k)$, or $P_{2}^{g}(k)$ are used as the observable, the index $i$ refers to specific $k$-bins. When $B^{g}_{0}$ is the observable, $i$ refers to specific triangle configurations $(k_{1},k_{2},k_{3})$. 

\subsection{Fisher Forecasts}
\label{sec:Fisherintro}

The Fisher forecast is arguably one of the most widely used statistical inference tools in  astronomy, due to both its simplicity and ease of implementation. For observed data vectors, $\vec{\sO}=\{\sO^{\ i}\}$ and a set of model parameters, $\vec{\theta}=\{\theta^{A}\}$ for which we assume uniform priors, the posterior distribution of model parameters given $\vec{\sO}$ is proportional to the likelihood function $\cal{L}(\vec{\theta} \vert \vec{\sO})$ of the model parameters given the data. The Fisher information matrix, $F_{AB}$, is then given by
\begin{equation}
F_{AB} =  - \left< \frac{\partial^{2} \mathrm{ln}\cal{L}}{\partial \theta^{A} \partial \theta^{B}}\right>,
\label{eq:FisherFormalDef}
\end{equation}
Under the assumption that the data $\vec{\sO}$ comes from a multivariate Gaussian distribution, for which the observable covariance matrix $C^{ij}$ is independent of the model parameters, the Fisher information matrix takes the simple form 
\begin{equation}
F_{AB} = d\sO^{\ i}_A C^{-1}_{ij} d\sO^{j}_B,
\label{eq:FisherDef}
\end{equation}
where $d\sO^{\ i}_{A}\equiv \frac{\partial \sO^{\ i}}{\partial \theta^A}$ is the partial derivative of the observable $\sO^{\ i}$ with respect to parameter $\theta^{A}$, evaluated for the fiducial set of parameters. 

The probability distribution for the parameters is given by the multivariate Gaussian distribution 
\begin{equation}
P(\vec{\theta}) \propto \mathrm{exp}\left(-\frac{1}{2} x^A F_{AB} x^{B}\right)
\label{eq:paramPdef}
\end{equation}
where $x^{A} = \theta^{A} - \theta^{A}_{0}$ is the change in parameter $\theta^{A}$ away from its fiducial value $\theta^{A}_{0}$, and
the exponent of (\ref{eq:paramPdef}) is related to $\chi^2$ by
\begin{equation}
\chi^2 = x^{A} F_{AB} x^{B}.
\label{eq:chi2}
\end{equation}

In considering the constraints on a single target parameter, $\theta^ A$, it is informative to consider both the unmarginalized 1$\sigma$ constraints, with $\chi^2=1$, given by
\begin{equation}
\sigma_{A,unmarg} = \sqrt{1/F_{AA}}
\label{eq:unmargSigma}
\end{equation}
and the constraints when fully marginalized over all parameters given by
\begin{equation}
\sigma_{A} = \sqrt{(F^{-1})^{AA}},
\label{eq:CramerRao}
\end{equation}
which reflects the Cramer-Rao bound \cite{cramer1946mathematical, Rao1992}.

Generalizing, we can consider joint constraints on a subset of target parameters after marginalizing over the others. For this, the $N$ cosmological parameters are split into $K$ ($K<N$) target parameters, for which most often we will consider $K=$1 or 2, and $N-K$ non-target parameters. Target parameters will be labeled using lower case Greek indices $\{\mu, \nu, \alpha, \beta, ...\}$ while lower case Latin indices from the set $\{a,b,c,d...\}$ denote the remaining set of non-target parameters. Arbitrary parameters from either group will continue to be denoted by upper case Latin indices from the set $\{A,B,C,D...\}$.

We can think of $F_{AB}$ as having the following block matrix structure
\begin{equation}
\left[F_{AB}\right] = \left[\begin{matrix} F_{\mu \nu} & F_{\mu b} \\ F_{a \nu} & F_{ab} \end{matrix} \right].
\end{equation}
To characterize constraints purely in terms of the target parameters, we can define a $K\times K$ marginalized Fisher information matrix, $\mathscr{F}_{\mu \nu}$, such that
\bea
\chi^2= x^{\mu}\mathscr{F}_{\mu \nu} x^{\nu}.
\label{eq:chiSimp_main}
\eea
Just as $F_{AB}$ provides a positive definite quadratic form for all of the parameters in the analysis, $\mathscr{F}_{\mu \nu}$ provides a positive definite quadratic form for only the target parameters. 

$\mathscr{F}_{\mu \nu}$ can be calculated from the full $F_{AB}$ using the block matrix inversion theorem:
\begin{equation}
\mathscr{F}_{\mu \nu} \equiv \left(\left({(F^{-1})^{\alpha \beta}}\right)^{-1}\right)_{\mu \nu} =(F_{\mu \nu} - F_{\mu a} (f^{-1})^{ab} F_{b \nu}).
\label{eq:margFisher_main}
\end{equation}
where $f_{ab}$ is the $N-K \times N-K$ block of $F$, and $(f^{-1})^{ab}$ its $N-K \times N-K$ matrix inverse, denoted with upper indices, such that $(f^{-1})^{ab}F_{bc} = (f^{-1})^{ab}f_{bc} = \delta^{a}_{c}$. This is different than $(F^{-1})^{ab}$, which inverts the whole $N \times N$ matrix and then takes the relevant ${ab}$ components. 

The relationship between marginalized and unmarginalized errors for the target parameters can then be written as:
\begin{equation}
\sigma^{-2}_{\mu} = \sigma_{\mu,unmarg}^{-2} -  F_{\mu a} (f^{-1})^{ab} F_{b \mu},
\label{eq:margAndUnmarg_main}
\end{equation}
Note $a$ and $b$ are both summed over, while $\mu$ is not. 

Analytic marginalization \cite{Taylor_2010} provides an alternative approach to block matrix inversion to obtain (\ref{eq:margFisher_main}). When analytically marginalizing over parameters, instead of explicitly integrating $P(\vec{\theta})$ over the ``uninteresting" non-target parameters, $\{\theta^{a}\}$, the set of $\{\theta^{a}\}$ are adjusted to achieve the lowest possible $\chi^2$ in response to deviations of the target parameters $\{\theta^{\mu}\}$ from their fiducial values, thereby maximizing the likelihood for each configuration of the $\{\theta^{\mu}\}$ and implicitly defining the profile likelihood. Expanding (\ref{eq:chi2}), we have 
\begin{equation}
\chi^{2} = x^{\mu} F_{\mu \nu} x^{\nu} + 2 x^{\mu} F_{\mu a} x^{a} + x^{a} F_{a b} x^{b}
\label{eq:chiExpanded}
\end{equation}
We now think of the $x^{\mu}$ as fixed at some arbitrary value, and adjust the $x^{a}$ so that $\chi^2$ is minimized. This requires setting the gradient of (\ref{eq:chiExpanded}) with respect to all of the $x^{a}$ equal to 0, which gives
\begin{equation}
0 = 2 x^{\mu} F_{\mu a}  + 2F_{a b} x^{b}\  \rightarrow \  x^{a} =  -(f^{-1})^{ab}  F_{b \mu}  x^{\mu} 
\label{eq:grad_xa}
\end{equation}
Setting $x^{a} =  -f^{a b}  F_{b \mu}  x^{\mu}$ in (\ref{eq:chiExpanded}) gives  
\begin{equation}
\chi^2 = x^{\mu} (F_{\mu \nu} - F_{\mu a} (f^{-1})^{ab} F_{b \nu}) x^{\nu}
\label{eq:chiSimp}
\end{equation}
equivalent to (\ref{eq:chiSimp_main}) and (\ref{eq:margFisher_main}) when the underlying probability distributions are all assumed to be Gaussian, which is explicitly the case in this work. 

From the framework of analytic marginalization, especially in the case of a single target parameter with associated 1D constraints, $\chi^{2}$ may be thought of a measure of how accurately a target function, here the derivative $\frac{\partial \sO^{\ i}}{\partial \theta^{\mu}}$, is able to be reproduced as a linear combination from a set of basis functions $\left\{\frac{\partial \sO^{i}}{\partial \theta^{a}} \right\}$. This may be explicitly seen by re-writing \ref{eq:chiSimp} as
\begin{equation}
\chi^{2} = \displaystyle \min_{x^{a}}\left[\left(x^{\mu}d\sO^{i}_{\mu} - x^{a} d\sO^{i}_{a}\right)C^{-1}_{ij}\left(x^{\nu}d\sO^{i}_{\nu} - x^{b} d\sO^{j}_{b}\right)\right],
\label{eq:explicitLinearCombination}
\end{equation}
with the optimal $x^{a}$ in this case given by $x^{a} =  +(f^{-1})^{ab}  F_{b \mu}  x^{\mu}$ since we have switched the sign relative to \ref{eq:chiExpanded}, to emphasize that $x^{a} d\sO^{i}_{a}$ is trying to minimize $\chi^{2}$ by canceling against $d\sO^{i}_{\mu}$ as efficiently as possible. If the optimal linear combination of the non-target parameter derivatives is successful in reproducing the target parameter derivative, then the 1D marginalized constraint $\sigma_{\mu}$ will be larger (worse) than if the non-target parameters were less successful. We will find this framework  provides a useful intuitive perspective to view the effect noise on the derivatives in the analysis.

\subsection{Numerical Derivative Estimation}
\label{sec:num_derivs}

It is imperative to accurately estimate the derivative, $d\sO^{\ i}_A$, in order to accurately estimate the Fisher matrix, $F_{AB}$ in (\ref{eq:FisherDef}). These derivatives are typically estimated from either analytic modeling or from simulations. In the latter, for Fisher analyses, distinct simulations are created in which one parameter from the cosmological parameter vector is increased and then decreased from its fiducial value, while holding all others constant.  Given parameter $\theta^{A}$, we typically have $N_{sims}$ pairs of simulation realizations, each pair with the same initial conditions, where $\theta^{A}$ is first increased and then decreased from its fiducial value, which we respectively label $\theta^{A}_{+}$ and $\theta^{A}_{-}$. From these simulations, a symmetric finite differencing scheme may be employed to calculate $d\sO^{\ i}_A$ for the $n^{th}$ realization:
\begin{eqnarray}
D0:\hspace{0.cm}& \left(\frac{\partial \sO^{\ i}_{n}}{\partial \theta^{A}}\right) &= \frac{\sO^{\ i}_{n}(\theta^A_{+}) - \sO^{\ i}_{n}(\theta^A_{-})}{2\Delta \theta^A}
\label{eq:s1_pm}
\end{eqnarray}
where $\theta^{A}_{\pm}= \theta^{A}_{0}\pm \Delta \theta^{A}$, with $\Delta \theta^A$ the change in parameter $\theta^{A}$ away from its fiducial value, $\theta^{A}_{0}$, and $\sO^{\ i}_{n}$ is the cosmological observable, such as the power spectrum at a specific wave number, $P(k^{i})$, or the monopole or quadrupole thereof, $P_{0}(k^{i})$ or $P_{2}(k^{i})$, calculated from the $n^{th}$ realization only. 

There are a number of cosmological parameters of distinct interest, however, for which a two-sided derivative is not as simply obtained. In some instances particular parameters which are intrinsically defined as positive may have an extremely small or zero fiducial value. This includes the sum of the neutrino mass, $M_\nu$, on which we focus here, but also $r$, the inflationary tensor to scalar ratio and $f_{NL}$, the amplitude of non-Gaussianity from inflation and large scale structure formation.  In the case of such parameter models, a variety of one-sided forward differentiation schemes may be employed, in which the derivative at the fiducial value is estimated by considering differences between models at only increased or only decreased parameter values relative to the fiducial.

The Quijote simulations, used in this work, include three neutrino models in addition to the fiducial model. The baseline stepsize for the simulations (away from the fiducial neutrino mass value of $M_{\nu}=0$) is $\Delta M_{\nu}=0.1\ \mathrm{eV}$, with the suite including realizations with $M_{\nu}=\{0,\ 0.1,\ 0.2,\ 0.4\}$eV i.e. 
$M_{\nu}=\{0,\ \Delta M_\nu,\ 2\Delta M_\nu,\ 4\Delta M_\nu\}$. $N_{sims}$ realizations are created for each model. For the $n^{th}$ realization, the same initial conditions are used across each model. 

We may calculate the numerical derivative of any observable, $\sO^{\ i}_{n}$, with respect to neutrino mass using differences calculated using any one of the following equations: 
\begin{eqnarray}
D1_{a}:\hspace{0.cm}& 
\left(\frac{\partial \sO^{\ i}_{n}}{\partial M_{\nu}}\right) 
&= \frac{\sO^{\ i}_{n}(4\ \Delta M_{\nu}) - \sO^{\ i}_{n,0}}{4\ \Delta M_{\nu}}
\label{eq:s1_ppp}
\\
D1_{b}:\hspace{0.cm}&
\left(\frac{\partial \sO^{\ i}_{n}}{\partial M_{\nu}}\right) 
&= \frac{\sO^{\ i}_{n}(2\ \Delta M_{\nu}) - \sO^{\ i}_{0,n}}{2\ \Delta M_{\nu}}
\label{eq:s1_pp}
\\
D1_{c}:\hspace{0.cm}&
\left(\frac{\partial \sO^{\ i}_{n}}{\partial M_{\nu}}\right) 
&= \frac{\sO^{\ i}_{n}(\Delta M_{\nu}) - \sO^{\ i}_{0,n}}{\ \Delta M_{\nu}}
\label{eq:s1_p}
\\
D2_{a}:\hspace{0.cm}&
\left(\frac{\partial \sO^{\ i}_{n}}{\partial M_{\nu}}\right) 
&= \frac{- \sO^{\ i}_{n}(4\ \Delta M_{\nu}) + 4\sO^{\ i}_{n}(2\ \Delta M_{\nu}) - 3\sO^{\ i}_{ 0,n}}{4\ \Delta M_{\nu}} 
\label{eq:f2_p_pp}
\\
D2_{b}:\hspace{0.cm}& 
\left(\frac{\partial \sO^{\ i}_{n}}{\partial M_{\nu}}\right) 
&= \frac{- \sO^{\ i}_{n}(2\ \Delta M_{\nu}) + 4\sO^{\ i}_{n}(\Delta M_{\nu}) - 3\sO^{\ i}_{0,n}}{2\ \Delta M_{\nu}} \ \ \ \ \ \
\label{eq:f2_pp_ppp}
\end{eqnarray}
and
\begin{eqnarray}
D3:\hspace{0.cm}
\left(\frac{\partial \sO^{\ i}_{n}}{\partial M_{\nu}}\right) &=& 
\frac{1}{12\ \Delta M_{\nu}} \left[\sO^{\ i}_{n}(4\ \Delta M_{\nu}) - 12\ \sO^{\ i}_{n}(2\ \Delta M_{\nu}) \right.\nonumber
\\ && \left. + 32 \sO^{i}_{n}(\Delta M_{\nu})- 21 \sO^{i}_{0,n}\right]
\label{eq:f3}
\end{eqnarray}

Since the observables are calculated from multiple realizations of numerical simulations with different initial conditions, the value of a given $\sO^{\ i}_{n}$ will statistically vary across different realizations with the same set of cosmological parameters. The estimate of $\sO^{\ i}$ used to calculate the derivatives, $d\sO^{\ i}_A$,  in the Fisher forecast (\ref{eq:FisherDef}) would therefore reflect the mean value across all simulations with the appropriate parameter values. 

In reality, we have a finite number of sims, and from these can obtain a sample mean for the derivative,
\begin{equation}
    d\overline{\sO}{}^{\ i}_{A}= \frac{1}{N_{sims}} \sum_{n=1}^{N_{sims}}\frac{\partial \sO^{\ i}_{n}}{\partial \theta^A}
    \label{eq:meanDeriv},
\end{equation}
where $\partial\sO^{\ i}_n/\partial\theta^{A}$ is obtained using either $D0$ for those which admit two-sided derivatives or one of $D1_{a}-D3$ for the one-sided derivative parameters. We use the bar in (\ref{eq:meanDeriv}) to indicate the mean value as calculated from the data.

In the  $N_{sims} \rightarrow \infty$ ideal limit of this mean, the true cosmological values of the derivative would be obtained as the limit,
\begin{equation}
    d\sO^{\ i}_A \equiv \lim_{N_{sims}\to\infty} \frac{1}{N_{sims}} \sum_{n=1}^{N_{sims}}\frac{\partial \sO^{\ i}_{n}}{\partial \theta^A} \equiv \left\langle\frac{\partial \sO^{\ i}_{n}}{\partial \theta^A}\right\rangle
    \label{eq:meanDerivExp}
\end{equation}
Where the angled brackets indicate an expectation value over all possible cosmological initial conditions.  

\subsection{Uncertainties in Derivative Estimation}
\label{sec:deriv_uncert}

There are two distinct types of errors related to the derivative estimation that we need to consider in constructing the Fisher matrix. The first is the inherent error introduced by the limitations of the finite differencing methods themselves. We refer to these as systematic errors. The second is due to the variation of observables across the simulated realizations. We refer to these as statistical uncertainties, noise, or standard uncertainty of the mean. The second would disappear in the limit of infinite simulated realizations, while the first would remain.

In relation to the systematic errors, the differencing approaches in $D1_{a}-D3$, given in (\ref{eq:s1_p})-(\ref{eq:f3}), each provide an estimate of $d\sO^{\ i}_{\M}$ to differing degrees of accuracy. 

Just as $\frac{f(x + \Delta x) - f(x)}{\Delta x}$ differs from the true value of $\frac{\mathrm{d} f}{\mathrm{d}x}$ for a finite value of $\Delta x$, none of these derivative methods would provide the true value of $d\sO^{\ i}_{\M}$, even if an infinite number of simulations could be used to calculate the numerical derivatives. 

We will refer collectively to (\ref{eq:s1_ppp})-(\ref{eq:s1_p}) as first-order differentiation schemes, (\ref{eq:f2_pp_ppp}) and (\ref{eq:f2_p_pp}) as second-order, and (\ref{eq:f3}) as a third-order differentiation scheme. These names refer to the fact that the error in derivative estimates for (\ref{eq:s1_ppp})-(\ref{eq:s1_p}) are $\propto \Delta M_{\nu}^1$, the error in (\ref{eq:f2_pp_ppp}) and (\ref{eq:f2_p_pp}) is $\propto \Delta M_{\nu}^{2}$, and in (\ref{eq:f3}) is $\propto \Delta M_{\nu}^{3}$. As such, in the presence of perfectly predicted observables ($N_{sims} \rightarrow \infty$), we would expect the third-order scheme $D3$ to be the most accurate predictor of the derivatives, followed by the second order and than the first-order schemes.

Separate from the intrinsic accuracy of the differencing scheme in approximating the partial derivative at the fiducial value, we also need to consider that there will be standard uncertainties in estimating the mean of an observable inherent to the finite sample size of simulated realizations $N_{sims}$. 

For a finite sample size of simulations, (\ref{eq:s1_pm})- (\ref{eq:f3}) will each have different levels of statistical uncertainty due to the different sets of cosmologies considered, different terms present in the numerator, and different step sizes in the denominator. In much the same way one characterizes the uncertainty in observables $\sO^{\ i}$ in the fiducial cosmology through the use of a covariance matrix, we can similarly define a \textit{mean} derivative covariance matrix that characterizes the standard error on the mean derivatives:
\begin{eqnarray}
    \Sigma^{ij}_{AB} 
    &=&\frac{1}{N_{sims}^2} \sum_{n=1}^{N_{sims}}\left(\frac{\partial \sO^{\ i}_{n}}{\partial \theta^A}- d\overline{\sO}{}^{\ i}_{A}\right) \left(\frac{\partial \sO^{\ j}_{n}}{\partial \theta^B}- d\overline{\sO}{}^{\ j}_{B}\right).
    \label{eq:Sigma}
\end{eqnarray}
$\Sigma^{ij}_{AB}$ can be considered as an $MN \times MN$ flattened covariance matrix for $A,B=1,N$ parameters and $i,j=1,M$ observables, 
with $\sigma\left(d\overline{\sO}{}^{\ i}_{A} \right) = \sqrt{\Sigma^{ii}_{AA}}$ the standard error on $d\overline{\sO}{}^{\ i}_{A}$ relative to 
the true derivative, $d \sO^{\ i}_{A}$.

While the derivatives, $d \sO^{\ i}_{A}$, are considered the true values of the numerical derivatives in terms of their convergence as $N_{sims} \rightarrow \infty$, note that the values of, say,  $\left(d\sO^{\ i}_{\M}\right)_{D3}$ and $\left(d\sO^{\ i}_{\M}\right)_{D1_{a}}$, while each entirely free of statistical uncertainty, would not necessarily be equal to each other due to differing amounts of systematic error in each respective differentiation scheme.

We also note that while (\ref{eq:s1_pm}) is a ``two-sided" scheme and all of (\ref{eq:s1_ppp}) through (\ref{eq:f3}) are one-sided schemes, it is actually the case that from a statistical uncertainty point of view, $D1_{a}$, $D1_{b}$, and $D1_{c}$ are more similar to $D0$ then they are to $D2_{a}$, $D2_{b}$, or $D3$ due to the fact that all of $D0$, $D1_{a}$, $D1_{b}$, and $D1_{c}$ are all first order differentiation schemes. Phrased differently, all of  $D1_{a}$, $D1_{b}$, and $D1_{c}$ may be interpreted as a ``symmetric" derivative simply around a different fiducial $\M$ value. Thus, we may expect these first-order differentiation schemes to have similar statistical behaviors, only differentiated by the different steps sizes, throughout the rest of this work.

In summary, when calculating numerical derivatives using finite differencing schemes from a finite set of simulations, there are two separate sources of uncertainty which must be taken into account. The first is the inherent error introduced by the limitations of the particular finite differencing method itself, and the second is the statistical variation of the observable derivatives across the different simulated realizations.  As previously stated, we will differentiate these two respective errors by referring to the first as systematic errors,  and the second as statistical uncertainties, standard uncertainty on the mean, or most commonly, simply statistical noise or noise. The second would disappear in the limit of infinite simulated realizations, while the first would remain.

\begin{figure*}[!t]
\includegraphics[width=0.9\linewidth]{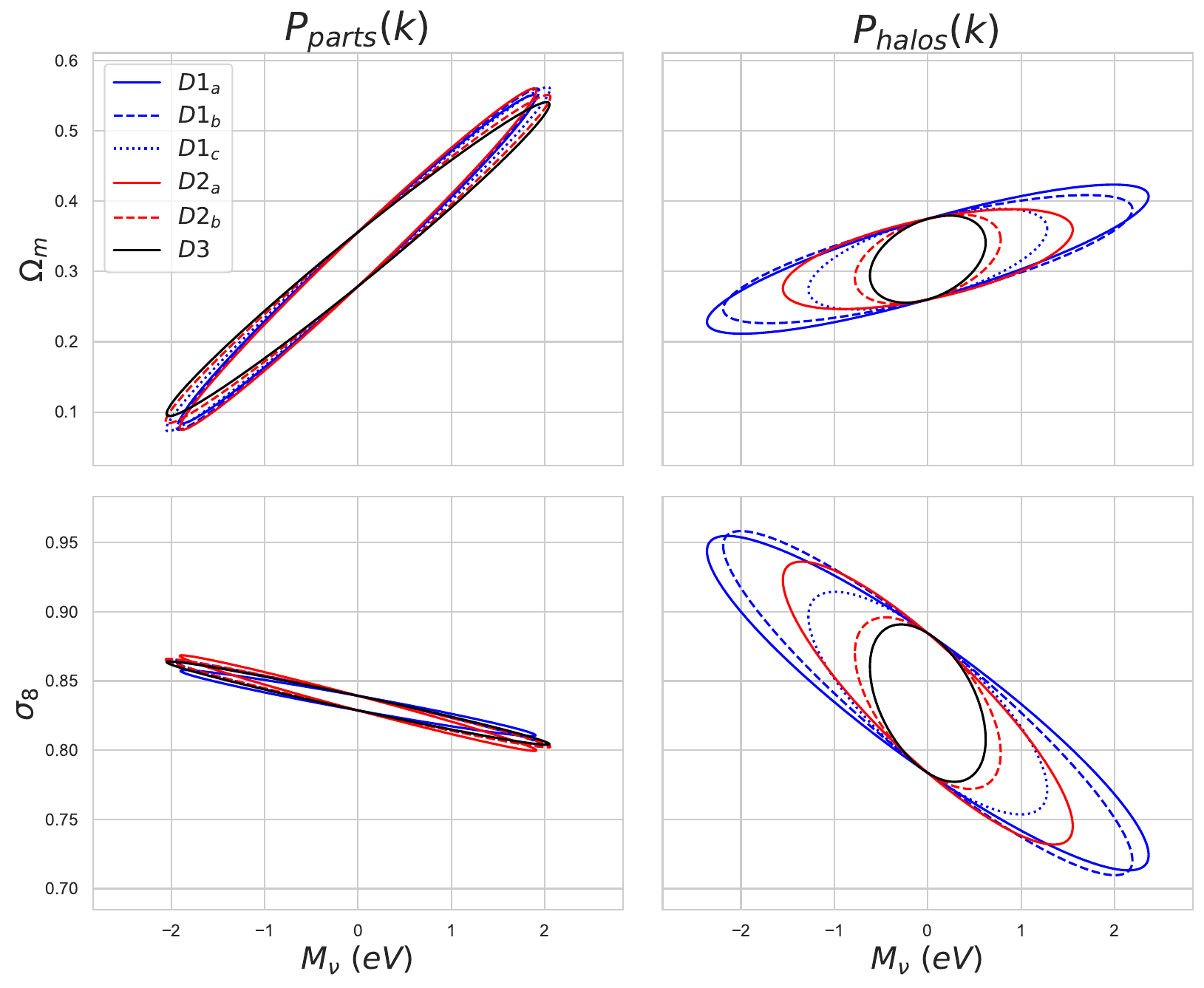}
\caption{95\% marginalized confidence ellipses in the [upper] $\{M_{\nu},\Omega_m\}$  and  [lower] $\{M_{\nu},\sigma_8\}$ parameter spaces  using the real space matter power spectrum from the Quijote simulations from [left] the particles, $P_{parts}$ and [right] halos, $P_{halos}$. Six different numerical differentiation schemes, $D1_{a}$ - $D3$, used to calculate the neutrino mass derivative $\frac{\partial P(k)}{\partial \M}$, are shown: first-order differentiation $D1_{a}$, $D1_{b}$ and $D1_{c}$ [blue: full, dashed and dotted], second-order, $D2_{a}$ and $D2_{b}$ [red: full  and dashed], and third-order, $D3$ [black, full]. }
\label{fig:partHaloDerivEllipsesBoth}
\end{figure*}

\section{Analysis}
\label{sec:noise_Fisher}

In section~\ref{sec:sim_fisher} the simulation-based Fisher forecasts for a subset of the cosmological parameters utilizing either $P_{parts}(k)$ or $P_{halos}(k)$ are presented. Then, in section~\ref{sec:modeling_stat_noise}, the effects of statistical noise within simulation-derived Fisher forecasts as a random variable are modeled and we derive how a $68\%$ confidence limit (CL) for the true cosmological constraints can be obtained. In section~\ref{sec:prac_imp}, we refine the theoretical model so that it can be practically implemented for available data, and briefly discuss the impacts of such refinements. In section \ref{sec:application}, we apply the model to the Fisher forecasts discussed in section~\ref{sec:sim_fisher} and discuss the implications, focusing mainly on the impact of the various $\M$ differentiation schemes. In section \ref{sec:mock_hod}, we apply the approach to redshift space 2-point and 3-point statistics with mock galaxy tracers, and discuss the implications for $\M$ constraints given the inclusion of HOD nuisance parameters. In section \ref{sec:notMnu}, we discuss the effects of statistical noise on parameter degeneracies, particularly between $\M$, $\Om$ and $\s8$, and how these effects manifest. 

\subsection{Simulation-Derived Fisher Analysis}
\label{sec:sim_fisher}

We utilize the Quijote simulations to study the application of simulation-derived Fisher techniques to obtain cosmological constraints. We focus initially on the matter power spectra, considering predictions from both the particles, $P_{parts}(k)$,  and the dark matter halos, $P_{halos}(k)$ for the 2D-constraints in the $\{\sigma_8,M_{\nu}\}$ and $\{\Omega_m,M_{\nu}\}$ parameter spaces. 

In Fig.~\ref{fig:partHaloDerivEllipsesBoth}, we present the joint marginalized confidence ellipses for $\M - \Om$ and $\M - \s8$ obtained from the Fisher analysis discussed in sections \ref{sec:Fisherintro} and \ref{sec:num_derivs}. The different ellipses are separated only by the particular differentiation scheme $D1_{a}-D3$, given in (\ref{eq:s1_ppp})-(\ref{eq:f3}), chosen for the $M_\nu$ derivative. The full set of cosmological parameters considered is $\left\{\M,\Om,\s8,\Omega_{b},h,n_{s} \right\}$.

\begin{figure*}[!t]
\includegraphics[width=0.9\linewidth]{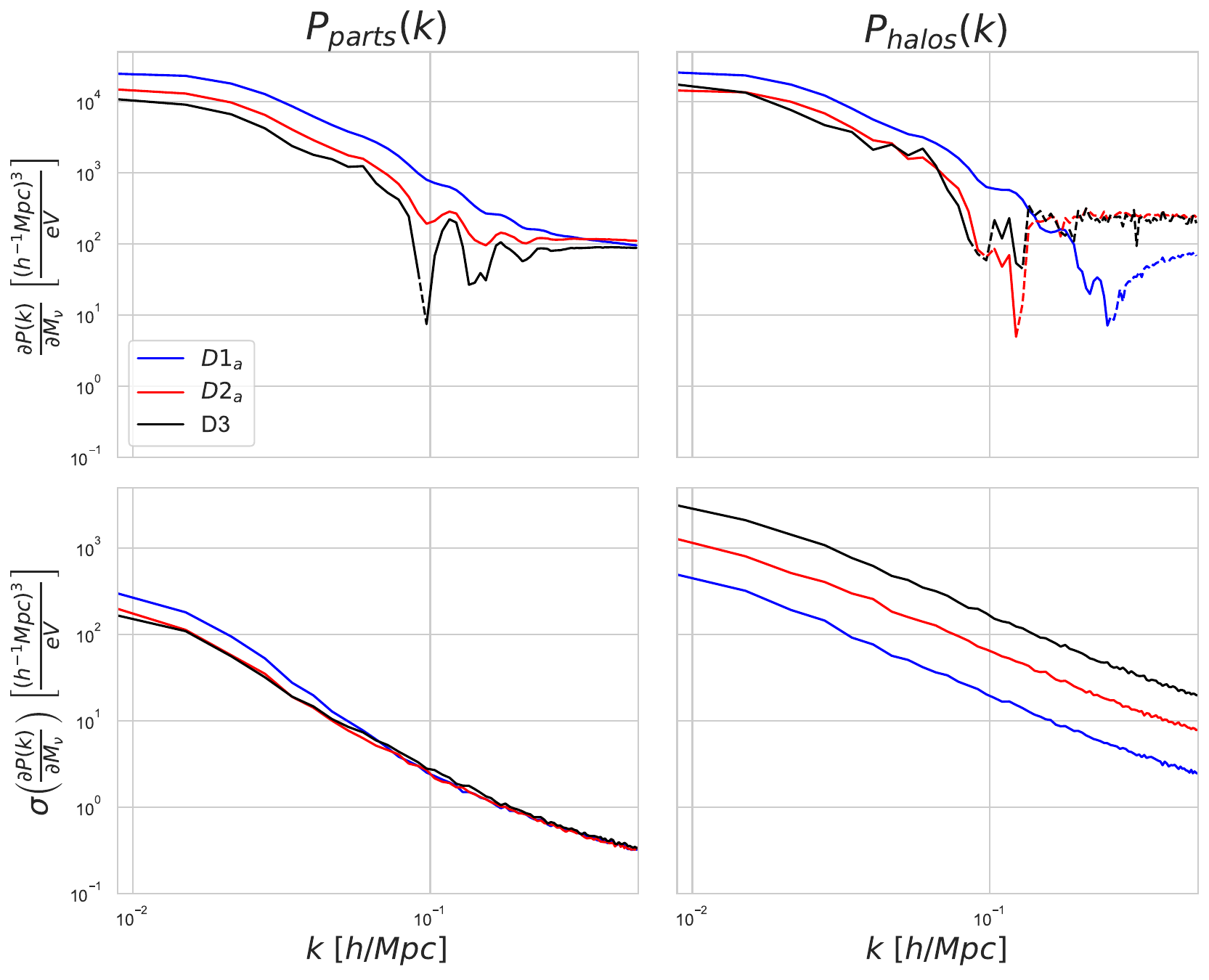}
\caption{[Upper] The mean and [lower] the standard error in the mean of the derivative of the real space matter power spectrum, $P(k)$, with respect to the neutrino mass, $M_\nu$, as calculated from the full set of the Quijote simulations with $N_{sims}=500$, from both [left] particles and [right] halos. Three different numerical differentiation schemes with respect to $M_{\nu}$ are shown: $D1_{a}$ [blue], $D2_{a}$ [red], and $D3$ [black]. Solid and dashed lines indicate positive and negative values respectively.}
\label{fig:derivsAndErrors}
\end{figure*}

The $P_{part}(k)$ confidence ellipses exhibit strong consistency across the various derivative schemes. Given that each differentiation method has different levels of both systematic and statistical uncertainties, the minor differences observed suggest that the total effect of both of these are negligible compared to the cosmological signal's impact on the derivative.

For the halos, however, the predicted ellipses are highly dependent on the numerical differentiation scheme used for $\M$. The tightest constraints on $M_{\nu}$ are given by the third-order differentiation scheme, $D3$ (\ref{eq:f3}), while the loosest constraints are from the first order schemes, $D1_{a}$ (\ref{eq:s1_ppp}) and $D1_{b}$ (\ref{eq:s1_pp}). This indicates that either the systemic error or the statistical uncertainty on the mean is causing the halo confidences ellipses to be highly dependent on the $M_{\nu}$ differentiation scheme used, rather than purely the underlying cosmological constraints.

To understand the possible effects of systematic error and statistical uncertainty in the derivatives on the Fisher forecast, Fig.~\ref{fig:derivsAndErrors} shows the mean and standard error on the mean for $\partial P(k)/\partial M_{\nu}$ versus $k$ for both the halo and particle power spectra using the $D1_{a}$, $D2_{a}$ and $D3$ differentiation schemes.

For the $M_{\nu}$ derivatives, we see that both $P_{parts}(k)$ and $P_{halos}(k)$ have somewhat similar systematic differences in their mean values of $\partial P(k)/\partial M_{\nu}$ between the three differentiation schemes. The main difference arises in comparing the statistical errors on the means. The statistical noise in the particle data is an order magnitude smaller than for the halos, and is also quite consistent across the three differentiation schemes. By contrast, for the halo data, the three differentiation schemes exhibit markedly different levels of statistical noise. This suggests that statistical noise within the numerical derivatives is at least partially responsible for the drastically different confidence ellipses observed in Fig.~\ref{fig:partHaloDerivEllipsesBoth}.

Comparing $\bar{\sigma}_{\M}$ and $\bar{\sigma}_{\M,unmarg}$ (barred quantities are those calculated from the simulations) in each differentiation scheme, we find remarkable consistency within the unmarginalized constraints, $\bar{\sigma}_{\M,unmarg}$, no matter the differentiation scheme for both $P_{parts}(k)$ and $P_{halos}(k)$. For $P_{parts}(k)$, the values of $\bar{\sigma}_{\M,unmarg}( \mathrm{eV})=\{0.09,\ 0.08,\ 0.09\}$ for $\{D3,\ D2_{a},\ D1_{a}\}$ respectively. For $P_{halos}(k)$, we find a similar level of consistency with $\bar{\sigma}_{\M,unmarg}( \mathrm{eV})=\{0.07,\ 0.06,\ 0.10 \}$ for $\{D3,\ D2_{a},\ D1_{a}\}$ respectively. It is only at the level of \textit{marginalized} constraints where constraints from $P_{halos}(k)$ begin to radically differ from each other, while constraints from $P_{parts}(k)$ still remain highly consistent with one another. For $P_{parts}(k)$, we find consistent values of $\bar{\sigma}_{\M}( \mathrm{eV})=\{0.82,\ 0.77,\ 0.76 \}$ for $\{D3,\ D2_{a},\ D1_{a}\}$ respectively, while for $P_{halos}(k)$, we find values differ significantly, with $\bar{\sigma}_{\M}( \mathrm{eV})=\{0.25,\ 0.62,\ 0.95\}$ again for $\{D3,\ D2_{a},\ D1_{a}\}$ respectively.

We therefore surmise that it is the interplay of statistical noise \textit{and} the marginalization process leading to the radically different $\M$ constraints in Fig.~\ref{fig:partHaloDerivEllipsesBoth}. Given this, we move forward with an investigation centered around understanding the various effects that statistical noise within simulation-based numerical derivatives can have on the results of the associated Fisher forecast, with a specific focus on the interaction of statistical noise with the marginalization process.

\subsection{Modeling the Effect of Statistical Gaussian Noise in a Fisher Forecast}
\label{sec:modeling_stat_noise}

Given the observations in the previous section, it is worthwhile investigating whether the differences in the confidence ellipses shown in Fig.~\ref{fig:partHaloDerivEllipsesBoth} are arising because of the differing levels of statistical noise in the $\M$ numerical derivatives.

We model the effects of statistical noise in the parameter derivatives on our Fisher forecasts through the equation
\begin{equation}
d\overline{\sO}{}^{\ i}_{A} =  d \sO^{\ i}_A + u^{i}_{A},
\label{eq:derivExpansion}
\end{equation}
where $d\overline{\sO}{}^{\ i}_{A} $ and $d\sO^{\ i}_{A}$ are defined in (\ref{eq:meanDeriv}) and (\ref{eq:meanDerivExp})  respectively. The random variable $u^{i}_{A}$ is used to model the the standard uncertainty in the mean of our numerical derivatives and is thus drawn from a multivariate Gaussian distribution, $u^{i}_{A} \sim  \mathcal{N}\left(0,\Sigma^{ij}_{AB}\right)$.

Our goal is to ultimately trace the effects of the noise term $u^{i}_{A}$ through to the Fisher calculation, (\ref{eq:chiSimp_main}). Inserting (\ref{eq:derivExpansion}) into the definition of $\overline{F}_{AB}$, (\ref{eq:FisherDef}), gives 
\bea
\overline{F}_{AB} &\equiv& d\overline{\sO}{}^{\ i}_{A} C^{-1}_{ij} d\overline{\sO}{}^{\ j}_{B} \nonumber \\
 &=& \left(d\sO^{i}_A + u^{i}_{A}\right) C^{-1}_{ij}\left(d \sO^{j}_B + u^{j}_{B}\right) \nonumber \\ &\equiv& F_{AB} + U_{AB} 
\label{eq:FplusU}
\eea
where we have implicitly defined the random variable $U_{AB}$ as
\begin{equation}
U_{AB}\equiv u^{i}_{A}C^{-1}_{ij} d\sO^{j}_B + d\sO^{\ i}_A C^{-1}_{ij}  u^{j}_{B} + u^{i}_{A}C^{-1}_{ij} u^{j}_{B}.
\label{eq:capitalU}
\end{equation}
Physically, $\overline{F}_{AB}$ is the value of the Fisher information matrix we calculate from the simulations in the typical straightforward manner. $F_{AB}$ is the true noise-free value of the Fisher information matrix in the limit $N_{sims} \rightarrow \infty$, and $U_{AB}$ is a random variable which functions as the noise contribution term to $F_{AB}$ due to statistical noise in the mean numerical derivatives from a finite number of simulations.

The effects of derivative noise on the constraints given by the marginalized Fisher Information for the target parameters, $\overline{\mathscr{F}}_{\mu \nu}$, can be seen by combining  (\ref{eq:margFisher_main}) and (\ref{eq:FplusU}),
\bea
\label{eq:margFisherTotal}
\overline{\mathscr{F}}_{\mu \nu} &\equiv& \overline{F}_{\mu \nu} - \overline{F}_{\mu a} (\overline{f}{}^{-1})^{ab} \overline{F}_{b \nu} \nonumber  
\\ 
&=& \left(F_{\mu \nu} + U_{\mu \nu} \right) \nonumber
\\
&-& \left(F_{\mu a} + U_{\mu a} \right) \left((f_{cd} + U_{cd})^{-1}\right)^{ab} \left(F_{b \nu} + U_{b \nu} \right).
\eea
Here the $cd$ indices indicate that the matrix inversion of $(f_{cd} + U_{cd})$ includes only the $(N-K) \times (N-K)$  non-target parameter portion of the relevant matrices, with $f_{cd} = F_{cd}$. 

In the limit $N_{sims} \rightarrow \infty$, we have that $\overline{\mathscr{F}}_{\mu\nu} \rightarrow \mathscr{F}_{\mu \nu}$.  For any finite value of $N_{sims}$, the  difference between $\overline{\mathscr{F}}_{\mu\nu}$ (\ref{eq:margFisherTotal}) and $\mathscr{F}_{\mu \nu}$ (\ref{eq:margFisher_main})
can be interpreted as the noise contribution to the marginalized Fisher information matrix, explicitly given by
\begin{eqnarray}
\label{eq:margFisherNoise}
\mathscr{F}^{Noise}_{\mu \nu} &\equiv& \overline{\mathscr{F}}_{\mu \nu} -  \mathscr{F}_{\mu \nu} \nonumber \\
& =& U_{\mu \nu} + F_{\mu a} \left(f^{-1}\right)^{ab}F_{b \nu}  \nonumber \\
  && -\left({F}_{\mu a} + U_{\mu a} \right) \left(\left(({f} + U)^{-1}\right)^{ab}\right) \left({F}_{b \nu} + U_{b \nu} \right). 
\end{eqnarray} 

For concreteness, in the $K=1$ case, with $\theta^{\mu}=M_{\nu}$ we can gain an intuition for the effects of noise on the $M_{\nu}$ marginalized 1D constraints. Combining (\ref{eq:margFisherNoise}) and (\ref{eq:chiSimp_main}), and setting $\chi^2=1$ for $1\sigma$ constraints gives us
\bea
\left(\bar{\sigma}_{M_{\nu}}\right)^{-2} &=& \overline{\mathscr{F}}_{M_{\nu} M_{\nu}} \nonumber \\
&=&  \mathscr{F}_{M_{\nu} M_{\nu}} +  \mathscr{F}^{Noise}_{M_{\nu} M_{\nu}} \nonumber \\
&=&  \sigma^{-2}_{M_{\nu}} +  \mathscr{F}^{Noise}_{M_{\nu} M_{\nu}}. \label{eq:TCP}
\eea
Physically, $\bar{\sigma}_{M_{\nu}}$ is the constraint we measure from simply performing a Fisher forecast with $N_{sims}$ paired sets of simulations used to calculate each mean numerical derivative while $\sigma_{M_{\nu}}$ is the constraint obtained in the limit $N_{sims} \rightarrow \infty$. $\mathscr{F}^{Noise}_{M_{\nu} M_{\nu}}$ is a random variable which characterizes the difference between the inverse squares of the two because of the derivative noise.

Unlike scenarios where noise terms may lead to subtle deviations in either direction with somewhat equal probability, in this instance, through (\ref{eq:TCP}), $\mathscr{F}^{Noise}_{M_{\nu} M_{\nu}}$ causes a systematic, and potentially sizeable, increase to the measured value of $\bar{\sigma}^{-2}_{\M}$, and therefore leads to tighter predicted Fisher constraints.

Our findings may be understood through the lens of analytic marginalization. The measured constraint, $\bar{\sigma}^{-2}_{\M}$, may be thought of a measure of how accurately the target function, $d\overline{\sO}{}^{\ i}_{\mu}$, is able to be reproduced as a linear combination from the set of non-target basis functions, $d\overline{\sO}{}^{\ i}_{a}$, as in the discussion surrounding (\ref{eq:explicitLinearCombination}). 

In the unmarginalized case, the derivative noise term $u^{i}_{\mu}$ simply adds ``random wiggles" to the target parameter derivative $d\overline{\sO}{}^{\ i}_{\mu} = d\sO^{\ i}_{\mu} + u^{i}_{\mu}$. As  we begin to add marginalizing parameters to the non-target set, here assumed to not have significant noise, these smooth parameter derivatives, $d\sO^{\ i}_{a}$, serve as smooth basis functions, that are typically highly efficient at reproducing, and therefore cancelling out, the smooth $d\sO^{\ i}_{\mu}$ contribution to $d\overline{\sO}{}^{\ i}_{\mu}$, but are \textit{highly inefficient} at reproducing the random wiggles caused by the random noise variable $u^{i}_{\mu}$. As we add more non-target parameters to our marginalizing set, our growing set of smooth basis functions $\left\{d\sO^{\ i}_{a}\right\}$ is increasingly able to reproduce and therefore cancel out the contribution arising from the smooth $d\sO^{i}_{\mu}$ to larger and larger degrees. However, this set fundamentally remains unable to meaningfully reproduce, and therefore cancel against, the random wiggles induced by $u^{i}_{\mu}$ in any regime where the number of parameters is small compared to the number of observables, which is almost universally the case. 
Furthermore, any noise within the non-target set \textit{reduces} the ability of the now noisy $d\overline{\sO}{}^{\ i}_{a} = d\sO^{\ i}_{a} + u^{i}_{a}$ to reproduce and cancel against the smooth portion of the target parameter derivative $d\sO^{\ i}_{\mu}$, which \textit{also} reduces the inherent degeneracy between the parameters and yields tighter noise-induced constraints.

Both effects are simultaneously captured by our noise term $\mathscr{F}^{Noise}_{\M \M}$ in the case $\theta^{\mu} = \M$. For a general target parameter $\theta^{\mu}$, $\mathscr{F}^{Noise}_{\mu \mu}$ increases the measured marginalized $\bar{\sigma}^{-2}_{\mu}$ relative to the true $\sigma^{-2}_{\mu}$, and therefore decreases the measured marginalized constraint, $\bar{\sigma}_{\mu}$, below the true value, $\sigma_{\mu}$. The above discussion also highlights why noise, which was previously contributing little at the level of unmarginalized constraints, becomes increasingly important and eventually dominant as more marginalizing parameters are added to the non-target set. As long as statistical noise within the new marginalizing parameters is not itself a new dominant source of noise, the additional parameters act to dilute the cosmological constraints, through degeneracies between the smooth components of the target and non-target derivatives, while leaving the noise contribution mostly unaffected. The cumulative effect of noise is to damp the intrinsic cosmological degeneracies between parameters, which while not impactful for unmarginalized constraints, does have a significant impact on the marginalized parameter constraints.

This result provides a useful mechanism to determine whether, for a specific set of simulations and differentiation schemes, statistical noise on the mean value of the parameter derivatives contributes significantly to the marginalized constraints on $M_{\nu}$ or on any other potential target parameter $\theta^{\mu}$. Given that $\mathscr{F}^{Noise}_{\mu\mu} \rightarrow 0$ as $N_{sims}\rightarrow \infty$, this result can also inform the minimum number of realizations required for the mean derivatives to be calculated with sufficient accuracy to avoid noise contamination of the parameter constraints. In regimes where the constraint is noise dominated, such that $\left<\mathscr{F}^{Noise}_{\mu \mu}\right> \simeq \bar{\sigma}_{\mu}^{-2}$, a higher value of $N_{sims}$ is needed before a meaningful conclusion can be inferred about the true underlying cosmological constraints.

Alternatively, within intermediate regimes, where the noise is still significant but not dominant, this result can help us diagnose the true constraint to within some confidence interval. For instance, in the $K=1$ single target parameter case, if $\overline{\mathscr{F}}_{\mu \mu} \equiv \bar{\sigma}_{\mu}^{-2} \gg \left<\mathscr{F}^{Noise}_{\mu \mu}\right>$, then we may conclude that $N_{sims}$ is large enough for our measured constraint to be trustworthy. In intermediate regimes, where $\left<\mathscr{F}^{Noise}_{\mu \mu}\right>$ is a sizeable fraction of $\bar{\sigma}_{\mu}^{-2}$ but their difference remains large with respect to $\sqrt{\mathrm{Var}\left(\mathscr{F}^{Noise}_{\mu \mu} \right)}$, we may calculate the $68\%$ confidence interval for the inverse square of the true constraint, $\sigma^{-2}_{\mu}$, as
\begin{equation}
\sigma_{\mu}^{-2} = \left(\bar{\sigma}_{\mu}^{-2} - \left< \mathscr{F}^{Noise}_{\mu \mu} \right>\right) \pm \sqrt{\mathrm{Var\left( \mathscr{F}^{Noise}_{\mu \mu} \right)}}.
\label{eq:trueConstApprox}
\end{equation}

The quantities $\left<\mathscr{F}^{Noise}_{\mu \mu} \right>$ and $\mathrm{Var}\left( \mathscr{F}^{Noise}_{\mu \mu}\right)$ may be calculated by sampling instances of the multivariate Gaussian random variable $u^{i}_{A}$ and computing the corresponding instance of $\mathscr{F}^{Noise}_{\mu \mu}$ via (\ref{eq:capitalU}) and (\ref{eq:margFisherNoise}). In Appendix \ref{sec:noiseLimits} we also compute approximations for these quantities in some physically relevant limits which we also employ below. 

In conclusion, our noise model explicitly incorporates the intricate relationship between derivative noise and the marginalization process as observed in \ref{sec:sim_fisher}, and provides a method to quantify the effects of noise, and from it, extract a confidence interval for the true 1D constraints.

\subsection{Implementing the Noise Model}
\label{sec:prac_imp}

We now develop a practical implementation of the noise model developed in \ref{sec:modeling_stat_noise} to estimate and understand the statistical uncertainties present in Fisher constraints estimated from a given set of simulations, and their impact on obtaining accurate estimates of the underlying cosmological constraints. The general procedure can be summarized as:
\begin{enumerate}
    \item Draw an instance of $u^{i}_{A}$ from the 0-mean multivariate Gaussian distribution characterized by $\Sigma^{ij}_{AB}$.
    \item Calculate $U_{AB}$ from (\ref{eq:capitalU}) for the instance of $u^{i}_{A}$. In theory, this requires knowing the $N_{sims} \rightarrow \infty$ values of the derivatives $d\sO^{\ i}_{A}$.
    \item Use $U_{AB}$ and (\ref{eq:margFisherNoise}) to calculate the associated instance of $\mathscr{F}^{Noise}_{\mu \nu}$.
    \item Repeat until enough instances of $\mathscr{F}^{Noise}_{\mu \nu}$ have been drawn that the distribution $p\left(\mathscr{F}^{Noise}_{\mu \nu}\right)$ is converged.
\end{enumerate}
While the above process is in theory exact, there are practical limitations. Primarily, (\ref{eq:capitalU}) and (\ref{eq:margFisherNoise}), used in steps 2 and 3, assume exact knowledge of the $N_{sims} \rightarrow \infty$ values of $d\sO^{i}_{A}$ and $F_{AB}$. If we had access to these quantities, then none of this analysis would be needed as we could simply run a Fisher forecast and be confident in it returning noise-free results! In the realistic case of finite sims, we need to approximate $d\sO^{\ i}_{A}$. However, in doing so, we must be careful to understand any biases that the approximations may cause to the resulting $p\left(\mathscr{F}^{Noise}_{\mu \nu}\right)$ relative to the true distribution. Below, we outline two different derivative approximation schemes each of which is practically useful for calculating $p\left(\mathscr{F}^{Noise}_{\mu \nu}\right)$. We find that each of our two approximation scheme does slightly bias the resulting $p\left(\mathscr{F}^{Noise}_{\mu \nu}\right)$, but that they do so in opposite ways 

To help gain an intuitive understand of this biasing, we will consider analytical expressions for $\left<\mathscr{F}^{Noise}_{\mu \nu}\right>$ and $\mathrm{Var}\left( \mathscr{F}^{Noise}_{\mu \nu}\right)$ in a highly relevant physical limit, that we refer to as the ``isolated noise" limit. In this limit, the statistical noise is confined solely to, or dominated by, the target parameter derivatives, in comparison to the non-target parameters, so that $u^{i}_{a} \rightarrow 0$. This is a well-motivated limit for many of the scenarios we consider with the neutrino mass, $\theta^{\mu}=M_{\nu}$, as the sole target parameter, for which we obtain:
\bea
\left<\mathscr{F}^{Noise}_{\M \M}\right>\big\vert_{u^{i}_{a}\rightarrow 0} &=& \Sigma^{ij}_{\M \M}C^{-1}_{ij} - d\sO^{i}_{a} C^{-1}_{ij} \Sigma^{jk}_{\M \M} C^{-1}_{kl} d\sO^{l}_{b} (f^{-1})^{ab}. \nonumber
\\
&&
\label{eq:INexp}
\eea
and
\bea
\mathrm{Var}\left(\mathscr{F}^{Noise}_{\M \M}\right)\big\vert_{u^{i}_{a}\rightarrow 0} &=& \left(4d\sO^{\ i}_{\M} d\sO^{\ j}_{\M} + 2\Sigma^{ij}_{\M \M}\right) \left(Q_{ik} \Sigma^{kl}_{\M \M} Q_{lj} \right).\nonumber
\\ 
&&
\label{eq:INvar}
\eea
where the auxiliary variable $Q_{ij}$ is defined as 
\begin{equation}
Q_{ij} = \big(C^{-1}_{ij} - C^{-1}_{ik} d\sO^{k}_{a} ({f^{-1}})^{ab} d\sO^{l}_{b} C^{-1}_{lj}\big).
\label{eq:defQ}
\end{equation}
The details of $\left<\mathscr{F}^{Noise}_{\mu \nu}\right>$ and $\mathrm{Var}\left( \mathscr{F}^{Noise}_{\mu \nu}\right)$ in the isolated noise limit are outlined in more detail in Appendix~\ref{sec:noiseLimits}.

Below we describe two pragmatic derivative approximation schemes and consider the potential limitations and biases introduced by each. The first is arguably the simplest approach possible, which we find primarily overestimates the variance in the noise. The second uses degeneracies between the parameters to null the cosmological signal from the target parameters so that the resulting Fisher is purely the result of noise, which we find underestimates the noise variance.

\subsubsection{Approximation I:  $F_{AB} = \overline{F}_{AB}$}

The first option is to use the mean value $d\overline{\sO}{}^{\ i}_{A}$ calculated from the simulations as a best estimate of the true value of $d\sO^{\ i}_A$ when calculating $\mathscr{F}^{Noise}_{\M \M}$. Explicitly, for both target and non-target parameters, when (\ref{eq:capitalU}) and (\ref{eq:margFisherNoise}) are used to calculate $\mathscr{F}^{Noise}$, all unbarred derivatives are approximated as
\begin{align}
d\sO^{\ i}_a &\rightarrow  d\overline{\sO}{}^{\ i}_{a}  \nonumber \\ 
d\sO^{\ i}_\mu &\rightarrow  d\overline{\sO}{}^{\ i}_{\mu}.
\label{eq:simApproxDeriv}
\end{align}
Although this approach is straightforward, we find that it can introduce bias into the results. Considering the isolated noise limit, and replacing the true value of the derivative $d\sO^{\ i}_{\M}$ with $d\overline{\sO}{}^{\ i}_{\M}$ in (\ref{eq:INvar}), leads to an increase in $\mathrm{Var}\left(\mathscr{F}^{Noise}_{\M \M} \right)\big\vert_{u^{i}_{a}\rightarrow 0}$ by on average $4\Sigma^{ij}_{\M \M} Q_{jk} \Sigma^{kl}_{M{\nu} M_{\nu}} Q_{li} \geq 0$. This results in the systematic overestimation of $\mathrm{Var}\left(\mathscr{F}^{Noise}_{\M \M} \right)\big\vert_{u^{i}_{a}\rightarrow 0}$, causing the reconstructed probability distribution $p\left(\mathscr{F}^{Noise}_{\M \M}\right)$ to be wider than the true distribution. Interestingly, in the isolated noise approximation, the expectation value of $\mathscr{F}^{Noise}_{\M \M}$ is wholly unchanged by Approx. I. This is because $d\sO^{\ i}_{\M}$ is absent from (\ref{eq:INexp}), and all other derivatives are assumed noiseless in this physical limit. 

\subsubsection{Approximation II: $F_{AB} = F^{degen}_{AB}$}

Given the overestimation of the variance inherent in the first approximation, we also consider a second approximation method that biases the width of the distribution in the opposite direction. This method artificially sets the ``true" target-parameter constraints to zero by approximating the true derivatives such that $\mathscr{F}_{\mu \nu} = 0$. For a lone target parameter such as $\theta^{\mu} = \M$, this gives $\sigma_{M_{\nu}}^{-2}=0$ so that the resulting $\bar{\sigma}^{-2}_{\M}$ will be entirely determined by the noise contribution $\mathscr{F}^{Noise}_{M_{\nu} M_{\nu}}$. In effect, this second approximation scheme calculates pure noise constraints, which we will now explain in more detail.

First, for the non-target parameters, we again assume $d\sO^{\ i}_{a}=d\overline{\sO}{}^{\ i}_{a}$. The true target parameter derivative(s) $d\sO^{\ i}_{\mu}$ are then approximated by a closely matched linear combination of the non-target parameter derivatives, $\{d\sO^{\ i}_a\}$. Explicitly, that is
\begin{align}
d\sO^{\ i}_a &\rightarrow  d\overline{\sO}{}^{\ i}_{a} \nonumber \\ 
d\sO^{\ i}_\mu &\rightarrow \overline{F}_{\mu b} (\bar{f}^{-1})^{ba} d\overline{\sO}{}^{\ i}_{a} 
\label{eq:degenDeriv}
\end{align}

where all quantities appearing on the right hand side of (\ref{eq:degenDeriv}) are values that would be calculated from the full set of simulations. We note that this is the same linear combination which appears in the discussion of analytic marginalization in \ref{sec:Fisherintro}, with the coefficients $\overline{F}_{\mu b} (\bar{f}^{-1})^{ba}$ equal to the same values of the $x^{a}$ values needed to minimize $\chi^{2}$ in (\ref{eq:explicitLinearCombination}). Explicitly, this linear combination of the non-target parameter derivatives $d\overline{\sO}{}^{\ i}_{a}$ reproduces the target parameter derivative(s) $d\overline{\sO}{}^{\ i}_{\mu}$ as closely as possible when using the inverse observable covariance matrix as a metric which characterizes the fit.

A degenerate Fisher matrix $F^{degen}_{AB}$ can then be constructed using this new set of linearly dependent derivatives which is exactly equal to $\overline{F}_{AB}$ except for the entries between target parameters, $F^{degen}_{\mu \nu}$:
\bea
F^{degen}_{\mu \nu} &=&  \overline{F}_{\mu a} (\bar{f}^{-1})^{ab} d\overline{\sO}{}^{\ i}_{b} C^{-1}_{ij}d\overline{\sO}{}^{\ j}_{c}(\bar{f}^{-1})^{cd}  \overline{F}_{d \nu} \nonumber\\ 
&=& \overline{F}_{\mu a} (\bar{f}^{-1})^{ab} \overline{F}_{b \nu} 
\label{eq:FisherDegenApprox1}
\\
F^{degen}_{\mu b} &=& \overline{F}_{\mu a} (\bar{f}^{-1})^{ac} d\overline{\sO}{}^{\ i}_{c} C^{-1}_{ij} d\overline{\sO}{}^{\ j}_{b} = \overline{F}_{\mu b} 
\label{eq:FisherDegenApprox2}
\\
F^{degen}_{a \nu} &=& d\overline{\sO}{}^{\ i}_{a} C^{-1}_{ij} d\overline{\sO}{}^{\ j}_{b} (\bar{f}^{-1})^{bc}  \overline{F}_{c \nu}= \overline{F}_{a \nu} 
\label{eq:FisherDegenApprox3}
\\
F^{degen}_{a b} &=& d\overline{\sO}{}^{\ i}_{a} C^{-1}_{ij} d\overline{\sO}{}^{\ j}_{b}= \overline{F}_{a b}.
\label{eq:FisherDegenApprox4}
\eea
When the approximation scheme in (\ref{eq:degenDeriv}) is used to calculate the true values of the numerical derivatives such that $F_{AB} = F^{degen}_{AB}$, $\mathscr{F}^{Noise}_{M_{\nu} M_{\nu}}$ takes on a new physical interpretation. From (\ref{eq:margFisher_main}) and (\ref{eq:TCP}), one can see that using $F_{AB} = F^{degen}_{AB}$ will give $\mathscr{F}_{\mu \nu}=0$, and therefore in the single target parameter case $\theta^{\mu} = \M$, we would have $\sigma^{-2}_{\M} = 0$ ($\sigma_{\M} = \infty$), which means in this instance, the hypothetical $\bar{\sigma}^{-2}_{\M}$ can be completely ascribed to the effects of statistical noise with $\bar{\sigma}^{-2}_{\M} = \mathscr{F}^{Noise}_{M_{\nu} M_{\nu}}$. Thus, the distribution $p\left(\mathscr{F}^{Noise}_{M_{\nu} M_{\nu}} \right)$ serves as a probability distribution of constraints $\bar{\sigma}^{-2}_{\M}$ generated entirely from noise. 

This second approximation method for the noise-free derivatives $d\sO^{\ i}_A$ also biases the reconstructed diagonal $p\left(\mathscr{F}^{Noise}_{\mu \mu}\right)$ to have a smaller variance compared to the true distribution. Considering the isolated noise limit with $\theta^{\mu} = \M$, using (\ref{eq:degenDeriv}) causes $\mathrm{Var}\left(\mathscr{F}^{Noise}_{\M \M} \right)\big\vert_{u^{i}_{a}\rightarrow 0}$ to moderately decrease in magnitude, as the terms in (\ref{eq:INvar}) involving $4 d{\sO}^{\ i}_{\M}d{\sO}^{\ j}_{\M}$ can be shown to now cancel exactly, whereas previously they contributed positively. Again, we find that in the physical limit of isolated noise, this second approximation also leaves the expectation value of $\mathscr{F}^{Noise}_{\M \M}$ entirely unchanged. Since approximation I and II disagree only in their assignments to the target parameter derivatives, (\ref{eq:INexp}) is unbiased by either approximation within the limit of isolated noise.

\subsubsection{Noise Distribution  Characterization \& Consistency Checks}

The ultimate aim of this analysis is to accurately characterize the statistical properties of the noise probability distribution $p\left(\mathscr{F}^{Noise}_{\mu \mu}\right)$ to determine if the effects of a single instance of the random variable $\mathscr{F}^{Noise}_{\mu \mu}$ can be disambiguated from the true cosmological signal, thereby allowing for the inference of the underlying cosmological constraints.

To this end, we can define three relevant noise probability distributions, $p\left( \mathscr{F}^{Noise}_{\mu\mu} \right)$, for a given target parameter $\theta^\mu$,
\bea
p^{True}&\equiv& p\left(\mathscr{F}^{Noise}_{\mu \mu} \vert\ d\sO \right),    \label{eq:p} \\
\bar{p} &\equiv&  p\left(\mathscr{F}^{Noise}_{\mu \mu} \vert\ d\overline{\sO} \right), \label{eq:pbar} \\
p^{degen} &\equiv&  p\left(\mathscr{F}^{Noise}_{\mu \mu} \vert\ d\sO^{degen} \right). \label{eq:pdegen}
\eea
Here, $p^{True}$ is the true noise distribution from which our noise term $\mathscr{F}^{Noise}_{\mu \mu}$ is actually drawn from, while $\bar{p}$ and $p^{degen}$ approximate $p^{True}$ through Approx. I and II respectively. Despite the presence of biasing, both $\bar{p}$ and $p^{degen}$ approximations to $p^{True}$ are practically useful. While each approach may bias the variance of $p\left(\mathscr{F}^{Noise}_{\mu \mu}\right)$ due to the different approximations for $d\sO^{\ i}_\mu$, they do so in opposite directions so that $p^{True}$ may be assumed to lie somewhere in-between in many cases of interest. We also find that in all cases considered, even in regimes where the isolated noise limit is not valid, both $\bar{p}$ and $p^{degen}$ yield highly similar values of $\left<\mathscr{F}^{Noise}_{\mu \mu}\right>$ calculated through numerical sampling. Thus, when these approximations for $p^{True}$ are used in conjunction with (\ref{eq:trueConstApprox}), both approximations largely agree on the center of our confidence interval. Given the presence of biasing to the variance however, we find that $\bar{p}$ (Approx. I) gives a wider, more conservative estimate of the $\sigma^{-2}_{\mu}$ confidence interval in comparison with $p^{degen}$ (Approx. II)

While agreement between various aspects of the reconstructed $p\left(\mathscr{F}^{Noise}_{\mu \mu}\right)$ between Approx. I and II is useful, it does not necessarily in itself guarantee that either $\bar{p}$ or $p^{degen}$ will be representative of $p^{True}$ in a general setting. For this reason, it is insightful to consider additional tests that assess how well either $\bar{p}$ or $p^{degen}$ may represent a hypothetical $p^{True}$, so that we may be confident in our use of (\ref{eq:trueConstApprox}). This matter is addressed in detail in Appendix \ref{sec:NumConvergence}.

As discussed in the appendix, in the isolated noise limit characterized by $u_a^i\rightarrow 0$,  we can characterize and compare the expectations and variances for the noise probability distributions in (\ref{eq:p})-(\ref{eq:pdegen}). Within the physical limit of isolated noise, the expectation values for Approx. I and Approx. II are unbiased,
 \begin{eqnarray}
    \left< p^{degen}\right>\vert_{u^{i}_{a}\rightarrow 0} = \left<p^{True}\right>\vert_{u^{i}_{a}\rightarrow 0} = \left<\bar{p}\right>\vert_{u^{i}_{a}\rightarrow 0}
\end{eqnarray}
and the variances bound the true one 
\begin{eqnarray}
    \mathrm{Var}\left(p^{degen}\right)\vert_{u^{i}_{a}\rightarrow 0} < \mathrm{Var}\left(p^{True}\right)\vert_{u^{i}_{a}\rightarrow 0} \lesssim \mathrm{Var}\left(\bar{p}\right)\vert_{u^{i}_{a}\rightarrow 0}.
\end{eqnarray}

Through the sets of consistency checks discussed in Appendix \ref{sec:NumConvergence}, we demonstrate that indeed $\bar{p}$ and $p^{degen}$ provide reliable estimates for $p^{True}$ and show where each has relative advantages. When noise within the target parameters is large we show in the appendix that $p^{degen}$ is often
more representative of $p^{True}$ than $\bar{p}$. For this reason we include both approximations in our analysis.

\begin{figure*}[!t]
\includegraphics[width=1.0\linewidth]{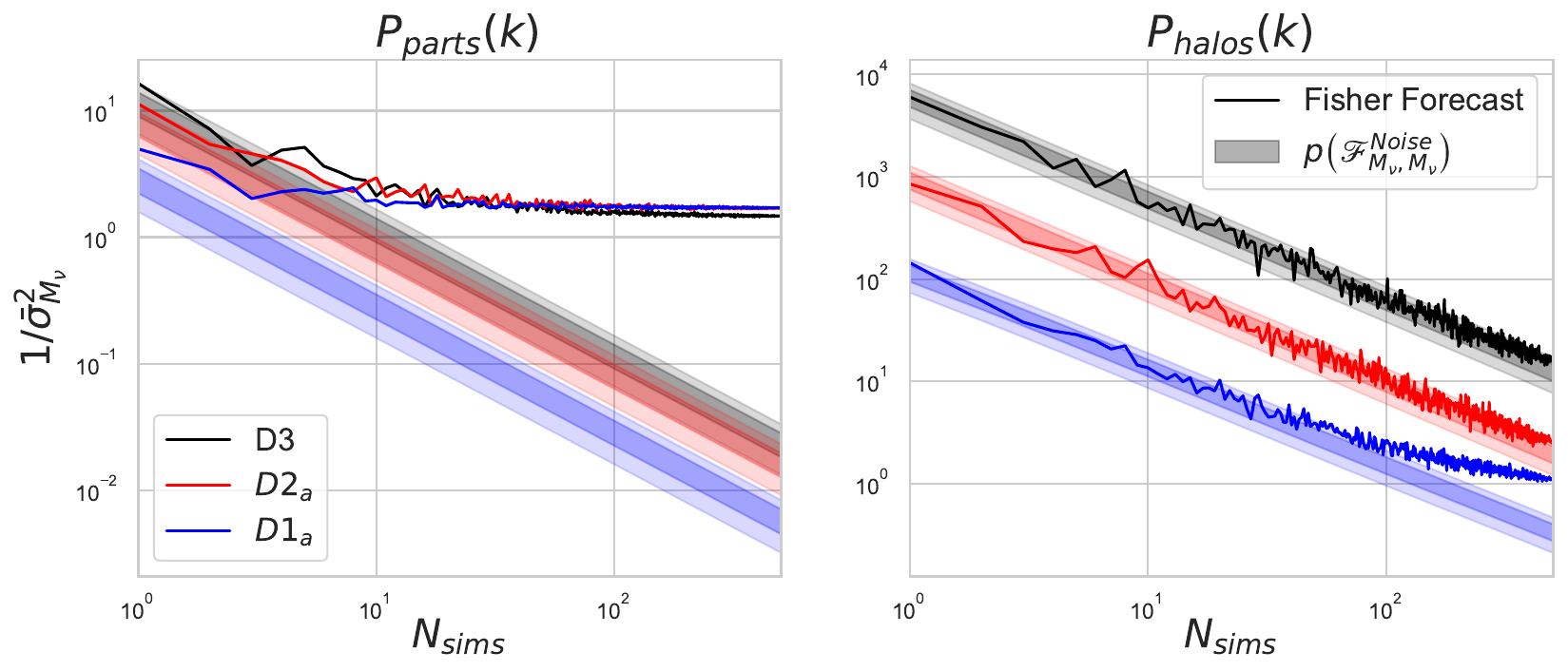}
\caption{The Fisher forecasted neutrino mass constraint, $1/\bar{\sigma}^2_{\M}$, as a function of the number of sets of simulations, $N_{sims}$, used to compute the numerical derivatives for three different numerical differentiation schemes: $D1_{a}$ [blue], $D2_{a}$ [red], and $D3$ [black] for the power spectrum $P(k)$ calculated from the particles [left] or from halos [right] to $k <$0.5$h$/Mpc. The $68\%$ and $95\%$ confidence intervals for the statistical noise distributions associated with each differentiation scheme, $p\left( \mathscr{F}^{Noise}_{\M \M} \right)$, are shown as shaded regions corresponding to the scheme's color. In all instances, $p\left( \mathscr{F}^{Noise}_{\M \M} \right)$ is approximated using $F_{AB}=F^{degen}_{AB}$ as outlined in section \ref{sec:prac_imp}.}
\label{fig:NoiseVsReals}
\end{figure*}

\subsection{Application to the Quijote Simulations $P(k)$}
\label{sec:application}

In this section we use Approximations I and II to investigate $p\left( \mathscr{F}^{Noise}_{\M \M} \right)$ for both $P_{parts}(k)$ and $P_{halos}(k)$ as a function of $N_{sims}$, and focus in on results for the maximum value of $N_{sims}=500$. We characterize the noise contributions to the various Fisher constraints for each power spectrum when using each of the $D1_{a}$, $D2_{a}$ and $D3$ differentiation schemes for the neutrino mass parameter.

In Figure~\ref{fig:NoiseVsReals} we compare the constraints calculated via Fisher forecast from the simulations, $\bar{\sigma}^{-2}_{\M},$ and also the associated 68\% and 95\% confidence limits for the $p\left(\mathscr{F}^{Noise}_{\M \M}\right)$ distribution. The latter is numerically evaluated by drawing $10^5$ instances of $u^{i}_{A}$ from the normal distribution defined by $\Sigma^{ij}_{AB}$ for each $\M$ derivative scheme and then calculating the forecast with different numbers of simulations between $N_{sims}$=1 and 500. For each case, (\ref{eq:margFisherNoise}) and (\ref{eq:degenDeriv}) are used to calculate the corresponding instance of $\mathscr{F}^{Noise}_{\M \M}$, using in Approx. II.

As described in (\ref{eq:TCP}), the random variable $\mathscr{F}^{Noise}_{\M \M}$ provides the difference between the measured $\bar{\sigma}^{-2}_{\M}$ and the $N_{sims}\rightarrow\infty$ limit, $\sigma^{-2}_{\M}$, the latter of which provides a measure of the cosmological constraining power up to the systematic errors from the differentiation scheme. Asymptotically, as discussed in appendix \ref{sec:noiseLimits}, $\mathscr{F}^{Noise}_{\M \M}$ goes as $1/N_{sims}$, meaning we expect this cosmological signal to be revealed as the number of simulation realizations increases and the $\mathscr{F}^{Noise}_{\M \M}$ noise floor falls below the cosmological signal's amplitude so that $\bar{\sigma}^{-2}_{\M}$ converges to $\sigma^{-2}_{\M}$.

For the power spectrum measured from particles, $P_{parts}(k)$, as $N_{sims}$ increases, the $\bar{\sigma}^{-2}_{\M}$ value for each differentiation scheme quickly exceed their respective 95\% confidence interval for $\mathscr{F}^{Noise}_{\M \M}$, and by the time $N_{sims}\sim 100$, all do so by over an order of magnitude. This indicates that for $N_{sims}=500$, the $\bar{\sigma}^{-2}_{\M}$ values have each converged to their respective true $\sigma^{-2}_{\M}$, and provide a clear estimate of the cosmological constraining power up to systematic errors inherent in each $\M$ differentiation scheme. Each of the $D3$, $D2_{a}$, and $D1_{a}$ values of $\bar{\sigma}^{-2}_{\M}$ asymptotes to a slightly different value of $\sigma^{-2}_{\M}$, which reflects the different levels of systematic error present in each finite differencing scheme. Since the $D3$ scheme has the lowest level of systematic error intrinsic to the scheme itself, its predicted value of $\bar{\sigma}_{\M,D3} = 0.83\ \mathrm{eV}$ may be assumed to be the most reflective of the true underlying cosmological constraint, rather than the $D2_{a}$ and $D1_{a}$ values of $\bar{\sigma}_{\M,D2_{a}} = 0.77\ \mathrm{eV}$ and $\bar{\sigma}_{\M,D1_{a}} = 0.76\ \mathrm{eV}$ respectively. The differences induced by the systematic errors are small, and fractionally, are at most a single digit percentage of any constraint. We also emphasize that this discussion of underlying systematic differences is only possible because when using $P_{parts}(k)$ from the underlying particle data, all three numerical differentiation schemes considered are virtually free of statistical noise as quantified by $p\left(F^{Noise}_{\M \M}\right)$ for $N_{sims}=500$.

When we consider $P_{halos}(k)$, we find a quite different set of circumstances. For the $D3$ and $D2_{a}$ differentiation schemes, the forecasted values of $\bar{\sigma}^{-2}_{\M}$ are indistinguishable from the  $\mathscr{F}^{Noise}_{\M \M}$ noise distribution for all values of $N_{sims}$ up to and including 500. For the $D1_{a}$ differentiation scheme, the forecasted $\bar{\sigma}^{-2}_{\M}$ has started to separate from the $\mathscr{F}^{Noise}_{\M \M}$ distribution by the time $N_{sims}\simeq 100$, but the amplitude of the forecasted constraint at  $N_{sims}=500$ is still of a comparable magnitude to $\left<\mathscr{F}^{Noise}_{\M \M}\right>$. We consider this constraint to be in an intermediate regime, meaning that for $D1_{a}$, while the measured $\bar{\sigma}^{-2}_{\M,D1_{a}} = 1.10\ \mathrm{eV}^{-2}$ is largely distinguishable from the distribution of noise, it is still moderately influenced by the noise term $\mathscr{F}^{Noise}_{\M \M}$ via (\ref{eq:TCP}). As such, it should be viewed as an approximate estimate of the potential cosmological constraint, rather than an accurate one. The differences between the true $\sigma_{\M,D1_{a}}$, $\sigma_{\M,D2_{a}}$, and $\sigma_{\M,D3}$ due to intrinsic systematic error in each differentiation scheme is subdominant to the statistical noise present in the respective measurements of $\bar{\sigma}^{-2}_{\M}$.

\begin{figure*}[!t]
\includegraphics[width=1.0\linewidth]{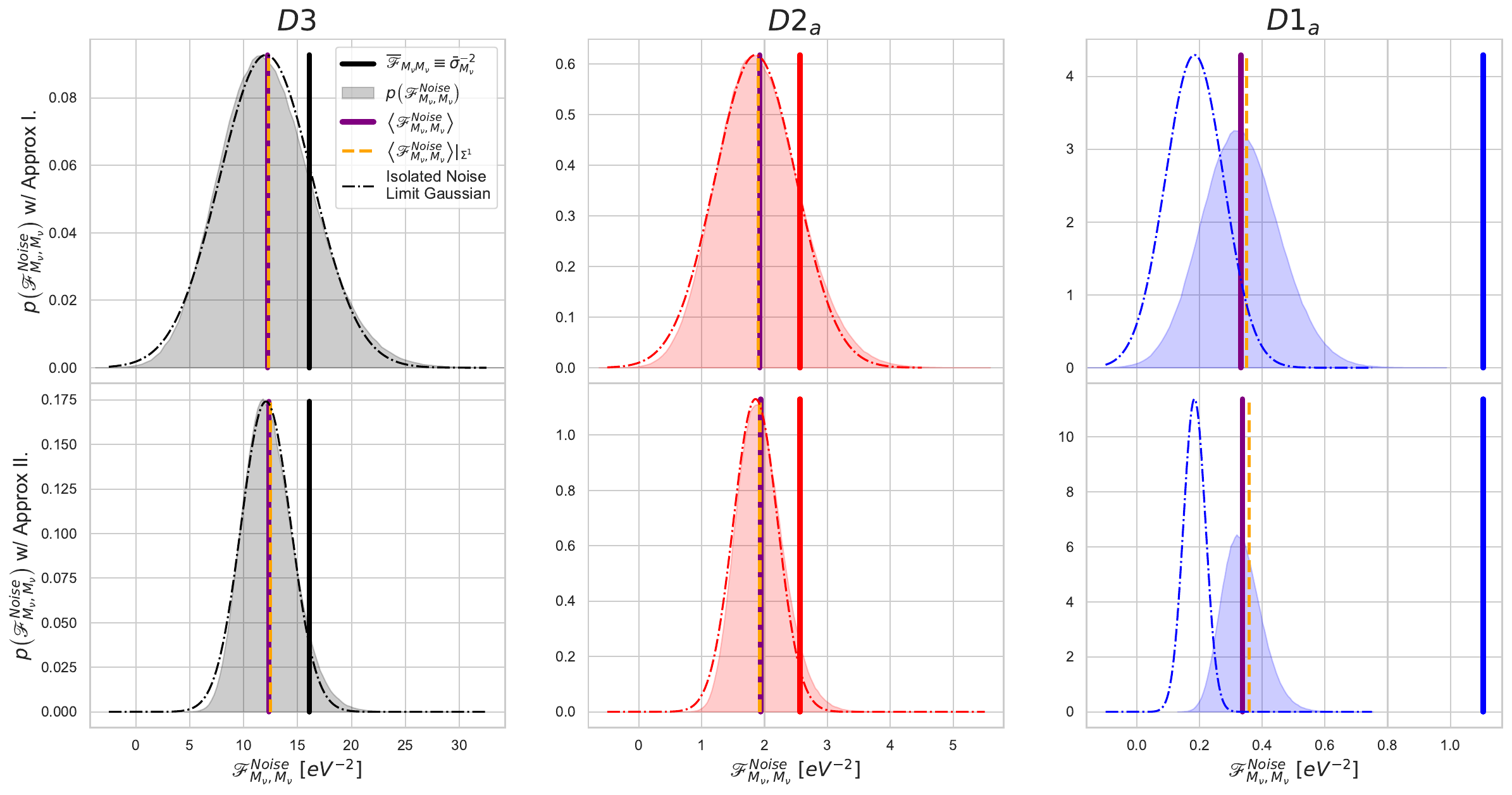}

\caption{The Fisher forecasted constraint, $\bar{\sigma}^{\ -2}_{\M}$, [thick vertical lines] for $P_{halos}(k)$ using $N_{sims}=500$ realization suites for $\M$ differentiation schemes $D3$ [left, black], $D2_{a}$ [center, red], and $D1_{a}$ [right, blue].  The associated noise distribution, $p\left(\mathscr{F}^{Noise}_{\M \M}\right)$ [shaded regions, same color], is also shown calculated with derivatives from [upper] Approx. I, $F_{AB}=\overline{F}_{AB}$, (\ref{eq:simApproxDeriv}) 
or [lower] Approx. II, ($F_{AB}=F_{AB}^{degen}$),(\ref{eq:degenDeriv}).
The  Gaussian approximation to $p\left(\mathscr{F}^{Noise}_{\M \M}\right)$ in the physical limit of isolated noise, (\ref{eq:INexp}) and (\ref{eq:INvar}), is shown [dot-dashed curves, same color] along with two estimates for the expectation of the noise Fisher, $\left<\mathscr{F}^{Noise}_{\M \M}\right>$: one calculated via numerical sampling [solid purple vertical line], and one using the perturbative approximation in the limit of large $N_{sims}$, in (\ref{eq:1/Nsims}) , $\left<\mathscr{F}^{Noise}_{\M \M}\right>\big\vert_{\Sigma^{1}}$ [dashed gold vertical line].
}
\label{fig:NoiseBothApprox}
\end{figure*}

We conclude from Fig.~\ref{fig:NoiseVsReals} that for $P_{halos}(k)$ when considering all differentiation schemes, one requires $N_{sims}\gg 500$ to be confident that the measured constraints could be considered truly converged to the underlying cosmological constraints. The results in Fig.~(\ref{fig:NoiseVsReals}) indicate that $D1_{a}$ gives the best estimate for $N_{sims}=500$, and will require a smaller total number of additional simulations than $D2_{a}$ or $D3$ to achieve this.

Figure~\ref{fig:NoiseBothApprox} shows the $N_{sims}=500$ slice of Fig.~\ref{fig:NoiseVsReals} for $P_{halos}(k)$, giving $\bar{\sigma}_{\M}^{-2}$ and $p\left(\mathscr{F}^{Noise}_{\M \M}\right)$ across each $\M$ differentiation scheme. The two rows compare the properties of the noise distributions $p\left( \mathscr{F}^{Noise}_{\M \M} \right)$ estimated using in Approx. II, (\ref{eq:degenDeriv}), as in Fig.~\ref{fig:NoiseVsReals}, and also calculated using Approx. I (\ref{eq:simApproxDeriv}). Two physical limits for the noise distributions are also considered: the limit of isolated noise and a perturbative expansion valid in the large $N_{sims}$ regime, both laid out in Appendix~\ref{sec:noiseLimits}. The main statistics for the noise distributions in Fig.~\ref{fig:NoiseBothApprox} are summarized in Table~\ref{tab:NoiseBothApprox}.

For all schemes, the noise probability distributions predicted using the more simple Approx. I (\ref{eq:meanDeriv}) are broader, and therefore a more conservative assessment of noise contamination, than those from the degenerate approximation (Approx. II), consistent with the discussion in \ref{sec:prac_imp}. 

Focusing first on the $D3$ $\M$ differentiation scheme in Fig.~\ref{fig:NoiseBothApprox} and summarized in Table~\ref{tab:NoiseBothApprox}, while the variance of the noise distribution depends on whether Approx. I or II is used, the means of both distributions are virtually identical. The measured constraint on $\M$, $\bar{\sigma}^{-2}_{\M,D3}= 16.1\ \mathrm{eV}^{-2}$, falls within the 68\% confidence interval of the Approx. I  distribution $\bar{p}$, and within the 95\% confidence interval of the Approx. II distribution $p^{degen}$. The fact that, for both approximations, $\bar{\sigma}^{-2}_{\M,D3} \simeq \mathscr{F}^{Noise}_{\M \M}$ means the measured value of $\bar{\sigma}^{-2}_{\M,D3}$ is potentially no different than a draw of the random variable $\mathscr{F}^{Noise}_{\M\M}$ and the $D3$ constraint therefor is clearly noise-dominated. Given $\bar{\sigma}^{-2}_{\M,D3} < \left<\mathscr{F}^{Noise}_{\M\M}\right> + \sqrt{\mathrm{Var}\left(\mathscr{F}^{Noise}_{\M \M} \right)}$ for the more conservative Approx. I, the estimated range for the noise-free constraint $\sigma^{-2}_{\M,D3}$ using (\ref{eq:trueConstApprox}) would include negative values, meaning there is no meaningful upper bound on $\sigma_{\M}$.

The $D3$ noise probability distributions for both  Approx. I or Approx II fall within the physical limit of isolated noise and are well characterized by the Gaussian distribution with the isolated noise expectation and variance, given by (\ref{eq:INexp}) and (\ref{eq:INvar}). The expectation values predicted from the perturbative large-$N_{sims}$ approach in (\ref{eq:1/Nsims}),  $\langle \mathscr{F}^{Noise}_{\M \M}\rangle\vert_{\Sigma^1}$, are also in excellent agreement with the means of each distribution.

\begin{table*}%
\begin{tabular}{|c|C{3em}|C{4em}|C{6em}|C{6em}|C{5em}|C{6em}||C{6em}|}
\hline
\multicolumn{2}{|c|}{\multirow{2}{*}{$eV^{-2}$}}
& \multirow{2}{*}{$\bar{\sigma}_{\M}^{-2}$} 
& \multicolumn{2}{c|}{$\left<\mathscr{F}^{Noise}_{\M\M} \right>\pm\sqrt{\mathrm{Var}\left(\mathscr{F}^{Noise}_{\M\M}\right)}$ }  
& \multirow{2}{*}{$\langle \mathscr{F}^{Noise}_{\M \M}\rangle \vert_{\Sigma^1}$ }
&  ${\sigma}_{\M}^{-2}$ 
&  ${\sigma}_{\M}$(eV) 
\\ \cline{4-5}\cline{7-8}
\multicolumn{2}{|c|}{} 
 & 
 & sampling
 & isol. noise
 &
 & sampling
 & sampling
\\ 
\hline
Approx. I
& $D3\ $ 
&  $16.1$
&  $12.2 \pm 4.4$
&  $12.1 \pm 4.3$
&  $12.3$
& $3.86 \pm 4.38$
& $0.51^{+\infty}_{-0.16}$
\\ 
$F_{AB}=\overline{F}_{AB}$ 
& $D2_{b}$   
&  $10.1$
&  $7.12 \pm 2.62$
&  $7.00\pm 2.56$
&  $7.17$ 
& $3.00 \pm 2.62$
& $0.58^{+1.05}_{-0.16}$
\\ 
& $D2_{a}$   
&  $2.56$
&  $1.92 \pm 0.65$
&  $1.86 \pm 0.65$
&  $1.90$ 
& $0.64 \pm 0.65$
& $1.25^{+\infty}_{-0.37}$
\\ 
& $D1_{c}$   
&  $3.78$
&  $2.29\pm 0.87$
&  $2.17\pm 0.84$
&  $2.31$ 
& $1.49 \pm 0.87$
& $0.82^{+0.45}_{-0.17}$
\\ 
& $D1_{b}$   
&  $1.29$
&  $0.69\pm 0.25$
&  $0.58\pm 0.23$
&  $0.68$ 
& $0.60 \pm 0.25$
& $1.29^{+0.40}_{-0.21}$
\\ 
& $D1_{a}$ 
&  1.10
&  $0.33 \pm 0.12$
&  $0.18 \pm 0.09$
& $0.35$ 
& $0.77 \pm 0.12$
& $1.14^{+0.10}_{-0.08}$
\\ 
  \hline
Approx. II
& $D3\ $
& $16.1$
& $12.3 \pm 2.3$ 
& $12.1 \pm 2.23$
& $12.5$ 
& $3.74 \pm 2.34$
& $0.52^{+0.33}_{-0.11}$
\\ 
$F_{AB}=F_{AB}^{degen}$ 
& $D2_{b}$   
&  $10.1$
&  $7.20\pm 1.37$ 
&  $7.00\pm 1.33$
&  $7.28$
& $2.92 \pm 1.37$
& $0.58^{+0.22}_{-0.10}$
\\ 
& $D2_{a}$
& $2.56$
& $1.94 \pm 0.37 $
& $1.86 \pm 0.35 $
& $1.93$
& $0.62 \pm 0.37$
& $1.27^{+0.71}_{-0.26}$
\\ 
& $D1_{c}$   
&  $3.78$
&  $2.31\pm 0.44$
&  $2.17\pm 0.41$
&  $2.35$ 
& $1.46 \pm 0.44$
& $0.83^{+0.16}_{-0.10}$
\\ 
& $D1_{b}$   
&  $1.29$
&  $0.70\pm 0.13$
&  $0.58\pm 0.11$
&  $0.60$ 
& $0.59 \pm 0.13$
& $1.30^{+0.17}_{-0.12}$
\\ 
& $D1_{a}$
&  $1.10$
&  $0.34 \pm 0.06$
& $0.18 \pm 0.04$ 
& $0.36$
& $0.77 \pm 0.06$
& $1.14^{+0.05}_{-0.04}$
\\ 
\hline
\end{tabular}
\caption{Summary of the Fisher forecast constraints for $P_{halos}(k)$, $\bar{\sigma}^{-2}_{\M}$,  along with the properties of their associated noise distributions using both Approx. I and Approx. II for all $\M$ differentiation schemes (\ref{eq:s1_ppp}) - (\ref{eq:f3}), as shown in Fig.~\ref{fig:NoiseBothApprox} for only $D3$, $D2_{a}$, and $D1_{a}$. Relevant quantities obtained from sampling as well as calculated analytically under the assumptions of ``isolated noise" and ``large $N_{sims}$", $\left<\mathscr{F}^{Noise}_{\M \M}\right> \vert_{\Sigma^1}$, are also summarized along with the inferred noise-free constraints, $\sigma_{\M}^{-2}$ and $\sigma_{\M}$, using (\ref{eq:trueConstApprox}).}
\label{tab:NoiseBothApprox}
\end{table*}

We find similar conclusions for the $D2_a$ scheme. The measured constraint, $\bar{\sigma}^{-2}_{\M,D2_{a}} = 2.56\ \mathrm{eV}^{-2}$, is still slightly greater than the mean noise $\left<\mathscr{F}^{Noise}_{\M\M} \right>$ for both Approx. I and Approx. II derivative approximations, but is just within the 68\% confidence interval using the more conservative Approx. I for $p\left(\mathscr{F}^{Noise}_{\M\M} \right)$. The isolated noise predictions for the full noise distributions and the large-$N_{sims}$ perturbative predictions for the expectation values both agree very well with the values coming from the full numerical sampling approach.  

The $D1_{a}$ scheme for $\M$ presents a different conclusion relative to $D2_a$ and $D3$. First, the measured constraint $\bar{\sigma}^{-2}_{\M,D1_{a}}= 1.10\ \mathrm{eV}^{-2}$ is significantly larger than the associated prediction for the noise distribution, $p\left(\mathscr{F}^{Noise}_{\M\M}\right)$ for both Approx. I and II. Through the lens of (\ref{eq:TCP}), this means that the true cosmological constraint $\sigma^{-2}_{\M,D1_{a}}$ is contributing significantly to the forecasted $\bar{\sigma}^{-2}_{\M,D1_{a}}$. This separation of $\bar{\sigma}^{-2}_{\M,D1_{a}}$ and $p\left(\mathscr{F}^{Noise}_{\M\M}\right)$ allows for the use of (\ref{eq:trueConstApprox}) to calculate a meaningful $68\%$ confidence interval for the true $\sigma^{-2}_{\M,D1_{a}}$, as shown in Table~\ref{tab:NoiseBothApprox}. 

From Fig.~\ref{fig:NoiseBothApprox} one can also see that $p\left(\mathscr{F}^{Noise}_{\M \M}\right)$ for the $D1_{a}$ scheme is no longer within the limit of isolated noise. Physically, this means that statistical noise is no longer solely dominated by $\M$ and noise from the non-target parameters is also meaningfully contributing to the overall noise distribution. Previously, this was not the case for either $D2_{a}$ or $D3$, as for both, non-target parameter noise contributed negligibly to $p\left(\mathscr{F}^{Noise}_{\M \M}\right)$. While the isolated noise limit no longer applies, we find  the perturbative expansion in the large-$N_{sims}$ limit, $\left<\mathscr{F}^{Noise}_{\M \M}\right>\vert_{\Sigma^1}$,  discussed in Appendix~\ref{sec:noiseLimits}, continues to provide a good estimate for the expectation value of the $D1_{a}$ noise distribution.

The accuracy of the large-$N_{sims}$ limit, $\left<\mathscr{F}^{Noise}_{\M \M}\right>\vert_{\Sigma^1}$, in reproducing $\left<\mathscr{F}^{Noise}_{\M \M}\right>$ for all scenarios considered in Fig.~\ref{fig:NoiseBothApprox} indicates, as discussed in appendix \ref{sec:noiseLimits}, that the mean noise contribution to $\bar{\sigma}^{-2}_{\M}$ is falling as $1/N_{sims}$ beyond $N_{sims}=500$. This provides us with a very useful tool to approximate how many more simulations would need to be run before the noise contribution to any given $\bar{\sigma}^{-2}_{\M}$ fell below a certain level. 

For $D1_a$ with $N_{sims}=$500, $\bar{\sigma}^{-2}_{\M,D1_{a}}$ is comprised of $\simeq 30\%$ noise, so that $\left<\mathscr{F}^{Noise}_{\M \M}\right> / \sigma^{-2}_{\M} \simeq 0.43$. For each additional order of magnitude we increase $N_{sims}$ beyond 500, this ratio will fall by an equivalent order of magnitude; for $N_{sims}=5000$, we would expect $\left<\mathscr{F}^{Noise}_{\M \M}\right> / \sigma^{-2}_{\M} \simeq 0.04$, and $\bar{\sigma}^{-2}_{\M}$ to only be $\sim 4\%$ noise. 

To apply this methodology to the heavily noise dominated $D2_a$ and $D3$, we may assume that, just as for $P_{parts}(k)$, the true noise-free values $\sigma^{-2}_{\M,D3}$ and $\sigma^{-2}_{\M,D2_{a}}$ can be well-approximated by the roughly converged value in $D1_a$, $\sigma^{-2}_{\M,D1_{a}} \simeq 0.8\ \mathrm{eV}$. Extrapolating the $N_{sims}=$ 500 result for $D3$ in Table~\ref{tab:NoiseBothApprox} to $N_{sims}=$ 5000 would mean that $\left<\mathscr{F}^{Noise}_{\M \M} \right> = 1.2$, which predicts $\bar{\sigma}^{-2}_{\M,D3}$ would still be over $50\%$ noise; one would require $N_{sims}\sim$ 70000 to achieve 10$\%$ noise! For $D2_{a}$ with $N_{sims}=$ 5000, we expect $\left<\mathscr{F}^{Noise}_{\M \M} \right> = 0.19$, so that $\bar{\sigma}^{-2}_{\M,D2_{a}}$ would contain $\sim20\%$ noise.

Finally, while we have been primarily focused on the $D1_a$, $D2_{a}$, and $D3$ differentiation schemes, we  summarize the results  for all six differentiation schemes in Table \ref{tab:NoiseBothApprox}. The three first-order schemes, $D1_a$, $D1_{b}$ and $D1_c$ differ only in the finite step size, $\Delta \M$, used in the differencing. Typically one expects a smaller step size to be preferable as it will have lower systematic uncertainties in estimating the true $\Delta \M\rightarrow 0$ derivative (as long as it is does not fall below the precision level of the analytic/simulation tool being used). As seen in Table \ref{tab:NoiseBothApprox},  however, we find that the constraints from first-order schemes with smaller steps sizes are also more contaminated by statistical noise - mirroring the higher-order differentiation schemes, which have lower systematic uncertainty but higher statistical noise. The larger noise translates into tighter predicted parameter constraints.  The order of the predicted noise levels in Table~\ref{tab:NoiseBothApprox} is aligned with the relative size of the confidence ellipses for the six derivative schemes in Fig.~\ref{fig:partHaloDerivEllipsesBoth}.

The key take-away from this analysis is that when balancing systematic accuracy versus statistical uncertainty both the order of the differentiation scheme and the finite step size must be taken into account. While the first-order schemes generally have lower noise than higher-order ones, the noise can still form a non-negligible fraction of their predicted constraints. As long as the noise is subdominant, (\ref{eq:trueConstApprox}) can be used to estimate a confidence interval for the noise-free constraint. 

\subsection{Mock Galaxy Tracers and HOD Parameter Inclusion}
\label{sec:mock_hod}

\begin{figure*}[!t]
\includegraphics[width=1.0\linewidth]{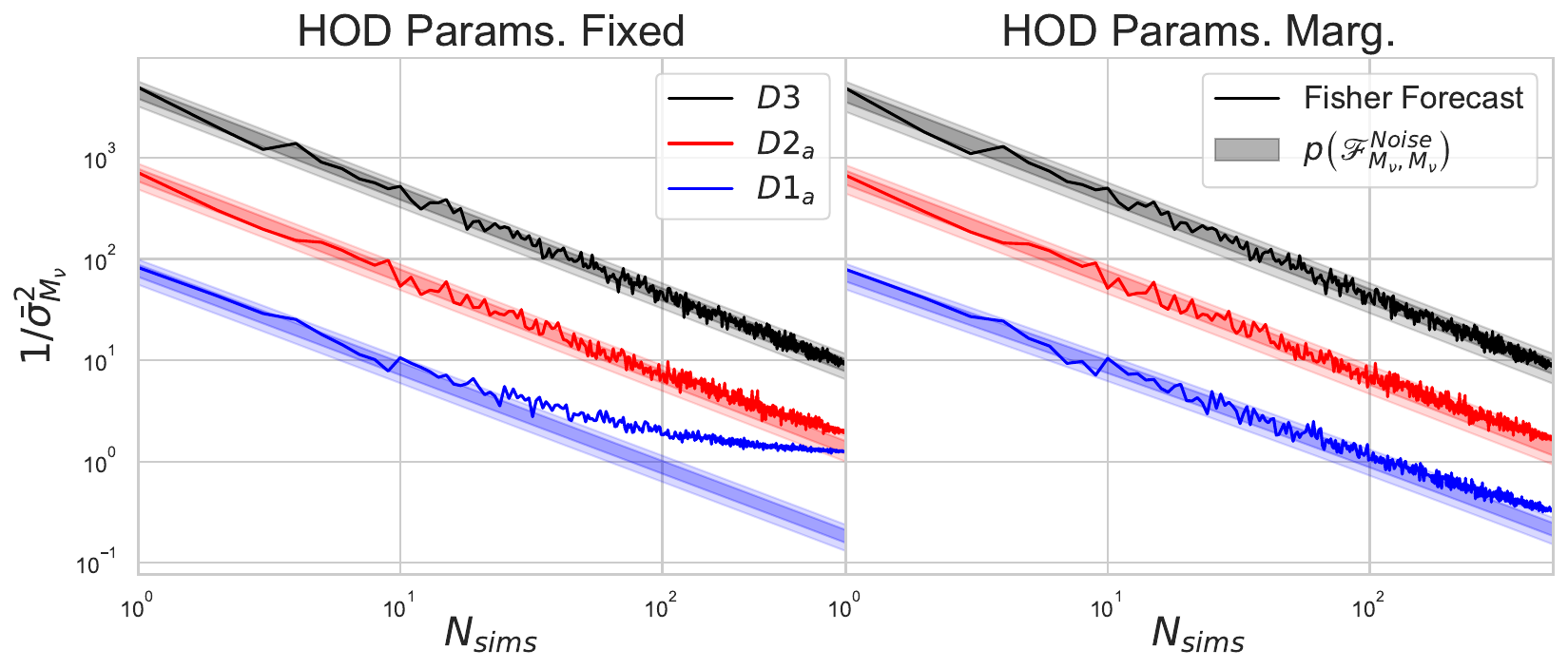}
\caption{The measured Fisher forecast neutrino mass constraint, $1/\bar{\sigma}^{2}_{M{\nu}}$, for the HOD-derived galaxy 2-point statistics $P^{g}_{0}(k)$ and $P^{g}_{2}(k)$ out to $k < 0.5 h / Mpc$,  as a function of $N_{sims}$. for three $\M$ numerical differentiation schemes: $D1_{a}$ [blue], $D2_{a}$ [red], and $D3$ [black]. The constraints are shown with the HOD fitting parameters either [left] fixed at their fiducial values or [right] included in the forecast as parameters to be marginalized over. The $68\%$ and $95\%$ confidence intervals for the statistical noise distributions associated to each differentiation scheme, $p\left( \mathscr{F}^{Noise}_{\M \M} \right)$, are shown as shaded regions corresponding to the scheme's color. In all instances, $p\left( \mathscr{F}^{Noise}_{\M \M} \right)$ is approximated using $F_{AB}=F^{degen}_{AB}$.}
\label{fig:NoiseVSreals_HOD_Pk}
\end{figure*}

\subsubsection{Galaxy 2-point statistics, $P^{g}_{0}(k)$ \& $P^{g}_{2}(k)$}

So far we have focused on constraints using the halo power spectrum as a prospective observable. In reality, galaxies, not halos, are used as the measurable tracer of the dark matter distribution. In this section, to better align with observations, we switch our focus from the various real space matter power spectra to the monopole and quadrupole of the redshift space galaxy power spectrum, $P^{g}_{0}(k)\ \&\ P^{g}_{2}(k)$, and use halo occupation distribution (HOD) derived mock galaxies from the Molino suite of catalogs to estimate the galaxy clustering statistics. 

Fig.~\ref{fig:NoiseVSreals_HOD_Pk} is equivalent to Fig.~\ref{fig:NoiseVsReals} except for the use of the redshift space galaxy power spectrum monopole and quadrupole as the observables rather than the various real space power spectra. It shows $\bar{\sigma}^{-2}_{\M}$ versus $N_{sims}$ using $P^{g}_{0}(k)$ and $P^{g}_{2}(k)$ as the joint observables for the $D3$, $D2_{a}$, and $D1_{a}$ differentiation schemes. In the analysis we consider predicted constraints with and without marginalization over the HOD nuisance parameters.  When marginalized, the HOD parameters are included amongst the non-target parameters alongside the non-$\M$ cosmological parameters. 

As with the dark matter halo power spectra, the constraints provided by the $D2_{a}$ and $D3$ differentiation schemes for the galaxy statistics are entirely noise dominated for $N_{sims}=500$, and have not begun to approach an asymptotic true cosmological value, regardless of the inclusion or exclusion of HOD parameters in the forecast.  When HOD parameters are fixed, and excluded from the marginalization, $D1_{a}$ provides an $\M$ constraint which exceeds the noise contribution, as indicated by the fact that $\overline{\sigma}_{\M,D1_{a}}^{-2}$ is $\sim$7 times larger than what noise alone on average could produce, given $p\left( \mathscr{F}^{Noise}_{\M \M} \right)$. This permits the use of (\ref{eq:trueConstApprox}) to calculate the $68\%$ confidence limit for the true $D1_{a}$ constraint. Using Approx. I (\ref{eq:meanDeriv}), we find $\sigma^{-2}_{\M,D1_{a}} = \left(1.06 \pm 0.07\right) eV^{-2}$ which translates to $\sigma_{\M,D1_{a}} = \left(0.97 \pm 0.03\right)\ \mathrm{eV}$. Comparing this estimate to the noise influenced estimate, $\bar{\sigma}_{\M,D1_{a}}=0.89\ \mathrm{eV}$, we can see that even small levels of statistical noise in the numerical derivatives can easily bias the results of the Fisher forecast by upwards of $10\%$. 

Our work provides an expected value and confidence interval for the true inverse square constraint $\sigma^{-2}_{\mu}$ through a full statistical analysis of the noise distribution $p\left(\mathscr{F}^{Noise}_{\mu\mu} \right)$. We note that this may be compared with the work of Coulton et al. \cite{coulton2023estimate} in which the authors derive a compressed Fisher matrix, $\overline{F}^{comp}$, which they find is biased low by noise. They propose a combined estimator, $\overline{F}^{comb}$, that is the geometric mean of $\overline{F}$ and $\overline{F}^{comp}$ to accelerate convergence towards the true constraint as $N_{sims}$ increases. In this regime for $D1_a$, in which noise is truly the subdominant effect, we find that applying the combined estimator of \cite{coulton2023estimate} provides an unbiased estimate within our confidence interval of $\bar{\sigma}^{-2}_{\M,comb}=1.05 eV^{-2}$. 

\begin{figure*}[t]
\includegraphics[width=1.0\linewidth]{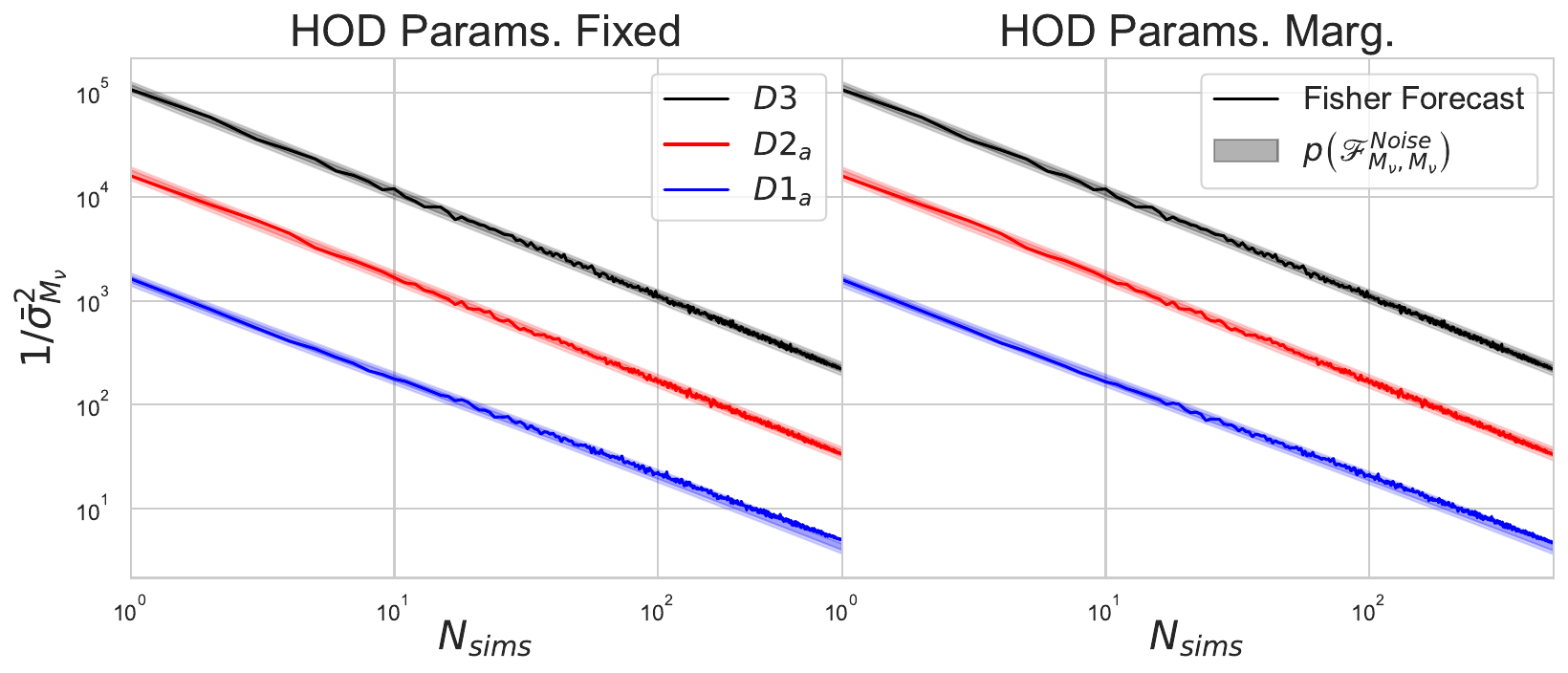}
\caption{As in Fig.~\ref{fig:NoiseVSreals_HOD_Pk} but for the galaxy redshift space bispectrum monopole, $B^{g}_{0}$, calculated from HOD derived mock galaxies to $k < 0.5 h / Mpc$.}
\label{fig:NoiseVSreals_HOD_Bk}
\end{figure*}

Figure~\ref{fig:NoiseVSreals_HOD_Pk} also considers the effect of marginalizing over the HOD parameters as nuisance parameters in  the Fisher forecast. In all cases, the inclusion of these new marginalizing parameters, causes the true value of $\sigma_{\M}^{-2}$ to fall, regardless of differentiation scheme, as expected. Even before the inclusion of the HOD parameters in the marginalization, the measured constraints for $D3$ and $D2_{a}$ were already noise-dominated, satisfying $\bar{\sigma}^{-2}_{\M} \sim \mathscr{F}^{Noise}_{\M \M}$.  As such, the decrease in the values of $\sigma^{-2}_{\M,D2_{a}}$ or $\sigma^{-2}_{\M,D3}$, without significant changes to the associated $p\left(\mathscr{F}^{Noise}_{\M\M}\right)$ distributions, does not significantly change $\bar{\sigma}_{\M^{-2}}$. Without the appreciation that this is a noise driven result, this could be falsely interpreted as the observables $P_{0}^{g}(k)$ \& $P_{2}^{g}(k)$ giving a cosmological constraint on $\M$ that is almost entirely immune or insensitive to the effects of HOD parameter marginalization! By contrast, for $D1_{a}$, the effect of the HOD parameters on the true constraints is clearly visible: their inclusion causes $\bar{\sigma}_{\M,D1_{a}}^{-2}$ to markedly decrease and mostly realign with the associated $p\left(\mathscr{F}^{Noise}_{\M\M}\right)$ noise distribution, which has yet to drop significantly below the new, lower true value of $\sigma_{\M,D1_{a}}^{-2}$. 

The broad conclusion from the discussion above is that irrespective of the differentiation scheme, all of the predicted $\M$ constraints using $P_{0}^{g}(k)\ \&\ P_{2}^{g}(k)$ are heavily noise influenced when HOD parameters are included in the Fisher forecast. When noise dominates the Fisher, it acts to throttle the normal effect that adding additional nuisance parameters would have, in typically introducing degeneracies that dilute marginalized constraints. The implication is that a greater number of simulations, $N_{sims}>$500, is required before one can utilize galaxy statistics from simulations to estimate the noise-free constraints, $\bar{\sigma}_{\M}^{-2}$, for even the most conservative of $\M$ differentiation schemes.

\subsubsection{The galaxy bispectrum, $B_{0}^{g}(k_{123})$}

When HOD parameters are included in the marginalization process, the complete Molino suite with $N_{sims}=500$, five independent HOD applications, and three independent lines of sight, is insufficient to produce converged Fisher $\M$ constraints for the two-point observables $P^{g}_{0} (k)\ \&\ P^{g}_{2}(k)$, even when employing $D1_{a}$, the most conservative $\M$ differentiation scheme. This suggests that it is also important to test the convergence of constraints arising from the Molino suite for higher-order clustering statistics, such as the redshift space bispectrum monopole,  as was studied in \cite{bispectrumII}. 

Figure~\ref{fig:NoiseVSreals_HOD_Bk} shows $\bar{\sigma}^{-2}_{\M}$ versus $N_{sims}$ for $B^{g}_{0}$ with and without marginalization over the HOD parameters. Immediately, it is clear that none of the constraints from the bispectrum, regardless of the $\M$ differentiation scheme, are distinguishable from the associated distribution of pure noise constraints given by $p\left(\mathscr{F}^{Noise}_{\M \M} \right)$, here calculated using Approx. II. Our analysis shows that not a single one of these constraints, are able to be used as cosmological constraints as they are all purely the result of numerical derivative noise.

We note that the authors of \cite{bispectrumII} refer to a $< 10\%$ variation in $\sigma_{\theta}$ for over the interval $N_{sims} \in \left[6000,7500\right]$ as a proof of convergence towards the cosmological constraint. This, however, is actually simply indicative of the fact that the calculated $\bar{\sigma}_{\theta}$ are primarily derived from noise. For a noise-dominated constraint in the limit of large $N_{sims}$, $\bar{\sigma}^{-2}_{\theta} \simeq  \left<\mathscr{F}^{Noise}_{\theta \theta}\right> \propto \Sigma^{1} \propto  1/N_{sims}$. Consequently, over the interval considered in \cite{bispectrumII}, one expects the forecasted $\bar{\sigma}_{\theta}$ would experience a change of approximately $\frac{\sqrt{7500} - \sqrt{6000}} { \sqrt{7500}} = 10.6\%$ if its primary contribution was coming from $\mathscr{F}^{Noise}_{\theta \theta}$, which is the same level as the authors of \cite{bispectrumII} report. As shown in Fig~\ref{fig:NoiseVsReals} for $P_{part}(k)$, and for $P_{halos}(k)$ (for $D1_{a}$ only), a truly or even partially converged $\bar{\sigma}_{\M}$ would vary by much less than $10\%$ on our equivalent $N_{sims} \in [400,500]$ interval, as the primary contribution to $\bar{\sigma}_{\M}$ would be derived from the true $\sigma_{\M}$, which is constant with respect to $N_{sims}$.

\subsection{Non-Neutrino Mass Parameters}
\label{sec:notMnu}

\subsubsection{Marginalized Constraints}

\begin{figure*}[!t]
\includegraphics[width=1.0\linewidth]{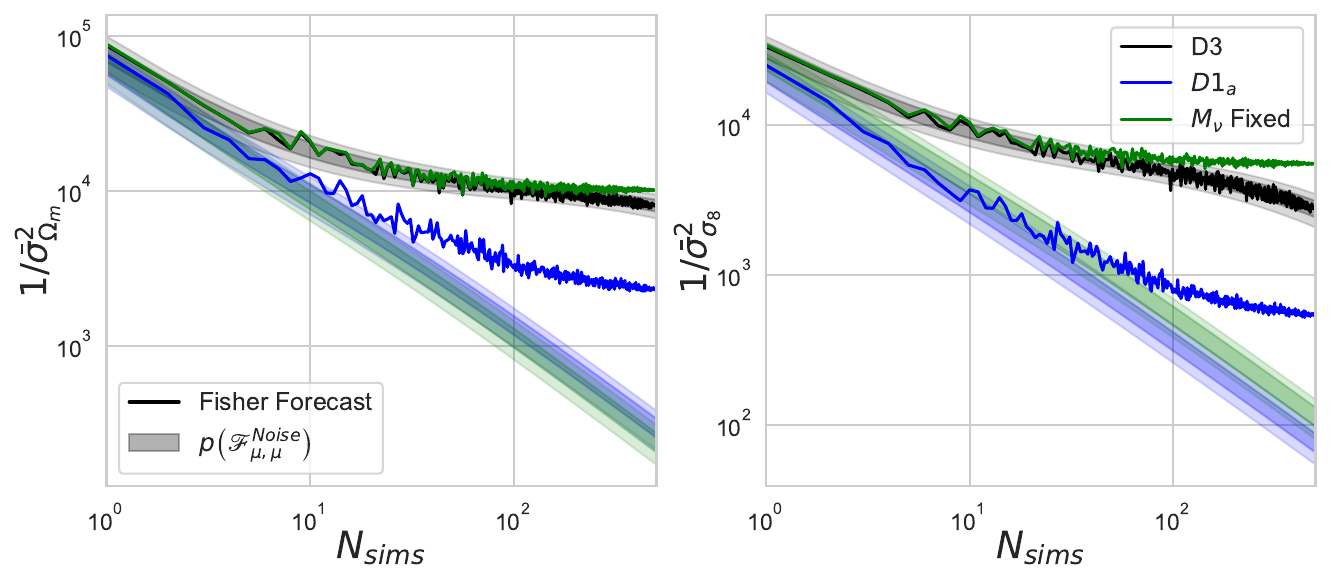}
\caption{Predicted constraints, $1/\bar{\sigma}^{2}_{{\mu}}$, for 
[left] $\theta^{\mu}=\Om$  and [right] $\theta^{\mu}=\s8$ 
as a function of $N_{sims}$, with $P^{g}_{0}(k)+P^{g}_{2}(k)$ as the observables and HOD parameters fixed. Two $\M$ differentiation scheme are shown, $D3$ [black], $D1_{a}$ [blue], plus with $\M$ fixed and excluded from the forecast [green]. Shaded regions in the same colors denote the $68\%$ and $95\%$ confidence intervals of the associated noise distribution, $p\left( \mathscr{F}^{Noise}_{\mu \mu}\right)$.}
\label{fig:NonMnuParams}
\end{figure*}

In the previous section, we focused on constraints for a single target parameter, $\theta^{\mu} =\M$. This was motivated by the very different $\bar{\sigma}_{\M}$ constraints obtained from the different $\M$ differentiation schemes (\ref{eq:s1_ppp}) - (\ref{eq:f3}) in Fig. \ref{fig:partHaloDerivEllipsesBoth}. Ultimately, the differences in the $\bar{\sigma}_{\M}$ constraints were traced to different levels of statistical noise in the respective  $\M$ derivatives, with the noise systematically tightening the measured $\bar{\sigma}_{\M}$constraints with respect to the $N_{sims} \rightarrow \infty$ values.

We now turn to consider the impact of derivative noise on non-$\M$ parameter constraints. There is no reason apriori to assume that the first-order derivatives used for non-$\M$ parameters, $D0$ in (\ref{eq:s1_pm}), should be considered entirely noise free. It also remains to be determined how the choice of $\M$ differentiation scheme impacts the constraints for the non-$\M$ parameters through intrinsic degeneracies. Accordingly, we will now examine constraints on two specific non-$\M$ target parameters,  $\Om$ and $\s8$, for which the cosmological impacts on large scale structure observables are typically quite degenerate with $\M$.  

Figure~\ref{fig:NonMnuParams} shows how $\bar{\sigma}^{-2}_{\Om}$ and $\bar{\sigma}^{-2}_{\s8}$ for the observables $P^{g}_{0}(k)\ \&\ P^{g}_{2}(k)$ with fixed HOD parameters evolve as a function of $N_{sims}$. Three treatments of $\M$ are considered: $\M$ is included in the Fisher using either $D1_{a}$ or $D3$, or excluded from the forecast entirely, i.e, $\M$ fixed. Shaded regions again represent the relevant $\mathscr{F}^{Noise}_{\mu \mu}$ noise distributions calculated using Approx. II (\ref{eq:degenDeriv}). 

To estimate the noise distributions for $\Om$ and $\s8$ in the presence of the noisy $D3$ $\M$ derivative, we use Approx. II, the degenerate Fisher approximation, but apply it {\it not} to the target parameter, $\Om$ or $\s8$, but to the $D3$ $\M$, as the noisiest parameter. As discussed in Appendix \ref{sec:NumConvergence}, when high levels of noise are confined to a single parameter, the degenerate Approx. II applied to that noisy parameter estimates the noise distribution very well. 

First consider the constraints on $\Om$ and $\s8$ when $\M$ is fixed and not marginalized over. For $N_{sims}=500$ simulations, the statistical noise term, $p\left(\mathscr{F}^{noise}_{\mu \mu}\right)$, is well below the cosmological constraint amplitude, $\sigma^{-2}_{\mu}$, for both $\Om$ and $\s8$. 

If we include $\M$ as a non-target parameter in the analysis, the degeneracies between $\M$ and $\Om$ or $\s8$ allow the statistical noise in the $\M$ derivative to bleed through into the marginalized constraints on $\Om$ and $\s8$. As we have seen, the statistical noise in $\M$ artificially tightens the constraints on $\M$. When considering the 2D joint constraints on $\{\Om, \M\}$ or $\{\s8, \M\}$, noise not only artificially tightens $\overline{\sigma}_{\M}$, but also acts to artificially break the degeneracies between the two parameters.

As shown in Fig.~\ref{fig:NonMnuParams}, this noise-based degeneracy breaking can lead to the (incorrect) inference, particularly acute for the noisiest differentiation scheme, $D3$, that the constraints on $\Om$ and $\s8$ when $\M$ is marginalized over are almost as tight as when $\M$ is fixed.  By comparison, the less noisy $D1_a$ $\M$ derivatives lead to significantly weaker constraints on $\Om$ and $\s8$, reflecting the intrinsic reduction in constraining power for $\Om$ and $\s8$ due to the cosmological degeneracies, but that are still, however, above the predicted noise levels. This is the same mechanism which leads to the findings in Fig.~\ref{fig:partHaloDerivEllipsesBoth} in which the differentiation schemes for $\M$, and their inherent noise, drive the parameter constraints not just for $\M$ but the other target parameters, with the tightest $\s8$ or $\Om$ parameter constraints from $P_{halos}(k)$ predicted for the noisiest $\M$ differentiation schemes.

In the context of analytic marginalization, random noise on the true smooth $\M$ derivative reduces its effectiveness as a basis function for reproducing the derivative of other target parameters ($\Om$ or $\s8$), thereby reducing the inherent degeneracies and tightening the constraints on the non-$\M$ target parameter in question. For $\Om$ especially, when $D3$ is used as the $\M$ derivative, $d\overline{\sO}{}^{\ i}_{\M}$ becomes almost \textit{entirely useless} as a basis function for reproducing $d\overline{\sO}{}^{\ i}_{\Om}$, and thus the measured constraint for $\Om$ when $\M$ is marginalized is effectively just as tight as when $\M$ is fixed and excluded from marginalization altogether. 

The main takeaway is that when considering the marginalized constraints of multiple parameters, it is necessary to  understand the statistical derivative noise of \textit{all} included parameters and determine if $N_{sims}$ is sufficiently large to allow the parameter with the greatest noise to have constraints above its noise floor. Practically, when derivative noise is subdominant for the noisiest parameter, then we find concerns about its impact on the other target parameters will be ameliorated.

Likewise, when combining a number of different observables, it is imperative to check that $N_{sims}$ is sufficient to reduce the derivative noise for all observables. If for one observable the constraints are noise-dominated, and therefore tighter than the true cosmological constraints, its combination with other observables (even if they are not themselves affected by statistical derivative noise) will lead to artificially tighter overall constraints and apparent degeneracy breaking that is purely an artifact of noise rather than cosmological signal. We suspect that this is the case for the ``degeneracy breaking" discussed in \cite{bispectrumI}, where the $D3$ differentiation scheme is used for $\M$ in a Fisher forecast of the halo bispectrum monopole from the Quijote simulations. Our analysis concludes that the true cosmological signal cannot be determined as the forecasted constraints are noise-dominated.

\subsubsection{Correlation Coefficients}

The noise-based (false) degeneracy breaking can also be quantified through considering the degree to which two parameters, $\theta^{\mu}$ and $\theta^{\nu}$, are correlated with each other. Given that the inverse of the Fisher information matrix $\left(F^{-1}\right){}^{AB}$ may be interpreted as the parameter covariance matrix, the correlation between two target parameters $\left\{\theta^{\mu},\theta^{\nu}\right\}$ is defined as 
\begin{equation}
\mathrm{Corr}(\theta^{\mu},\theta^{\nu})\equiv\frac{(F^{-1})^{\mu \nu}}{\sqrt{(F^{-1})^{\mu \mu}(F^{-1})^{\nu \nu}}}=\frac{-\mathscr{F}_{\mu \nu}}{\sqrt{\mathscr{F}_{\mu \mu}\mathscr{F}_{\nu \nu}}}.
\label{eq:Corr}
\end{equation}
As written, (\ref{eq:Corr}) applies only in the limit $N_{sims} \rightarrow \infty$. To include the effects of statistical noise, we modify this to include the noise contaminated quantity, $\overline{\mathscr{F}}_{\mu \nu} = \mathscr{F}_{\mu \nu} + \mathscr{F}^{Noise}_{\mu \nu}$, of dimension $2 \times 2$, given in (\ref{eq:margFisherNoise}),
\begin{equation}
\overline{\mathrm{Corr}}(\theta^{\mu},\theta^{\nu})=\frac{-\left(\mathscr{F}_{\mu \nu} + \mathscr{F}^{Noise}_{\mu \nu}\right)}{\sqrt{\left(\mathscr{F}_{\mu \mu} + \mathscr{F}^{Noise}_{\mu \mu}\right)\left(\mathscr{F}_{\nu \nu} + \mathscr{F}^{Noise}_{\nu \nu}\right)}}.
\label{eq:CorrNoise}
\end{equation}
To illustrate this, consider $P_{0}^{g}(k)\ \&\ P_{2}^{g}(k)$  with $\{\theta^{\mu},\theta^{\nu}\} = \{\M,\Om\}$ as our target parameter set, and the remaining set of cosmological parameters $\{h,n_{s},\Omega_{b},\Om \}$ as the non-target set which are marginalized over. 

The $\mathscr{F}^{Noise}_{\M \M}$ noise term in the denominator is positive and potentially large compared to the true $\mathscr{F}_{\M \M}$. Thus, noise acts to systematically increase the denominator of (\ref{eq:CorrNoise}). By contrast, the effects of noise on the numerator, are less pronounced; the off diagonal noise terms $\mathscr{F}^{Noise}_{\mu \nu}$ can be positive or negative and are typically smaller in magnitude than the diagonal noise terms when compared relative to their respective $\overline{\mathscr{F}}$ entries. This can be seen most easily in the isolated noise limit in Appendix~\ref{sec:noiseLimits} for which both $\left<\mathscr{F}^{Noise}_{\mu \nu} \right>$ and $\mathrm{Var}\left(\mathscr{F}^{Noise}_{\mu \nu}\right)$ become relatively simple expressions of the mean derivative covariance matrix $\Sigma^{ij}_{\mu\nu}$. When $\mu\neq\nu$, the expectation, $\left<\mathscr{F}^{Noise}_{\mu \nu} \right>$, in (\ref{eq:INexpMixed}), depends only on the smaller, off-diagonal entries $\Sigma^{ij}_{\mu \nu}$. The $\mu\neq\nu$ variance, $\mathrm{Var}\left(\mathscr{F}^{Noise}_{\mu \nu}\right)$, given in (\ref{eq:INvarMixed}), however, also depends on the larger diagonal entries, $\Sigma^{ij}_{\mu \mu}$ and $\Sigma^{ij}_{\nu \nu}$. This means that typically, for $\mu\neq\nu$, $\mathscr{F}^{Noise}_{\mu \nu}=0$ lies somewhere near the center of the full distribution $p\left(\mathscr{F}^{Noise}_{\mu \nu}\right)$, so that individual instances of $\mathscr{F}^{Noise}_{\mu \nu}$ can be positive or negative without too large of a preference for either sign. 

All together, while the numerator of (\ref{eq:CorrNoise}) is randomly increased or decreased by individual instances of noise, the denominator is systematically increased over its $N_{sims}\rightarrow \infty$ value, and drastically so in many cases of interest. Thus, as $N_{sims}$ increases, $\overline{\mathrm{Corr}}\left(\theta^{\mu},\theta^{\nu}\right)$ should initially grow in magnitude as the denominator decreases and eventually asymptote to a constant value, $\mathrm{Corr}\left(\theta^{\mu},\theta^{\nu}\right)$. 

\begin{figure}[!t]

\includegraphics[width=1.0\linewidth]{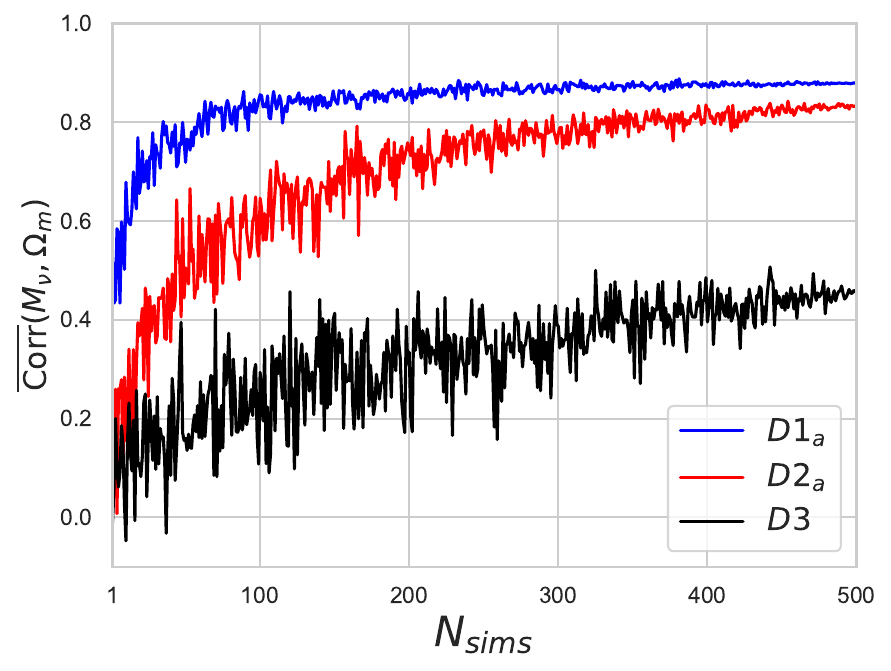}
\caption{The correlation between \{$\M,\Om$\} using $P^{g}_{0}(k)+ P^{g}_{2}(k)$ as the observable, as a function of the number of sets of simulations $N_{sims}$ and the $\M$ derivative scheme: $D1_{a}$ [blue], $D2_{a}$ [red], and $D3$ [black].}
\label{fig:CorrPlot}
\end{figure}

Figure \ref{fig:CorrPlot} shows how the trends in $\overline{\mathrm{Corr}}$ for $\{\M, \Om\}$, for mock galaxies with HOD parameters fixed, as $N_{sims}$ increases are consistent with the findings above. 

When $\M$ is one of the two target parameters, in any regime where $\overline{\mathrm{Corr}}$ remains noise-dominated, the relative ordering of the magnitudes of $\overline{\mathrm{Corr}}$ for the different  differentiation schemes, $D1_a$ through $D3$, should be the reverse order of the amount of statistical noise in each scheme. Thus, the $D1_{a}$ differentiation scheme features the largest absolute value of $\overline{\mathrm{Corr}}$ and $D3$ the smallest, due to the respective amounts of statistical noise in the $\M$ numerical derivatives. The relative correlations shown in Fig.~\ref{fig:CorrPlot} are  consistent with the 2D contour ellipses in Fig.~\ref{fig:partHaloDerivEllipsesBoth}. The $D3$ results, in which the statistical noise acts as if to break degeneracies between parameters, have contours that are inherently more circular and less correlated than those from $D1_a$, 
which has lower statistical noise.

\section{Conclusions}
\label{sec:conclusion}

In this work, we conduct a sytematic and thorough examination of the various impacts of noise in numerical derivatives on inferences from simulation-based Fisher forecasts. We determine that it is statistical noise within the numerical derivatives, and not differing levels of systematic error, that is responsible for the markedly different Fisher-derived confidence ellipses in Fig.~\ref{fig:partHaloDerivEllipsesBoth}. Principally, we find that the coupling between the marginalization process and statistical noise in numerical derivatives from a finite sample size, $N_{sims}$, biases the associated Fisher forecast. As a result, the predicted marginalized constraints are artificially tightened by noise relative to the true cosmological ones obtained in the $N_{sims} \rightarrow \infty$ limit.

We perform a detailed analysis, modeling the effects of statistical noise in the numerical derivatives through the multivariate Gaussian random variable $u^{i}_{A}$. Tracing the effects of this random variable through our forecast allows us to separate the measured marginalized Fisher matrix, $\overline{\mathscr{F}}_{\mu \nu}$, into a  term, $\mathscr{F}_{\mu \nu}$, that characterizes the true underlying cosmological constraints and a random variable noise term, $\mathscr{F}^{Noise}_{\mu \nu}$, encapsulating the effects of noise arising from the finite number of simulations used to compute the numerical derivatives. 

When we restrict to the case of a single target parameter $\theta^{\mu}$ (most commonly $\theta^{\mu}=\M$) so that $1D$ marginalized constraints may be measured, this noise term $\mathscr{F}^{Noise}_{\M \M}$ quantifies the difference between the inverse squares of the true underlying cosmological constraint, $\sigma^{-2}_{\mu}$, and the measured noise influenced constraint $\bar{\sigma}^{-2}_{\mu}$, as shown in (\ref{eq:TCP}). We find for all scenarios considered that the expectation value, $\left<\mathscr{F}^{Noise}_{\mu \mu}\right>$, is positive and can be large enough to provide the dominant contribution to the forecasted $\bar{\sigma}^{-2}_{\mu}$. This significantly biases $\bar{\sigma}_{\mu}$ below the true $\sigma_{\mu}$ leading to an artificially tighter constraint than the true cosmological one. In intermediate regimes, in which noise is significant but not dominant, we propose a method to estimate the $68\%$ C.L. for the true cosmological $\sigma_{\mu}^{-2}$, given in (\ref{eq:trueConstApprox}). 

For constraints on $\M$, we find that the level of statistical noise is heavily dependent on the exact $\M$ forward-differentiation scheme selected. While higher-order differentiation schemes such as $D2_{a}$ (\ref{eq:f2_pp_ppp}) or $D3$ (\ref{eq:f3}) may provide more systematic accuracy in the limit of $N_{sims} \rightarrow \infty$, for a finite number of simulations they experience larger amounts of statistical noise and therefore require a higher value of $N_{sims}$ before their respective constraints can be considered to not be significantly influenced by noise. For the Quijote simulations, for which $N_{sims}=500$, we find this ill-convergence is a problem even for commonly considered statistics such as $P(k)$ calculated from the dark matter halos. We also find that for differentiation schemes of the same order, such as $D1_{a}$, $D1_{b}$, and $D1_{c}$, the finite step size used in the differentiation scheme plays a similar role in determining the levels of statistical noise within the measured constraint as does the exact order of the differentiation scheme. We emphasize that for $\M$, both the order of the differentiation schemes and the finite $\Delta \M$ step size are what contribute to the level of noise contamination in the associated $\bar{\sigma}^{-2}_{\M}$, rather than the fact that the differentiation scheme is only one-sided.

Transitioning our analysis to more observationally relevant mock galaxies, we find that when HOD parameters are included in the marginalization process, the number of simulations needed to calculate derivatives with sufficient accuracy to avoid noise contamination of the constraints increases dramatically. This arises because the probability distribution for $\mathscr{F}^{Noise}_{\mu \mu}$ is only modestly affected by the additional parameters (if their derivatives are not a new source of significant noise), while the true value of $\sigma^{-2}_{\mu}$ is markedly reduced due to intrinsic degeneracies introduced by the inclusion of new parameters. The decrease to the true cosmological constraint, $\sigma^{-2}_{\mu}$, without an accompanying decrease to the associated noise floor, can cause a simulation-derived constraint, $\bar{\sigma}^{-2}_{\mu}$, that had exceeded the noise floor prior to the inclusion of the additional nuisance parameters to become comparable with the $\mathscr{F}^{Noise}_{\mu \mu}$ noise floor post inclusion. It is interesting to note that in cases where the measured $\bar{\sigma}^{-2}_{\mu}$ was noise-dominated prior to the introduction of the HOD parameters, their addition into the marginalization has little impact on the inferred constraints. Without consideration of the noise, this could be incorrectly inferred to mean that these cosmological constraints are immune to marginalization over additional nuisance parameters! In these cases, the measured $\bar{\sigma}^{-2}_{\mu}$ is already dominated by noise with the true cosmological constraints contributing negligibly, and thus any changes to $\bar{\sigma}^{-2}_{\mu}$ are primarily due to changes in the $p\left(\mathscr{F}^{Noise}_{\mu \mu}\right)$ distribution, which are typically small for the observables and parameters we have considered here.

When we extend our analysis to non-$\M$ parameters, we find that the choice of $\M$ differentiation scheme significantly influences the results for non-$\M$ parameters such as $\s8$ or $\Om$ through the marginalization process. In the noise-free limit, there are inherent degeneracies between $\M$, $\Om$ and $\s8$ that dilute the constraints on each individual parameter. By contrast, when a noise-prone differentiation scheme, such as $D3$, is used for $\M$, the tightening of the $\M$ constraints due to the noise is the dominant effect. It acts to break degeneracies between the parameters and can be so pronounced that it is equivalent to obtaining $\s8$ or $\Om$ parameter constraints in the limit that $\M$ is assumed to be perfectly known, with a fixed value and therefore entirely excluded from the marginalization. We show through considering a correlation function that this same phenomena is also responsible for reducing the magnitude of the measured parameter correlations $\overline{\mathrm{Corr}}\left(\theta^{\mu}, \theta^{\nu} \right)$, as demonstrated in Fig.~\ref{fig:CorrPlot}, and implicitly shown by the eccentricity of the confidence ellipses in Fig.~\ref{fig:partHaloDerivEllipsesBoth}

This analysis provides crucial insight into understanding how statistical noise may artificially influence and limit complex inter-dependencies within parameter estimation processes. Our findings therefore underscore the critical need for careful consideration of the effects of statistical noise in simulation-based forecasts. If overlooked, these effects can introduce significant biases and result in overly optimistic forecasts through the artificial tightening of parameter constraints, thereby exaggerating the precision of the forecast. The methodology developed in this work enables researchers to quantify the number of simulations required to effectively overcome this noise contamination inherent in numerical derivatives. This quantification is crucial, as the requisite $N_{sims}$ varies significantly depending on the specific observables and cosmological parameters under study both in the target and non-target sets. By providing a framework to estimate these needs, our approach helps ensure that simulation resources are allocated efficiently, maximizing the fidelity of the resulting cosmological inferences, and ensuring that Fisher forecasts remain reliable and reflective of the true uncertainties in cosmological models.

Forecasts using simulations are becoming a common method employed to evaluate survey strategies and cosmological capabilities, making our findings and our methods to mitigate noise-induced biases both timely and pertinent for key cosmological surveys such as DESI, Euclid, LSST, Roman Space Telescope, and SPHEREx.

Our work was motivated by the effects of selecting between the various neutrino mass differentiation schemes, (\ref{eq:s1_ppp}) - (\ref{eq:f3}), and the findings imply parallel considerations for the measurement of, and constraints on, other parameters for which higher order forward differentiation schemes are commonly used, such as the non-Gaussianity parameter $f_{\mathrm{NL}}$ and the inflationary scalar-to-tensor ratio $r$. Our results are completely general, however, and the sensitivity to both the order of the differentiation scheme and step size means that careful consideration of derivative noise should be part of simulation-based Fisher forecast analyses for any cosmological parameters.

Finally, the use of large scale cosmological simulation suites, designed to model realistic galaxy samples and survey characteristics for forthcoming observational campaigns, extends beyond Fisher forecasts. These simulations also underpin emulators used in Monte Carlo Markov Chain (MCMC) analyses, comparing simulated theories to real data, and in training machine learning tools for model inference. Characterizing the potential effects of statistical noise within these simulation-derived tools will also be crucial to exploring their effectiveness and potential limitations given the extensive computational resources which underlie their foundations, and should be the subject of future work. This consideration will enhance the robustness of both predictive models and the empirical analyses they support.

\begin{acknowledgments}

The work of CW and RB is supported by NSF grant AST-2206088,  NASA ATP grant 80NSSC18K0695, and NASA ROSES grant 12-EUCLID12-0004. CW would like to thank Tom Loredo for useful discussions related to analytic marginalization and the profile likelihood, and John Drew Wilson for useful discussions throughout, especially related to Appendix \ref{sec:NumConvergence}.

\end{acknowledgments}

\appendix
\section*{APPENDIX}
In Appendix~\ref{sec:noiseLimits} we consider two physically relevant limits of the derivative noise distribution. In Appendix ~\ref{sec:NumConvergence} we outline an analytic approach for estimating the true probability distribution $p^{True}$ and assessing the quality of agreement of our different approximation schemes, (\ref{eq:meanDeriv}) and (\ref{eq:degenDeriv}). 

\section{Useful Physical Limits for $\mathscr{F}^{Noise}_{\mu \nu}$ }
\label{sec:noiseLimits}

In this appendix, we explore two distinct physical limits relevant for calculating the probability distribution, $p\left(\mathscr{F}^{Noise}_{\mu \nu}\right)$, of the statistical noise in the Fisher matrix, $\mathscr{F}^{Noise}_{\mu \nu}$, for target parameters $\theta^\mu$ and $\theta^\nu$. This noise term arises from the process of calculating finite difference derivatives of observables with respect to the forecast parameters from a finite number of simulations, and characterizes the difference between the noise influenced $\overline{\mathscr{F}}_{\mu \nu}$ and the noise free $\mathscr{F}_{\mu \nu}$. We specifically focus on the expectation and variance as the key moments for this noise distribution. For all expressions, we will include the general case for arbitrary parameters $\theta^{\mu}$ and $\theta^{\nu}$ as well as the diagonal case of the most relevance to the main text with explicitly $\theta^{\mu} = \theta^{\nu} = \M$.

We first provide some integral equations from the main text that provide relevant context without having to look back through the paper. Defining the ideal, noise-free derivative of the observable $\sO^{\ i}$ with respect to parameter $\theta^A$, as  $d\sO^{\ i}_{A} \equiv \frac{\partial \sO^{\ i}}{\partial \theta^A}$,  we can express the derivative obtained from simulations, $d\overline{\sO}{}^{\ i}_{A}$, in terms of $d\sO^{\ i}_{A}$ (eq.(\ref{eq:derivExpansion}) in the main text) as
\begin{equation}
d\overline{\sO}{}^{\ i}_{A} =  d \sO^{\ i}_A + u^{i}_{A}.
\label{eq:app_derivExpansion}
\end{equation}
Here, $u^{i}_{A}$ is a random instance of derivative noise drawn from a Gaussian distribution with zero mean and variance based on the covariance of the mean (eq.(\ref{eq:Sigma}) in the main text),
\begin{eqnarray}
    \Sigma^{ij}_{AB} &=&\frac{1}{N_{sims}^2} \sum_{n=1}^{N_{sims}}\left(\frac{\partial \sO^{\ i}_{n}}{\partial \theta^A}- d\overline{\sO}{}^{\ i}_{A}\right) \left(\frac{\partial \sO^{\ j}_{n}}{\partial \theta^B}- d\overline{\sO}{}^{\ j}_{B}\right).
    \label{eq:app_Sigma}
\end{eqnarray}
The contribution from the derivative noise to the marginalized Fisher matrix for the target parameters can then be written as (eq. (\ref{eq:margFisherNoise}) in the main text),
\begin{eqnarray}
\label{eq:app_margFisherNoise}
\mathscr{F}^{Noise}_{\mu \nu} &\equiv& \overline{\mathscr{F}}_{\mu \nu} -  \mathscr{F}_{\mu \nu} \nonumber \\
& =& U_{\mu \nu} + F_{\mu a} \left(f^{-1}\right)^{ab}F_{b \nu}  \nonumber \\
  && -\left({F}_{\mu a} + U_{\mu a} \right) \left(\left(({f} + U)^{-1}\right)^{ab}\right) \left({F}_{b \nu} + U_{b \nu} \right).  \ \ \ 
\end{eqnarray}
where $\{\mu,\nu\}$ indicate target parameters and Latin indices from the set $\{a,b,c,d,e,g\}$, here and below, are used for the set of non-target parameters which are marginalized over. Latin indices from the set $\{i,j,k,l\}$ are used to indicate observable indices. We emphasize that indices are only summed on when a matching upper and lower index both appear. Otherwise, matching pairs of lower indices are not summed over, and simply function as labels.

In the diagonal case, the mean and variance of the noise contribution to the marginalized fisher, $\mathscr{F}^{Noise}_{\mu \mu}$, for a specific target parameter, $\theta^\mu$, can then be used to estimate the uncertainty on the true cosmological constraint, $\sigma_{\mu}^{-2}$, on $\theta^\mu$ based on that measured from the simulations, $\bar{\sigma}_{\mu}^{-2}$, (eq. (\ref{eq:trueConstApprox}) in the main text),
\begin{equation}
\sigma_{\mu}^{-2} = \left(\bar{\sigma}_{\mu}^{-2} - \left< \mathscr{F}^{Noise}_{\mu \mu} \right>\right) \pm \sqrt{\mathrm{Var\left( \mathscr{F}^{Noise}_{\mu \mu} \right)}}.
\label{eq:app_trueConstApprox}
\end{equation}

\subsection*{Limit $\# 1$: Isolated Noise}

We first consider a simple but highly applicable approximation, the isolated noise limit, which is particularly relevant when $D3$ or $D2_{a}$ is used for $\theta^{\mu}=M_\nu$ differentiation. 

In this limit, we assume that noise predominantly affects only the target parameter derivatives (denoted with Greek indices), meaning for the non-target parameters, $u^{i}_{a} =0$.  This greatly simplifies (\ref{eq:app_margFisherNoise}), so that for general target parameters $\theta^{\mu}$ and $\theta^{\nu}$, we have 
\begin{equation}
\left<\mathscr{F}^{Noise}_{\mu \nu}\right>\big\vert_{u^{i}_{a}\rightarrow 0} = \Sigma^{ij}_{\mu \nu}C^{-1}_{ij} - d{\sO}^{\ i}_{a} C^{-1}_{ij} \Sigma^{jk}_{\mu \nu} C^{-1}_{kl} d{\sO}^{l}_{b} ({f^{-1}})^{ab} 
\label{eq:INexpMixed}
\end{equation}
and in the diagonal case of $\theta^{\mu} = \theta^{\nu} = \M$, we have
\begin{equation}
\left<\mathscr{F}^{Noise}_{\M \M}\right>\big\vert_{u^{i}_{a}\rightarrow 0} = \Sigma^{ij}_{\M \M}C^{-1}_{ij} - d{\sO}^{\ i}_{a} C^{-1}_{ij} \Sigma^{jk}_{\M \M} C^{-1}_{kl} d{\sO}^{l}_{b} ({f^{-1}})^{ab}.
\label{eq:app_INexp}
\end{equation}
where $\Sigma^{ij}_{\mu\nu}$ is defined in (\ref{eq:app_Sigma}).

The isolated noise limit also makes calculating $\mathrm{Var}\left(\mathscr{F}^{Noise}_{\mu \nu}\right)$ analytically tractable. We start by defining the auxiliary variable:
\begin{equation}
Q_{ij} = \big(C^{-1}_{ij} - C^{-1}_{ik} d\sO^{k}_{a} ({f^{-1}})^{ab} d\sO^{l}_{b} C^{-1}_{lj}\big)
\label{eq:app_defQ}
\end{equation}
so that in the isolated noise limit, for general parameters $\theta^{\mu}$ and $\theta^{\nu}$ we may compactly write 
\begin{equation}
\begin{split}
\mathrm{Var}\left(\mathscr{F}^{Noise}_{\mu \nu} \right)\big\vert_{u^{i}_{a}\rightarrow 0} &= \left(d\sO^{i}_{\mu} d\sO^{j}_{\mu} + \frac{1}{2}\Sigma^{ij}_{\mu \mu}\right) \left(Q_{ik} \Sigma^{kl}_{\nu \nu} Q_{lj} \right) \\
&+\left(2d\sO^{i}_{\mu} d\sO^{j}_{\nu} + \Sigma^{ij}_{\mu \nu}\right) \left(Q_{ik} \Sigma^{kl}_{\nu \mu} Q_{lj} \right) \\
&+\left(d\sO^{i}_{\nu} d\sO^{j}_{\nu} + \frac{1}{2}\Sigma^{ij}_{\nu \nu}\right) \left(Q_{ik} \Sigma^{kl}_{\mu \mu} Q_{lj} \right).
\label{eq:INvarMixed}
\end{split}
\end{equation}
For the diagonal case with $\theta^{\mu} = \theta^{\nu} = \M$ this gives simply
\begin{equation}
\begin{split}
\mathrm{Var}\left(\mathscr{F}^{Noise}_{\M \M}\right)\big\vert_{u^{i}_{a}\rightarrow 0} &= \left(4d\sO^{i}_{\M} d\sO^{j}_{\M} + 2\Sigma^{ij}_{\M \M}\right) \left(Q_{ik} \Sigma^{kl}_{\M \M} Q_{lj} \right).
\label{eq:app_INvar}
\end{split}
\end{equation}
Together, these expressions for the mean and variance constitute the isolated noise Gaussian limit for the general case of $\mathscr{F}^{Noise}_{\mu \nu}$ through equations (\ref{eq:INexpMixed}) and (\ref{eq:INvarMixed}), and for our specific case study of $\mathscr{F}^{Noise}_{\M \M}$ through equations (\ref{eq:app_INexp}) and (\ref{eq:app_INvar}) and are frequently referred to in the main text due to the relative simplicity of the above equations.

\subsection*{Limit $\# 2$: Large $N_{sims}$}

As discussed in the main text, there are some scenarios in which the isolated noise scenario does not serve as a good approximation, for example when using the $D1_{a}$ derivatives for $\M$. In this case, we can consider an alternative, less simple but still tractable, approximation which provides an analytic estimate for moments of $\mathscr{F}^{Noise}_{\mu \nu}$ in the limit of large $N_{sims}$.

In this limit, we assume that all of the $u^{i}_{A}$ terms are small and can thus be treated perturbatively, which makes calculating the expectation value of $\mathscr{F}^{Noise}_{\mu \nu}$ analytically tractable. Since $\Sigma^{ij}_{AB} \rightarrow 0$ as $N_{sims} \rightarrow \infty$, we may treat $\Sigma^{ij}_{AB}$ as our small parameter when series expanding (\ref{eq:app_margFisherNoise}). Notably, expanding in $\Sigma^{ij}_{AB}$ is equivalent to expanding in $1/N_{sims}$ since all entries of $\Sigma^{ij}_{AB}$ explicitly go as $1/N_{sims}$ as $N_{sims} \rightarrow \infty$, providing a direct link between the number of simulations and the perturbation expansion. For general target parameters $\theta^{\mu}$ and $\theta^{\nu}$, we have to lowest order in $\Sigma$, that:
\begin{align}
\label{eq:1/Nsims_munu}
\left<\mathscr{F}^{Noise}_{\mu \nu} \right>
_{\Sigma^{1}}
&= \Sigma^{ij}_{\mu \nu}C^{-1}_{ij} -  {F}_{\mu a} ({f^{-1}})^{ab} \Sigma^{ij}_{b \nu} C^{-1}_{ij} \nonumber \\
& - \Sigma^{ij}_{\mu a} C^{-1}_{ij}({f^{-1}})^{ab}  {F}_{b \nu}  \nonumber \\
&- d{\sO}^{i}_{a} C^{-1}_{ij} \Sigma^{jk}_{\mu b} C^{-1}_{kl} d{\sO}^{l}_{\nu} ({f^{-1}})^{ab} \nonumber \\
&- d{\sO}^{i}_{a} C^{-1}_{ij} \Sigma^{jk}_{\mu \nu } C^{-1}_{kl} d{\sO}^{l}_{b} ({f^{-1}})^{ab} \nonumber \\
&- d{\sO}^{i}_{\mu} C^{-1}_{ij} \Sigma^{jk}_{a b} C^{-1}_{kl} d{\sO}^{l}_{\nu} ({f^{-1}})^{ab} \nonumber \\
&- d{\sO}^{i}_{\mu} C^{-1}_{ij} \Sigma^{jk}_{a \nu} C^{-1}_{kl} d{\sO}^{l}_{b} ({f^{-1}})^{ab} \nonumber \\
&+ {F}_{\mu a}({f^{-1}})^{ab}\left(\Sigma^{ij}_{bc}C^{-1}_{ij}\right)({f^{-1}})^{cd}{F}_{d\nu} \nonumber \\
&+ {F}_{\mu a}({f^{-1}})^{ab}({f^{-1}})^{cd} C^{-1}_{ij} C^{-1}_{kl} \nonumber \\
&\times \Big[ d{\sO}^{i}_{c} \Sigma^{jk}_{bd} d{\sO}^{l}_{\nu} + d{\sO}^{i}_{d} \Sigma^{jk}_{b \nu} d{\sO}^{l}_{c} \nonumber \\
&+ d{\sO}^{i}_{\nu} \Sigma^{jk}_{cd} d{\sO}^{l}_{b} + d{\sO}^{i}_{b} \Sigma^{jk}_{c \nu} d{\sO}^{l}_{d} \Big] \nonumber \\
&+ {F}_{\nu a}({f^{-1}})^{ab}({f^{-1}})^{cd} C^{-1}_{ij} C^{-1}_{kl} \nonumber \\
&\times \Big[ d{\sO}^{i}_{c} \Sigma^{jk}_{bd} d{\sO}^{l}_{\mu} + d{\sO}^{i}_{d} \Sigma^{jk}_{b \mu} d{\sO}^{l}_{c} \nonumber \\
&+ d{\sO}^{i}_{\mu} \Sigma^{jk}_{cd} d{\sO}^{l}_{b} + d{\sO}^{i}_{b} \Sigma^{jk}_{c \mu} d{\sO}^{l}_{d} \Big] \nonumber \\
&-{F}_{\mu a}({f^{-1}})^{ab}({f^{-1}})^{cd} ({f^{-1}})^{eg} {F}_{g \nu} C^{-1}_{ij} C^{-1}_{kl}\nonumber \\
&\times \Big[d{\sO}^{i}_{b} \Sigma^{jk}_{ce} d{\sO}^{l}_{d} + 
d{\sO}^{i}_{b} \Sigma^{jk}_{cd} d{\sO}^{l}_{e} \nonumber \\
&+ d{\sO}^{i}_{c} \Sigma^{jk}_{be} d{\sO}^{l}_{d} +  d{\sO}^{i}_{c} \Sigma^{jk}_{bd} d{\sO}^{l}_{e} \Big].
\end{align}
where the Latin indices from the set $\{i,j,k,l\}$ are used to sum over the observables. 

This expression simplifies in the diagonal case where we set $\theta^{\mu} = \theta^{\nu} = \M$ for concreteness. To lowest order in $\Sigma$, we have that 
\begin{align}
\label{eq:1/Nsims}
\left<\mathscr{F}^{Noise}_{\M \M} \right>
_{\Sigma^{1}}&= \Sigma^{ij}_{\M \M}C^{-1}_{ij} - 2 {F}_{\M a} ({f^{-1}})^{ab} \Sigma^{ij}_{b \M} C^{-1}_{ij} \nonumber \\
&- 2 d{\sO}^{i}_{\M} C^{-1}_{ij} \Sigma^{jk}_{a \M} C^{-1}_{kl} d{\sO}^{l}_{b} ({f^{-1}})^{ab} \nonumber \\ 
&- d{\sO}^{i}_{a} C^{-1}_{ij} \Sigma^{jk}_{\M \M} C^{-1}_{kl} d{\sO}^{l}_{b} ({f^{-1}})^{ab} \nonumber \\
&- d{\sO}^{i}_{\M} C^{-1}_{ij} \Sigma^{jk}_{ab} C^{-1}_{kl} d{\sO}^{l}_{\M} ({f^{-1}})^{ab} \nonumber \\
&+ {F}_{\M a}({f^{-1}})^{ab}\left(\Sigma^{ij}_{bc}C^{-1}_{ij}\right)({f^{-1}})^{cd}{F}_{d\M} \nonumber \\
&+ 2{F}_{\M a}({f^{-1}})^{ab}({f^{-1}})^{cd} C^{-1}_{ij} C^{-1}_{kl} \nonumber \\
&\times \Big[ d{\sO}^{i}_{c} \Sigma^{jk}_{bd} d{\sO}^{l}_{\M} + d{\sO}^{i}_{d} \Sigma^{jk}_{b \M} d{\sO}^{l}_{c} \nonumber \\
&+ d{\sO}^{i}_{\M} \Sigma^{jk}_{cd} d{\sO}^{l}_{b} + d{\sO}^{i}_{b} \Sigma^{jk}_{c \M} d{\sO}^{l}_{d} \Big] \nonumber \\
&-{F}_{\M a}({f^{-1}})^{ab}({f^{-1}})^{cd} ({f^{-1}})^{eg} {F}_{g \M} C^{-1}_{ij} C^{-1}_{kl}\nonumber \\
&\times \Big[2 d{\sO}^{i}_{b} \Sigma^{jk}_{ce} d{\sO}^{l}_{d} + 
d{\sO}^{i}_{b} \Sigma^{jk}_{cd} d{\sO}^{l}_{e} \nonumber \\
&+ d{\sO}^{i}_{c} \Sigma^{jk}_{be} d{\sO}^{l}_{d} \Big]
\end{align}
This expression may be used to calculate the expected level of noise to lowest order in $1/N_{sims}$ present in the measured value of $\bar{\sigma}^{-2}_{\M}$ and is compared to other approaches in Fig.~\ref{fig:NoiseBothApprox}.

\section{Noise Distribution Convergence Tests}
\label{sec:NumConvergence}

In this appendix, we identify when and how we can actually consider our diagonal noise distributions, $p\left(\mathscr{F}^{Noise}_{\mu \mu} \right)$, as calculated with either Approx I: $F_{AB} = \overline{F}_{AB}$, or approx II: $F_{AB} = F^{degen}_{AB}$, as representative of the true noise distribution, calculated with the true $N_{sims} \rightarrow \infty$ values of the derivatives $d\sO^{i}_{A}$. As in the main text, we refer to three distributions: 
\bea
p^{True}&\equiv& p\left(\mathscr{F}^{Noise}_{\mu \mu} \vert\ d\sO \right) \label{eq:app_p} \\
\bar{p} &\equiv&  p\left(\mathscr{F}^{Noise}_{\mu \mu} \vert\ d\overline{\sO} \right) \label{eq:app_pbar}\\
p^{degen} &\equiv&  p\left(\mathscr{F}^{Noise}_{\mu \mu} \vert\ d\sO^{degen} \right) \label{eq:app_pdegen}
\eea
where $p^{True}$ refers to the true distribution, $\bar{p}$ refers to the distribution calculated with Approx. I, and $p^{degen}$ refers to the calculation with Approx. II. We find that in practice, all of the noise distributions are highly Gaussian, and thus we may ask the simplified question of when $\left<\bar{p}\right>$ and $\mathrm{Var}\left( \bar{p}\right)$ or $\left<p^{degen}\right>$ and $\mathrm{Var}\left( p^{degen}\right)$ are representative of the true $\left<p^{True}\right>$ and $\mathrm{Var}\left(p^{True}\right)$. 

We begin with the simpler of the two approximations, Approx I: $F_{AB} = \overline{F}_{AB}$. In accordance with (\ref{eq:app_derivExpansion}), we assume that the mean derivatives we have from the simulations differ from the true derivatives by a single unknown instance of the random variable $u^{i}_{A}$. This lets us write 
\begin{equation}
\bar{p} = p\left(\mathscr{F}^{Noise}_{\mu \mu} \vert\ d\sO^{i}_{A} + \tilde{u}^{i}_{A} \right)
\end{equation}
where $\tilde{u}^{i}_{A}$ is the \textit{actual} instance of noise drawn from the multivariate Gaussian distribution characterized by $\Sigma^{ij}_{AB}$ which separates the true derivatives $d\sO^{\ i}_{A}$ and the derivatives $d\overline{\sO}{}^{\ i}_{A}$ as measured from the simulations. Ideally, we would like to determine the exact effect that $\tilde{u}^{i}_{A}$ had in order to turn $p^{True}$ into $\bar{p}$ so that we could reverse it, or at the very least determine if $\bar{p}$ is representative of $p^{True}$ in terms of mean and variance so that we can reliably use (\ref{eq:app_trueConstApprox}). 

While a precise reversal is impossible for an individual instance of noise, we can consider this in statistical context, and reverse the average effects of noise on $\bar{p}$ to estimate a hypothetical $p^{True}$. We do this by first adding \textit{additional} instances of noise to $d\overline{\sO}{}^{\ i}_{A}$. Explicitly, we define the ``$n$ times barred" noise distribution $\overline{p}^{[n]}$ as 
\bea
\left.d\overline{\sO}^{[n]}\right.^{i}_{A}&
\equiv& d\sO^{\ i}_{A} + \sqrt{n} u^{i}_{A} \label{eq:dO_Xbar}\\
\overline{p}^{[n]}  &\equiv&   p\bigg(\mathscr{F}^{Noise}_{\mu \mu} \vert\ d\overline{\sO}^{[n]} \bigg)  \label{eq:p_X}
\eea
where $u^{i}_{A}$ is sampled from the multivariate Gaussian distribution characterized by $\Sigma^{ij}_{AB}$. The $\sqrt{n}$ term multiplying $u^{i}_{A}$ ensures the variance of the new noise term is $n$ times that of a single instance. Since $u^{i}_{A}$ is a multivariate Gaussian random variable, $n$ independent instances of $u^{i}_{A}$ added together is statistically equivalent of one instance of $u^{i}_{A}$ multiplied by $\sqrt{n}$. This allows for a continuous $n \in \mathbb{R}_{\geq 0}$ rather than just over the set of positive integers. 

From (\ref{eq:p}) and (\ref{eq:p_X}), $p^{True} \equiv \overline{p}^{[0]}$, and from (\ref{eq:pbar}), $\bar{p}$ is one particular instance of 
$\overline{p}^{[1]}$ if the instance of $u^{i}_{A}$ drawn in defining $\overline{p}^{[1]}$ was equal to $\tilde{u}^{i}_{A}$ 

For fixed $n$, we may calculate a different instance of the distribution $\overline{p}^{[n]}$ for each instance of the random variable $u^{i}_{A}$, so that a \textit{population} of slightly different $\overline{p}^{[n]}$ probability distributions may be constructed each with its own mean and variance. Computing the expectation values of these distributions of means and variances then will give us the average mean, $\mu(n)$ and the average variance, $\sigma^2(n)$, across the set of distributions
$\left\{\overline{p}^{[n]}\right\}$ for a given $n$. Both of these quantities, $\mu(n)$ and $\sigma^2(n)$, are continuous and differentiable functions in $n$ and thus may be Taylor expanded. Given that $\mu(0)= \left<p^{True}\right>$ and $\sigma^{2}(0) = \mathrm{Var}\left( p^{True}\right)$, so that through this approach, the goal of estimating  $p^{True}$ may be reformulated as estimating  $\mu(0)= \left<p^{True}\right>$ and $\sigma^{2}(0) = \mathrm{Var}\left( p^{True}\right)$ from extrapolating $\{\mu(n)$\} and $\{\sigma^{2}(n)\}$ for $n>0$ .  

As an example, we may perform this process analytically in the limit of isolated noise, where the noise in the non-target parameters is negligible, $u^{i}_{a} = 0$, by assumption. In this limit, a random instance of the probability distribution $\overline{p}^{[n]}$ has as a mean 
\begin{align}
\left<\overline{p}^{[n]}\big\vert_{u^{i}_{a}\rightarrow 0}\right> &=\left<p^{True}\right>\label{eq:ExppXbar}
\end{align}
due to the fact that (\ref{eq:app_INexp}) is independent of $d\sO^{\ i}_{\mu}$ (substituting out the $\M$ index in (\ref{eq:app_INexp}) for a general target parameter $\theta^{\mu}$), and has as a variance
\begin{align}
\mathrm{Var}\left(\overline{p}^{[n]}\right)\big\vert_{u^{i}_{a}\rightarrow 0} &= \mathrm{Var}\left(p^{True}\right)\big\vert_{u^{i}_{a}\rightarrow 0} \\
&+\left(4 n u_{\mu}^{i} u_{\mu}^{j} +8 \sqrt{n} u_{\mu}^{i} d\sO^{j}_{\mu}\right)\left(Q_{ik} \Sigma^{kl}_{\mu \mu} Q_{lj} \right).\label{eq:VarpXbar}
\end{align}
Taking the expectation value of the above equations across all possible $\overline{p}^{[n]}$ probability distributions simply amounts to an integral over the multivariate Gaussian distribution used to define $u^{i}_{\mu}$, so that in the limit of isolated noise, we have 
\begin{align}
\mu(n)\big\vert_{u^{i}_{a}\rightarrow 0}&=\left<p^{True}\right>\label{eq:ExpExppXbar}
\end{align}
and
\begin{align}
\sigma^{2}(n)
\vert_{u^{i}_{a}\rightarrow 0}= \mathrm{Var}\left(p^{True}\right)\big\vert_{u^{i}_{a}\rightarrow 0} +4 n\left(\Sigma^{ij}_{\mu \mu} Q_{ik} \Sigma^{kl}_{\mu \mu} Q_{lj} \right) \label{eq:ExpVarpXbar}
\end{align}
which are a constant and a linear function of $n$ respectively. While these expressions become more complicated outside of this limit, all of the above quantities may still be calculated numerically via sampling.

There is one small issue with the above approach, and that is that in general, we cannot calculate $d\overline{\sO}^{[n]}$ in an unbiased manner due to the baked-in single instance of noise $\tilde{u}^{i}_{A}$. However, we can modify the above approach to deal with this fact by using 
\begin{equation}
\left.d\overline{\sO}^{[n]}\right.^{i}_{A}\approx d\sO^{\ i}_{A} +\tilde{u}^{i}_{A}+ \sqrt{n-1} u^{i}_{A} \label{eq:dO_XbarApprox}
\end{equation}
When (\ref{eq:dO_XbarApprox}) is used for $d\overline{\sO}^{[n]}$ instead of (\ref{eq:dO_Xbar}), it is in general no longer exactly true, but only approximately true, that $\mu(0) \simeq \left<p^{True} \right>$ and $\sigma^{2}(0) \simeq \mathrm{Var}\left(p^{True} \right)$. For example, when using (\ref{eq:dO_XbarApprox}) instead of (\ref{eq:dO_Xbar}) in the limit of isolated noise, \ref{eq:ExpVarpXbar}) becomes 
\begin{align}
\sigma^{2}(n)\vert_{u^{i}_{a}\rightarrow 0}&= \mathrm{Var}\left(p^{True}\right)\big\vert_{u^{i}_{a}\rightarrow 0} \nonumber \\
&+4\left((n-1)\Sigma^{ij}_{\mu \mu} + \tilde{u}^{i}_{\mu} \tilde{u}^{j}_{\mu} + 2\tilde{u}^{i}_{\mu} d\sO^{\ j}_{\mu}   \right)  \left(Q_{ik} \Sigma^{kl}_{\mu \mu} Q_{lj} \right) \label{eq:ExpVarpXbarApprox}
\end{align}
so that 
\begin{align}
\sigma^{2}(0)\vert_{u^{i}_{a}\rightarrow 0}&= \mathrm{Var}\left(p^{True}\right)\big\vert_{u^{i}_{a}\rightarrow 0} \nonumber \\
&+4\left(\tilde{u}^{i}_{\mu} \tilde{u}^{j}_{\mu} - \Sigma^{ij}_{\mu \mu} + 2\tilde{u}^{i}_{\mu} d\sO^{\ j}_{\mu} \right)  \left(Q_{ik} \Sigma^{kl}_{\mu \mu} Q_{lj} \right) \label{eq:ExpVarpXbarApprox0}
\end{align}
Here, we highlight that on average, $\left<\tilde{u}^{i}_{\mu} \tilde{u}^{j}_{\mu} \right> = \Sigma^{ij}_{\mu \mu}$ and $\left<\tilde{u}^{i}_{\mu} d\sO^{\ j}_{\mu}\right> = 0$ so that even when approximating (\ref{eq:dO_Xbar}) with (\ref{eq:dO_XbarApprox}), it is still the case that $\sigma^{2}(0)\vert_{u^{i}_{a}\rightarrow 0}\simeq \mathrm{Var}\left(p^{True}\right)\big\vert_{u^{i}_{a}\rightarrow 0}$.
Given this, the procedure for verifying numerical convergence is as follows: 
\begin{itemize}
\item Construct the distributions $\bar{p}$, using Approx. I, and $p^{degen}$, using Approx. II, via numerical sampling, as in \ref{sec:prac_imp}.

\item Construct populations of $\left\{ \overline{p}^{[n]} \right\}$ using  (\ref{eq:dO_XbarApprox}) and (\ref{eq:p_X}) across some set of $\{n\geq1\}$ values and from these, calculate $\mu(n)$ and $\sigma^{2}(n)$ for each $n$ in preparation to Taylor expand. Care must be taken when sampling and choosing finite difference step sizes in $n$ so that the derivatives used in the expansion are converged. We use $\Delta n = 0.25$, and require that our sets $\left\{ \overline{p}^{[n]} \right\}$ consist of 30000 distributions, each of which have been sampled 10000 times to determine an accurate measure of $\mu(n)$ and $\sigma^{2}(n)$ for each $n$.

\item Find the lowest ($k^{th}$) order Taylor expansions of $\mu(n)$ and $\sigma^{2}(n)$ centered at $n = 2$, with finite difference derivatives calculated using points at $n \geq 2$, which can reliably recover $\mu(1) = \left<\bar{p}\right>$ and $\sigma^{2}(1) =\mathrm{Var}\left( \bar{p}\right)$. In practice, we restrict ourselves to $k \leq 2$ order expansions at maximum to avoid over fitting. 

\item  Perform the same $k^{th}$ order Taylor expansion of $\mu(n)$ and $\sigma^{2}(n)$ centered at $n=1$ to approximate $\mu(0) = \left< p^{True} \right>$ and $\sigma^{2}(0) = \mathrm{Var}\left( p^{True} \right)$. 

\item
Compare the reconstructed Gaussian approximation for $p^{True}$, using $\mu(0)$ and $\sigma^{2}(0)$, to $\bar{p}$ and $p^{degen}$ to assess how representative they each are.

\end{itemize}

\begin{figure}[!t]
\includegraphics[width=1.0\linewidth]{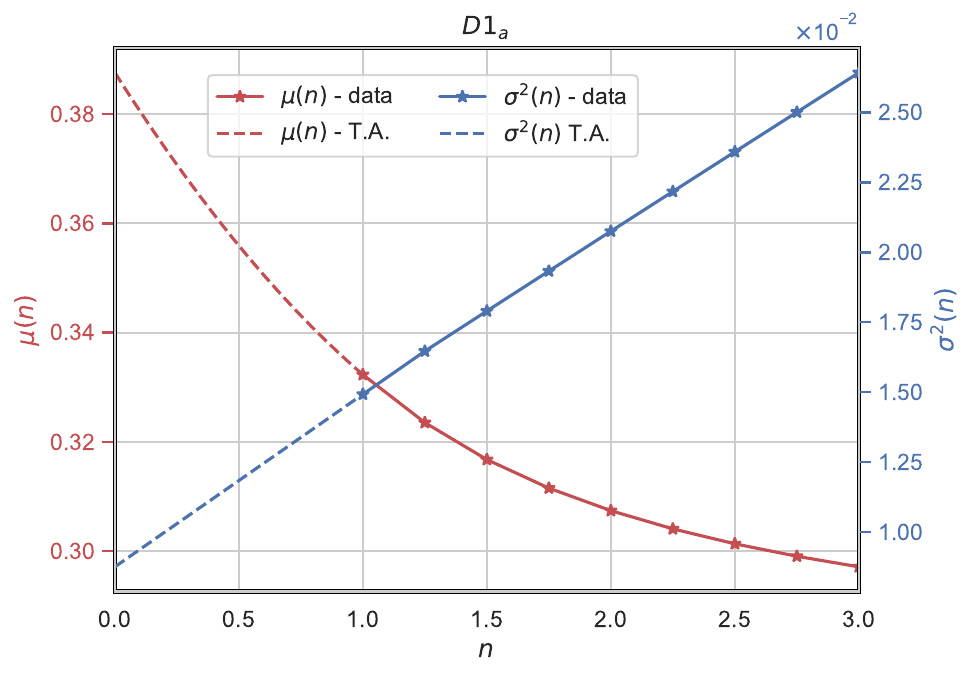}
\caption{Taylor expansion approach to estimating statistics for $p^{[0]}$ for $P_{halos}$ for $D1_a$ shown in Fig.~\ref{fig:appendixFigHalos}.  $\mu(n)$ [red, left axis] and $\sigma^{2}(n)$ [blue, right axis] are shown evaluated using (\ref{eq:dO_XbarApprox}) and (\ref{eq:p_X}) via numerical sampling from $n=1$ to $n=3$ with $\Delta n=0.25$, and extended down to $n=0$ via Taylor expansion. 
}
\label{fig:appendixFigTA}
\end{figure}

Figure~\ref{fig:appendixFigTA} shows the results of this process to estimate $\mu(0)$ and $\sigma^{2}(0)$ for target parameter $\M$ using the $D1_{a}$ differentiation scheme and observable $P_{halos}(k)$. In accordance with Fig.~\ref{fig:NoiseBothApprox}, we can see that $p\left(\mathscr{F}^{Noise}_{\M \M} \right)$ for this differentiation scheme and observable is not within the physical limit of isolated noise, and thus we should expect both $\mu(n)$ and $\sigma^{2}(n)$ to vary as a continuous function of $n$. 

We find a first-order and second-order Taylor expansion can be used to reliably reproduce the $\sigma^{2}(n)$ and $\mu(n)$ functions respectively, and use these to extrapolate to $n=0$,  calculating $\mu(0) \simeq 0.39 \mathrm{eV}^{-2}$, a small increase to the $\bar{p}$ value (Approx. I) of $\mu(1) \simeq 0.33 \mathrm{eV}^{-2}$, and $\sigma^{2}(0) = 8.8\times 10^{-3} \mathrm{eV}^{-2}$ instead of the $\bar{p}$ value of $\sigma^{2}(1) = 1.5 \times 10^{-2} \mathrm{eV}^{-2}$. If we use the hypothetical $p^{True}$ defined by these Taylor expanded quantities to modify the confidence limit prediction of $\sigma^{-2}_{\M,D1_{a}}$ from (\ref{eq:app_trueConstApprox}), we find now that $\sigma^{-2}_{\M,D1_{a}} = 0.72 \pm 0.09 \mathrm{eV}^{-2}$, which is slightly different than the $\bar{p}$ interval of $\sigma^{-2}_{\M,D1_{a}} = 0.77 \pm 0.12 \mathrm{eV}^{-2}$ quoted in Table~\ref{tab:NoiseBothApprox} of the main text. 

\begin{figure*}[!t]
\includegraphics[width=0.99\linewidth]{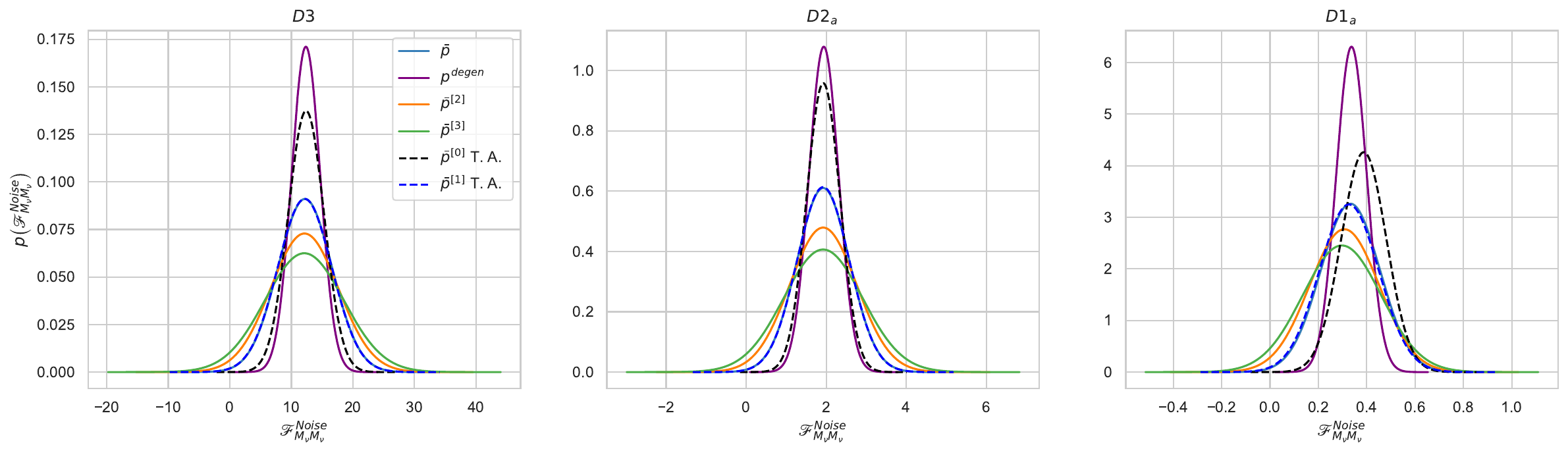}
\caption{Gaussian approximations for $\bar{p}$ [light blue, solid], and $p^{degen}$ [purple, solid], the average $\overline{p}^{[2]}$ [orange, solid], the average $\overline{p}^{[3]}$ [green, solid] and the $1^{st}$ order Taylor approximations for $\overline{p}^{[1]} \simeq \bar{p}$ [blue, dashed] and $\overline{p}^{[0]} \simeq p^{True}$ [black, dashed] using $P_{Halos}(k)$ as the observable for target parameter $\M$ with the $D3$ [left], $D2_{a}$ [center] or $D1_{a}$ [right] differentiation schemes.}
\label{fig:appendixFigHalos}
\end{figure*}

Figure~\ref{fig:appendixFigHalos} shows the results of applying this process to check all noise distributions previously shown in Fig.~\ref{fig:NoiseBothApprox}. Focusing first on the $D1_{a}$ differentiation scheme, we can see the combination of $\mu(n)$ and $\sigma^{2}(n)$ shown in Fig.~\ref{fig:appendixFigTA} represented as actual probability distributions for $n \in \{3,2,1,0\}$. As n decreases from 3 down to 0, the distributions become narrower with the mean slightly but steadily increasing, just as shown in Fig.~\ref{fig:appendixFigTA}. It is this small difference between $\bar{p}$ and our hypothetical $p^{True}=p^{[0]}$ which slightly modifies our confidence interval for $\sigma^{-2}_{\M,D1_{a}}$ as discussed above. 

For $D3$ and $D2_{a}$ differentiation schemes, we find that $p^{degen}$ is more representative of our hypothetical $p^{True}$ than $\bar{p}$. Since according to Fig.~\ref{fig:NoiseBothApprox} these noise distributions are all well within the physical limit of isolated noise, this can be understood analytically through (\ref{eq:ExpExppXbar}) and (\ref{eq:ExpVarpXbar}), where $\mu(n)$ is independent of $n$ and $\sigma^{2}(n)$ is linear in $n$. Thus, while $\left<\bar{p}\right>$ and $\left<p^{degen} \right>$ will both agree with $\mu(0) = \left< p^{True}\right>$, for the variances we have
\begin{align}
\mathrm{Var}\left( p^{degen}\right)\vert_{u^{i}_{a}\rightarrow 0} = \mathrm{Var}\left(p^{True}\right)\big\vert_{u^{i}_{a}\rightarrow 0} -2\left(d\sO^{i}_{\mu} d\sO^{j}_{\mu} Q_{ik} \Sigma^{kl}_{\nu \nu} Q_{lj} \right)
\end{align}
and (dropping subdominant terms linear in $\tilde{u}^{i}_{\mu}$)
\begin{align}
\mathrm{Var}\left(\bar{p}\right)\vert_{u^{i}_{a}\rightarrow 0} \simeq \mathrm{Var}\left(p^{True}\right)\big\vert_{u^{i}_{a}\rightarrow 0} +4\left(\tilde{u}^{i}_{\mu} \tilde{u}^{j}_{\mu}Q_{ik} \Sigma^{kl}_{\nu \nu} Q_{lj} \right)
\end{align}
so that now
\begin{equation}
\mathrm{Var}\left(p^{degen}\right)\vert_{u^{i}_{a}\rightarrow 0} < \mathrm{Var}\left(p^{True}\right)\vert_{u^{i}_{a}\rightarrow 0} \lesssim \mathrm{Var}\left(\bar{p}\right)\vert_{u^{i}_{a}\rightarrow 0}
\end{equation}
When noise within the target parameters is large, it is often the case that the term separating $\mathrm{Var}\left( p^{degen}\right)\vert_{u^{i}_{a}\rightarrow 0}$ and  $\mathrm{Var}\left(p^{True}\right)\vert_{u^{i}_{a}\rightarrow 0}$ is much smaller than that which separates $\mathrm{Var}\left(\bar{p}\right)\vert_{u^{i}_{a}\rightarrow 0}$ and $\mathrm{Var}\left(p^{True}\right)\big\vert_{u^{i}_{a}\rightarrow 0}$, so that $p^{degen}$ is more representative of $p^{True}$ than $\bar{p}$ is. For $D2_{a}$, this correspondence is almost exact between $p^{degen}$ and the reconstructed $p^{True}$. 

Finally, we note that even in cases when noise is isolated to a single parameter which is \textit{not} within the target parameter set, we find that $p^{degen}$ may still be highly representative of a hypothetical $p^{True}$ if the assumptions of (\ref{eq:degenDeriv}) are relaxed so that the \textit{noisiest} parameter derivative as defined by $\Sigma^{ij}_{A,A}C^{-1}_{ij} / \overline{F}_{A,A}$ is reconstructed as a degenerate linear combination of the others. 
This application of Approx. II to the noisiest parameter, rather than the target parameter, allows the degeneracy's between the true parameter derivatives to be more faithfully represented in scenarios in which noise is almost exclusively concentrated within a single non-target parameter. This is demonstrated in Fig.~\ref{fig:NonMnuParams} showing predictions for $p\left(\mathscr{F}^{Noise}_{\mu \mu}\right)$ on $\Om$ and $\s8$ when reconstructing the noisy D3 $\M$ derivative as a degenerate linear combination of the other parameters.

\newpage
\bibliographystyle{apsrev}
\bibliography{references}

\end{document}